\documentclass[a4paper,11pt]{article}
\pdfoutput=1 
\usepackage{jheppub} 
\usepackage[T1]{fontenc}
\usepackage{fontawesome}
\usepackage[table]{xcolor}
\usepackage[caption=false]{subfig}
\usepackage{epsfig}
\usepackage{blkarray}
\usepackage{colortbl}
\usepackage[T1]{fontenc} 
\usepackage{tikz-feynman}
\usepackage{hepnames,rotating}
\usepackage{adjustbox}
\usepackage{multirow}
\usepackage{enumitem}
\usepackage{slashed}
\usepackage{graphicx}
\usepackage{lipsum}
\usepackage{chngcntr}
\usepackage{bigints}
\usepackage[normalem]{ulem}
\usepackage{amsmath}
\usepackage{textcomp}
\usepackage{cleveref}
\usepackage{physics}
\usepackage{pdflscape}
\crefname{figure}{fig\,.}{figs\,.} 
\crefname{equation}{eq\,.}{eqs\,.} 
\usepackage[most]{tcolorbox}
\newtcbox{\mymath}[1][]{nobeforeafter, math upper, tcbox raise base, enhanced, colframe=blue!30!black, colback=blue!30, boxrule=1pt,  #1}
\crefname{figure}{fig\,.}{figs\,.} 
\crefname{equation}{eq\,.}{eqs\,.}
\newcommand{\stkout}[1]{\ifmmode\text{\sout{\ensuremath{#1}}}\else\sout{#1}\fi}
\definecolor{calpolypomonagreen}{rgb}{0.12, 0.3, 0.17}
\definecolor{gx11g}{rgb}{0.0, 1.0, 0.0}
\definecolor{airforceblue}{rgb}{0.36, 0.54, 0.66}
\DeclareSymbolFont{myletters}{OML}{ztmcm}{m}{it}
\DeclareMathSymbol{\uplambda}{\mathord}{myletters}{"15}
\newcommand{\bea}{\begin{eqnarray}}
\newcommand{\eea}{\end{eqnarray}}
\usepackage{xcolor}
\newcommand\ccg[1]{\cellcolor{blue!5!red!5!green!5}{#1}} 
\newcommand\ccw[1]{\cellcolor{blue!5}{#1}}

\usepackage[font=small]{caption}
\usepackage[font={small},labelfont={bf}]{caption}
\usepackage{url}
\definecolor{MyDarkBlue}{rgb}{0.1, 0.1, 0.8}
\definecolor{SBlue}{rgb}{0.2, 0.4, 0.7} 
\definecolor{MyLightBlue}{rgb}{0.22,0.51,0.9}
\definecolor{MyGreen}{rgb}{0.0, 0.5, 0.0}
\definecolor{BrickRed}{rgb}{0.8, 0.25, 0.33}
\usepackage{hyperref}
\hypersetup{colorlinks, citecolor=BrickRed,linkcolor=MyDarkBlue, urlcolor=MyGreen}
\title{Multiparticle scalar dark matter with $\mathcal{Z}_N$ symmetry}

\author[a]{Subhaditya Bhattacharya, }
\author[a]{Lipika Kolay, }
\author[a]{and Dipankar Pradhan.}

\affiliation[a]{Department of Physics, Indian Institute of Technology Guwahati,\\North Guwahati, Assam-781039, India,}

\emailAdd{subhab@iitg.ac.in}
\emailAdd{klipika@iitg.ac.in}
\emailAdd{d.pradhan@iitg.ac.in}
\abstract{More than one dark sector particle transforming under the same symmetry provides one stable 
dark matter (DM) component which undergoes co-annihilation with the heavier particle(s) decaying to DM. 
Specific assumptions on the kinematics and on the coupling parameters may render the heavier component(s) stable and 
contribute as DM. The choices of the charges of the dark sector fields under transformation
 play a crucial role in the resultant phenomenology. 
In this paper, we systematically address the possibility of obtaining two scalar DM components under $\mathcal{Z}_N$ 
symmetry. We consider both the possibilities of DM being weakly interacting massive particle (WIMP) or 
pseudofeebly interacting massive particle (pFIMP). We elaborate upon $\mathcal{Z}_3$ 
symmetric model, confronting the relic density allowed parameter space with recent most direct and indirect search bounds and prospects. We also highlight the 
possible distinction of the allowed parameter space in single component and two component cases, as well as between WIMP-WIMP and WIMP-pFIMP scenarios.}
\keywords{Models for Dark Matter, Particle Nature of Dark Matter, Specific BSM Phenomenology.}
\makeatletter
\gdef\@fpheader{}
\makeatother
\begin{document}
\maketitle
\flushbottom
\newpage
\section{Introduction}
\label{sec:intro}
What constitutes the non-luminous dark matter (DM) prevailing in the Universe is one of the most critical questions today
\cite{Zwicky:1933gu, Zwicky:1937zza}. Different astrophysical and cosmological observations have confirmed its 
existence \cite{vanAlbada:1984js, Trimble:1987ee, Sofue:1999jy, Sofue:2000jx,  Clowe:2006eq, Girardi:2008jp, Girardi:2015xoa, Dacunha:2021vdf,  Springel:2007tu, Komatsu:2014ioa} and suggested that DM constitutes around 26.8\% of the total energy budget, 
expressed in terms of relic density, $\rm\Omega_{ DM} h^2=0.1200\pm 0.0012$ \cite{Planck:2018vyg}, although a laboratory detection is still awaited. 
Absent a suitable candidate, the Standard Model (SM) of particle physics needs to be augmented to realise the presence of DM. 
The broad characteristics associated with DMs are massive, stable and having no (or feeble) electromagnetic interactions. 
Other properties remain unconstrained, giving rise to plethora of possibilities. In this article, we focus on multiparticle DM. Such 
frameworks are motivated by several phenomenological advantages, the most important one stems from the fact that they allow depletion of DM via 
interaction between two dark sector particles, which helps alleviating direct, indirect and collider search limits. There is almost no constraint on multipartite 
features of DM sector, excepting for the collisionlessness of DM particles, coming from Bullet cluster/Abel cluster observations, $\frac{\sigma}{m_{\rm DM}}\lesssim 1 ~{\rm cm^2/gm}$ \cite{Clowe:2003tk,Markevitch:2003at,Randall:2008ppe,Kahlhoefer:2015vua}.

In order to satisfy the relic density observed by the anisotropies in CMBR experiments, the DM needs to be stable at the scale of the Universe's age. 
The stability of a fundamental particle is governed by an unbroken symmetry. The lightest dark particle transforming under a symmetry 
(discrete or continuous) does not possess any tree or loop-level decay terms with the visible sector. 
Then ideally, for having two stable DM particles, two different symmetries are required. However, it is also possible to have one long lived DM component, where the 
decay terms are assumed very small and the life time is larger than the age of the Universe, which does not necessitate a protection via symmetry. However, the 
large decay life time requires either a kinematic suppression or the couplings involved in the process to be small or both.
We study such possibilities here with a single $\mathcal{Z}_N~(N=2,3,4,...)$ symmetry, where the heavier dark sector particle is long lived and serves as the second 
DM component. Our intention will be to bring out the consequent phenomenology in such cases, especially their implications in direct and indirect search aspects.

DM classification is mostly done based on its production mechanism, whether it remains in thermal bath or out of equilibrium, leading to possibilities like 
thermal WIMP \cite{Gondolo:1990dk, Balazs:2014rsa}, non-thermal FIMP \cite{Yaguna:2011qn, Baer:2014eja, Biswas:2016iyh, Bernal:2017kxu, Curtin:2018mvb, Chakrabarty:2022bcn}, SIMP \cite{Hochberg:2014dra, Hochberg:2014kqa, Hochberg:2015vrg, Tulin:2017ara,  Choi:2018iit, Bhattacharya:2019mmy, Cirelli:2020bpc, Barman:2021ugy, Kamada:2022zwb} etc. In multicomponent DM frameworks, one can have all kinds of combinations, like WIMP-WIMP \cite{Bhattacharya:2016ysw}, WIMP-FIMP \cite{DuttaBanik:2016jzv, Bhattacharya:2021rwh}, WIMP-SIMP, SIMP-SIMP \cite{Ho:2022erb, Choi:2021yps}, FIMP-FIMP \cite{Pandey:2017quk, PeymanZakeri:2018zaa} etc. 
In recent studies \cite{Bhattacharya:2022dco, Bhattacharya:2022vxm} we showed that a DM having feeble interaction with the visible sector can thermalise and freeze out via interaction with another thermal DM component, called pFIMP \cite{Bhattacharya:2022dco}. pFIMP has some distinguishable phenomenological features. In this work, how pFIMPs can arise in multicomponent scenarios having single $\mathcal{Z}_N$ symmetry is addressed. We specifically highlight the cases of $\mathcal{Z}_2$ and $\mathcal{Z}_3$ symmetric scenarios and compare it to $\mathcal{Z}_2\otimes \mathcal{Z}^{\prime}_2$ and $\mathcal{Z}_3\otimes \mathcal{Z}^{\prime}_3$ models. We show the consequent allowed parameter space of the models goes beyond the so-called Higgs resonance region \cite{Bhattacharya:2022dco}. 

Amongst DM search strategies, recoil spectrum of detector nuclei (or electron) in possible scattering with DM via direct detection (DD), gamma-ray (and anti particle) spectrum at the galactic centre via indirect detection (ID), and missing energy (or missing transverse momentum) signal at the collider experiments are the major ones. Null results from these experiments so far have imposed bounds on DM parameters, specifically DM-SM coupling. For example, DD puts a bound on DM-nucleon/electron scattering cross-section as a function of DM mass. As we focus here on the GeV-TeV scale cold DM (which in turn support structure formation of the universe), the most relevant DD bounds are obtained from PandaX-4T \cite{PandaX:2023xgl}, XENONnT \cite{XENON:2023cxc}, LUX-ZEPLIN \cite{LZ:2022lsv}, DARWIN/XLZD \cite{Baudis:2024jnk} etc. ID provides a bound on DM self-annihilation and semi-annihilation cross-section as a function of DM mass, the ones relevant for us comes from Fermi-LAT \cite{Fermi-LAT:2015att,Fermi-LAT:2016afa,Fermi-LAT:2016uux}, H.E.S.S \cite{HESS:2016mib}, CTA \cite{CTAO:2024wvb,CTAConsortium:2017dvg,CTA:2020qlo,Silverwood:2014yza} observations. The relevance of these bounds and search prospects with future sensitivities are scrutinised here in context of the multi component DM framework under $\mathcal{Z}_N$ symmetry.

This paper is organised as follows: in sec\,.~\ref{sec:zn} we discussed the possibility of getting two stable scalar DM components transforming under a single $\mathcal{Z}_N$ symmetry and in sec\,.~\ref{sec:z2} we discuss briefly the case of $\mathcal{Z}_2$ symmetry. The case pertaining to $\mathcal{Z}_3$ symmetry is elaborated for both WIMP-WIMP and WIMP-pFIMP combinations in sec\,.~\ref{sec:z3}. We finally summarise in sec\,.~\ref{sec:summary}. Appendix \ref{sec:decay}, and \ref{fig:z3-feynman} provide some necessary details omitted in the main text.
\section{Generic discussion on two component DM under $\mathcal{Z}_N$ symmetry}
\label{sec:zn}
Two stable DM components require two discrete symmetries, as studied in many different contexts \cite{Zurek:2008qg, Profumo:2009tb, Medvedev:2013vsa, Bian:2013wna, Bhattacharya:2013hva, Esch:2014jpa, Aoki:2016glu}, while some possibilities of having two DMs using a single discrete symmetry \cite{Yaguna:2019cvp, Yaguna:2021vhb, Yaguna:2021rds, Belanger:2020hyh} have also been discussed. However, a systematic study of obtaining all such possibilities under a 
single symmetry still requires attention. We address it via appropriate assumptions on the coupling parameters of the parent Lagrangian respecting 
$\mathcal{Z}_N$ symmetry. Our study is limited to scalar DM, while a similar study with vector and/or fermion DM is possible. 
Similarly, one can also have more than two DM components in a trivial extension of what we present here.

Specifically, we extend the SM with two scalar fields $\Phi_{1}$ and $\Phi_{2}$ transforming under a single $\mathcal{Z}_N$ 
symmetry. The fields $\Phi_1$,$\Phi_2$  might be real or complex, depending on their transformation under the $\mathcal{Z}_{N}$ symmetry while $\Phi_1$ and $\Phi_2$ are assumed to be gauge singlets under the SM gauge groups. 
The lighter component is automatically stable, while the heavier one becomes (kinematically) stable after making some coupling parameters vanishingly small in the non-degenerate case. 

In general, the heavier dark sector (DS) particle can have $(n+n^{\prime})$ body decay as,
\bea
\text{heavier} ~{\rm DS} \to n~ \text{lighter}~{\rm DS} \,+\, n^{\prime}~ \text{SM}\,,
\eea
with $n,~n^{\prime}=0,1,2,3...,{\rm~and~}~n+n^{\prime}\ge 2$; where the exact values of $n(n^{\prime})$ depend on the model.
Now, the tree or loop level decay of this particle can be stopped by appropriate choices of the model parameters:
\begin{enumerate}[label=(\alph*)]
\item For $n^{\prime}=0$, the heavier DS particle is stable when 
\bea 
n~m_{\rm lighter}>m_{\rm heavier}>m_{\rm lighter}\,.
\label{eq:heavy2light}
\eea
\item When $n^{\prime}\neq 0$, a kinematic condition $  n~m_{\rm lighter}+n^{\prime}~m_{\rm SM}>m_{\rm heavier}>n~m_{\rm lighter}$
can stop the on-shell decay, but off-shell decays are still possible. The only way to make the heavier DS component stable is to choose  
appropriate couplings leading to the decay adequately small leading to a long lived particle (LLP) DM.

\end{enumerate}

Further, the DMs can be WIMP, FIMP or SIMP depending on the strength of their couplings to SM. WIMP-WIMP combination is more constrained from 
direct search than WIMP-FIMP, given that FIMP is mildly constrained from DD/ID experiments (unless the DM-nucleon/electron scattering occurs via a light mediator). 
On the other hand, pFIMP receives milder constraints than WIMP having feeble DM-SM coupling, but still remains accessible for future sensitivities of DD/ID search 
via WIMP loop, having sizeable interaction with WIMP.  Therefore, we are more interested in obtaining WIMP-pFIMP limit of this model, having either the lighter particle 
or the heavier one as pFIMP \cite{Bhattacharya:2022dco, Bhattacharya:2022vxm}.
In order to study two-component DM, let us first write the renormalizable scalar potential, $V(\Phi_1,\Phi_2,H)$ ($H$ represents the SM Higgs isodoublet), 
\begin{eqnarray}
\rm V(\Phi_1,\Phi_2,H)\supset  \uplambda_{1}|\Phi_1|^2H^{\dagger}H  & & +\uplambda_{2}|\Phi_2|^2H^{\dagger}H
+\uplambda_{3}(\Phi_1\Phi_2+\Phi_1\Phi_2^*+h.c\,.)H^{\dagger}H \nonumber \\  & &
+(\uplambda_{4} \Phi_{1}^2 + \uplambda_{5} \Phi_2^2 + h.c.) H^{\dagger}H +V^{\rm int}(\Phi_1,\Phi_2)\,,
\label{eq:self-zn0}
\end{eqnarray}
where $\uplambda_{1},~\uplambda_{2},~\uplambda_{3},~\uplambda_{4},~\uplambda_{5}$ are dimensionless Higgs portal couplings, and $V^{\rm int}(\Phi_1,\Phi_2)$ contains 
all possible renormalizable interaction terms (with mass dimension $\mathcal{M}$[2], $\mathcal{M}$[3] and $\mathcal{M}$[4]) between $\Phi_1$ and $\Phi_2$ as, 
\bea\begin{split}
\mathcal{M}[2]:&~\{\Phi_1\Phi_2,~\Phi_1\Phi_2^*\}\,+\,h.c\,;\\
\mathcal{M}[3]:&~\{\Phi_1|\Phi_2|^2,~\Phi_2|\Phi_1|^2,~\Phi_1\Phi_2^2,~\Phi_1{\Phi_2^*}^2,~\Phi_1^2\Phi_2\,,~\Phi_1^2\Phi_2^*\}\,+\,h.c\,;\\
\mathcal{M}[4]:&~\{\Phi_1^2\Phi_2^2,~\Phi_1^2{\Phi_2^*}^2,~\Phi_1^2|\Phi_2|^2,~|\Phi_1|^2\Phi_2^2,|\Phi_1|^2|\Phi_2^2,~\Phi_1\Phi_2^3,~\Phi_1^*\Phi_2^3,~ \Phi_1^3\Phi_2,
\\&~ \Phi_1^3 \Phi_2^* ,~\Phi_1\Phi_2|\Phi_1|^2,~\Phi_1\Phi_2|\Phi_2|^2,~\Phi_1\Phi_2^*|\Phi_1|^2,~\Phi_1\Phi_2^*|\Phi_2|^2\}\, +\,h.c.
\label{eq:self-zn}
\end{split}\eea
In eqs\,.~\eqref{eq:self-zn0} and \eqref{eq:self-zn}, we wrote all possible interaction terms \footnote{However, the self-interaction terms will also be present:  $\Phi_1^2,~\Phi_2^2,~|\Phi_1|^2,~|\Phi_2|^2, ~\Phi_1^3,~\Phi_2^3,~\Phi_1 |\Phi_1|^2,~\Phi_2 |\Phi_2|^2,~ \Phi_1^4,$ $\Phi_2^4,~\Phi_1^2 |\Phi_1|^2,~\Phi_2 |\Phi_2^2|$.}, however the choice of the 
symmetry restricts them. It is obvious that if both $\Phi_1$ and $\Phi_2$ contribute as DM, then to prevent the tree-level decay of the heavier component, 
$\uplambda_{3}$ in eq\,.~\eqref{eq:self-zn0} needs to be sufficiently small\footnote{Additionally, for WIMP/pFIMP nature, $\uplambda_{1},\uplambda_{2}$ can be chosen moderate (or small).}, further restrictions arise depending on the possible interaction terms that one can write. 
If, $\Phi_1 \to \omega_N^{q_1} \Phi_1$ and $\Phi_2 \to \omega_N^{q_2} \Phi_2$ under 
$\mathcal{Z}_{N}$ symmetry, where $\omega_N^q=e^{i2\pi(q/N)}$ and $q_{1,2}$ define the charges of the fields, then the choice of $q_{1,2}$ 
also plays a crucial role in deciding the interaction terms, as shown in \cref{tab:zn}. The interaction terms are broadly classified into four cases:
\bea
&(i)&~q_1\,=\,q_2, {\rm ~and }~ q_1+q_2\,=\, N\,, \nonumber \\
&(ii)&~q_1\,=\,q_2, {\rm ~but }~ q_1+q_2\,\neq\, N\,, \nonumber \\
&(iii)&~ q_1\,\neq\,q_2, {~\rm but }~ q_1+q_2\,=\, N\,, \nonumber\\
&(iv)&~q_1\,\neq\,q_2, {\rm ~and ~}q_1+q_2\,\neq\, N\,.
\label{eq:class}
\eea

For the first three cases in eq\,.~\eqref{eq:class}, we have to consider both kinematic constraints and choose some couplings vanishingly small 
for making the heavy DS particle stable.  The heavy DS particle in all these cases decay to SM particle plus DM, falling in the case (b) mentioned before.  
For the fourth condition in eq\,.~\eqref{eq:class}, i.e. $q_1\,\neq\,q_2, {\rm ~and ~}q_1+q_2\,\neq\, N$, 
a two-component DM scenario can arise just by imposing kinematic constraints, as here the decay of heavy DS particle occurs 
within the DS {\em only}, falling into category (a) discussed above; for some example analysis of such scenarios, 
see \cite{Belanger:2014bga, Yaguna:2019cvp, Belanger:2020hyh, Yaguna:2021vhb, Yaguna:2021rds, Qi:2023egb}. 
\begin{table}[htb!]\centering
\scalebox{0.625}{
\renewcommand{\arraystretch}{1.5}
\begin{tabular}{|c |c |c |c |}\hline
\rowcolor{gray!30}\multicolumn{3}{|c|}{\textbf{Two dark sector scalar fields $\Phi_1$ and $\Phi_2$ and interaction terms under $\mathcal{Z}_{\rm N}$ symmetry}}\\\hline
\cellcolor{gray!25}Symmetry&\cellcolor{gray!25}Transformation&\cellcolor{gray!25}Interaction terms between DS particles \\ \hline\hline
\multirow{12}{*}{$\mathcal{Z}_{\rm N}$}&\cellcolor{gray!20} $\omega_N^{q_1},~\omega_N^{q_2}~(q_1\,=\,q_2) $ &\cellcolor{gray!20}\\
&\cellcolor{gray!20} ${\rm and ~}q_1+q_2\,=\, N $ & \multirow{-2}{*}{\cellcolor{gray!20}${\Phi_1^2,~ \Phi_2^2,~\Phi_1^4,~ \Phi_2^4,~\color{black}\Phi_1\Phi_2},~\Phi_1^2 \Phi_2^2,~{\color{black}\Phi_1^3\Phi_2,~\Phi_2^3\Phi_1}$}  \\
&\cellcolor{gray!15}$\omega_N^{q_1},~\omega_N^{q_2}$ & \cellcolor{gray!15} \\
&\multirow{-1}{*}{\cellcolor{gray!15}$(q_1\,=\,q_2\,=\,q)$}  &\multirow{-2}{*}{\cellcolor{gray!15} ${|\Phi_1|^2,~ |\Phi_2|^2,~\left[\Phi_1^3,~{\color{black}\Phi_2^3}\right]_{\dfrac{3q}{N}\in \mathbb{N}},~|\Phi_1|^4,~ |\Phi_2|^4,~\color{black}\Phi_1\Phi_2^*},~|\Phi_1\Phi_2^*|^2,~(\Phi_1{\Phi_2^*})^2,~{\color{black}\Phi_1\Phi_2^*(|\Phi_1|^2+|\Phi_2|^2)},$}\\
&\cellcolor{gray!15} &\cellcolor{gray!15}  \\ &\multirow{-3}{*}{\cellcolor{gray!15}${\rm but ~}q_1+q_2\,\neq\, N $ } & \multirow{-2}{*}{\cellcolor{gray!15} $\left[\Phi_1^2 \Phi_2^2,\Phi_1^3\Phi_2,~\Phi_2^3\Phi_1,~\Phi^4_1,~{\color{black}\Phi^4_2}\right]_{\dfrac{4q}{N}\in \mathbb{N}},~\left[{\color{black}\Phi_1^2\Phi_2,~\Phi_2^2\Phi_1}\right]_{\dfrac{3q}{N}\in \mathbb{N}},~\left[\Phi_1^3\Phi_2^*,~\Phi_2^3\Phi_1^*\right]_{\dfrac{2q}{N}\in \mathbb{N}},~\left[{\color{black}\Phi_1^2\Phi_2^*,~\Phi_2^2\Phi_1^*}\right]_{\dfrac{q}{N}\in \mathbb{N}}$}  \\
&\cellcolor{gray!10} & \cellcolor{gray!10} \\
&\cellcolor{gray!10} $\omega_N^{q_1},~\omega_N^{q_2}~(q_1\,\neq\, q_2) $ &\multirow{-2}{*}{${|\Phi_1|^2,~ |\Phi_2|^2,~\left[\Phi_2^3\right]_{\dfrac{3q_2}{N}\in \mathbb{N}},~|\Phi_1|^4,~ |\Phi_2|^4,~\color{black} \Phi_1\Phi_2,~\Phi_1^2 \Phi_2^2},~|\Phi_1 \Phi_2|^2,~\Phi_1\Phi_2({\color{black}|\Phi_1|^2+|\Phi_2|^2}),~$ \cellcolor{gray!10}}\\
&\cellcolor{gray!10}${\rm but ~}q_1+q_2\,=\, N $ & \cellcolor{gray!10} \\
&\cellcolor{gray!10}  &\multirow{-2}{*}{\cellcolor{gray!10}$[{\Phi_1^3,~\color{black}\Phi_1^2 \Phi_2^*,~\Phi_2^2 \Phi_1^*}]_{\dfrac{3q_1}{N}\in \mathbb{N}} ,~[\Phi_1^3 {\Phi_2^*},~\Phi_2^3 {\Phi_1^*},~\Phi_1^2 {\Phi_2^*}^2,~{\Phi^4_1,~\color{black}\Phi^4_2}]_{\dfrac{4q_1}{N}\in \mathbb{N}} $ }  \\
& $\cellcolor{gray!5}\omega_N^{q_1},~\omega_N^{q_2}~(q_1\neq q_2)  $ &\cellcolor{gray!5} \\
&\cellcolor{gray!5}  ${\rm and ~}q_1+q_2\,\neq\, N $ &\multirow{-2}{*}{\cellcolor{gray!5}$\left[\Phi_1^m \Phi_2^n\right]_{\dfrac{mq_1+nq_2}{N}\in \mathbb{N}}$,~$\left[\Phi_1^m {\Phi_2^*}^n\right]_{\dfrac{mq_1-nq_2}{N}\in \mathbb{N}}$ where \{$m,~n=1,2,3$ and $q_1^{\rm max},~q_2^{\rm max}=N-1$\} {\color{black}}}  \\\cline{1-3}
\multicolumn{3}{|c|}{Examples of some specific symmetries: $\mathcal{Z}_2$, $\mathcal{Z}_3$ and $\mathcal{Z}_4$}\\\cline{1-3}
\multicolumn{3}{|c|}{$q_1\,=\, q_2~{\rm and} ~q_1+q_2\,=\, N $ }\\\hline
$\mathcal{Z}_2$& $\omega_2,~\omega_2 $ & $\Phi _1^2,\Phi _2^2,\Phi _1^4,\Phi _2^4$,~${\color{black}\Phi_1\Phi_2},~\Phi_1^2\Phi_2^2,~{\color{black}\Phi_1^3\Phi_2,~\Phi_1\Phi_2^3}$  \\\cline{1-3}
$\mathcal{Z}_4$& $\omega_4^2,~\omega_4^2 $ & $\Phi _1^2,\Phi _2^2,\Phi _1^4,\Phi _2^4$,~${\color{black}\Phi_1\Phi_2},~\Phi_1^2\Phi_2^2,~{\color{black}\Phi_1^3\Phi_2,~\Phi_1\Phi_2^3}$  \\\cline{1-3}
\multicolumn{3}{|c|}{$q_1\,=\, q_2~{\rm but} ~q_1+q_2\,\neq\, N $ }\\\cline{1-3}
\multirow{2}{*}{$\mathcal{Z}_3$}& $\omega_3,~\omega_3 $&\multirow{2}{*}{$\left| \Phi _1\right|^2,\left| \Phi _2\right|^2,\Phi _1^3,{\color{black}\Phi _2^3},\left| \Phi _1\right|^4,\left| \Phi _2\right|^4,{\color{black}\Phi _1 \Phi _2^*},\left(\Phi _1\Phi _2^*\right)^2,\left| \Phi _1 \Phi _2\right|^2,\Phi _1 \Phi _2^2,~{\color{black}\Phi _1^2 \Phi _2,~\Phi _1 \Phi _2^*(\left| \Phi _1\right|^2+\left| \Phi _2\right|^2)}$} \\\cline{2-2}
& $\omega_3^2,~\omega_3^2 $& \\\cline{1-3}
\multirow{2}{*}{$\mathcal{Z}_4$}
& $\omega_4,~\omega_4$&\multirow{2}{*}{$\left| \Phi _1\right|^2,\left| \Phi _2\right|^2,{\color{black}\Phi _1^4,\Phi _2^4},\left| \Phi _1\right|^4,\left| \Phi _2\right|^4,{\color{black}\Phi _1 \Phi _2^*},\left(\Phi _1\Phi _2^*\right)^2,{\color{black}(\Phi _1 \Phi _2)^2},\left| \Phi _1 \Phi _2\right|^2,\Phi _1^3 \Phi _2, \Phi _2^3\Phi _1,{\color{black}\Phi _1 \Phi _2^*(\left| \Phi _1\right|^2+\left| \Phi _2\right|^2)}$} \\\cline{2-2}
& $\omega_4^3,~\omega_4^3 $& \\\cline{1-3}
\multicolumn{3}{|c|}{$q_1\,\neq\, q_2~{\rm but} ~q_1+q_2\,=\, N $ }\\\cline{1-3}
\multirow{2}{*}{$\mathcal{Z}_3$}& $\omega_3^2,~\omega_3 $&\multirow{2}{*}{$\left| \Phi _1\right|^2,\left| \Phi _2\right|^2,\Phi_1^3,~{\color{black}\Phi_2^3},~|\Phi_1|^4,~|\Phi_2|^4,~{\color{black}\Phi_1\Phi_2},~\Phi_1^2\Phi_2^2,~|\Phi_1\Phi_2|^2,~\Phi_2^2\Phi_1^*,~{\color{black}\Phi_1^2\Phi_2^*},~{\color{black}\Phi_1\Phi_2(|\Phi_1|^2+|\Phi_2|^2)}$} \\\cline{2-2}
& $\omega_3,~\omega_3^2 $& \\\cline{1-3}
\multirow{2}{*}{$\mathcal{Z}_4$}
& $\omega_4,~\omega_4^3 $&\multirow{2}{*}{${\left| \Phi _1\right|^2,\left| \Phi _2\right|^2,~{\color{black}\Phi_1^4,~\Phi_2^4},|\Phi_1|^4,~|\Phi_2|^4,~\color{black}\Phi_1\Phi_2},~(\Phi_1\Phi_2)^2,{\color{black}(\Phi_1\Phi_2^*)^2},~|\Phi_1\Phi_2|^2,~{\color{black}\Phi_1^3\Phi_2^*,~\Phi_2^3\Phi_1^*},~{\color{black}\Phi_1\Phi_2(|\Phi_1|^2+\Phi_2|^2) }$} \\\cline{2-2}
& $\omega_4^3,~\omega_4 $& \\\cline{1-3}
\hline
\end{tabular}}
\caption{Self interaction and interactions between two scalar fields $\Phi_1$ and $\Phi_2$ which transform under a discrete symmetry $\mathcal{Z}_{N}$ as 
$\Phi_1 \to \omega_N^{q_1} \Phi_1$ and $\Phi_2 \to \omega_N^{q_2} \Phi_2$ with $\omega_N^{q_i}=e^{i2\pi(q_i/N)}$ and $q_{1,2}$ being the integer charges of $\Phi_{1,2}$ 
with $q_{1,2}=1,~2,~3..,~N-1$. Depending upon the choice of $q_{1,2}$, the interaction terms are shown. $ \mathbb{N}$ denotes set of integer numbers.
}
\label{tab:zn}
\end{table}

Let us take a closer look into the possible terms that one can write involving $\Phi_1$ and $\Phi_2$ depending on their charges under 
$\mathcal{Z}_{N}$ transformation as shown in \cref{tab:zn}. The first sub-row of \cref{tab:zn} shows the case when they have same charge: 
$q_{1}=q_{2}=q$, with $q=N/2$. As $q$ needs to be an integer, this requires $N$ to be even. In such a situation, $\Phi_1$ and $\Phi_2$ can be two real scalar fields, 
and only select few terms are allowed amidst all the possibilities listed in eq\,.~\eqref{eq:self-zn}. The second sub-row in \cref{tab:zn}, shows the possible terms when 
$q_{1} = q_{2}$, but $q_{1} + q_{2} \neq N $, which restricts us from writing $\Phi_{1} \Phi_{2}$ term, but retaining a term like $\Phi_{1} \Phi_{2}^{*}$ is possible, allowing 
complex scalar fields only. It is obvious but worth reminding that the complex conjugate fields possess $-q_i$ charges as $\Phi_i^{*} \to {\omega^{q_i}_N}^{*} \Phi_i^{*}$ 
under $\mathcal{Z}_{N}$. Apart, terms like $\Phi_{1,2}^3$ can only be written if $\dfrac{3q}{N}$ is an integer, i.e., 
$\dfrac{3q}{N}\in \mathbb{N}$, where $ \mathbb{N}$ denotes the set of integer numbers. The third and fourth sub-rows 
show the interaction terms when $q_{1} \neq q_{2}, q_{1} + q_{2} = N$, and  $q_{1} \neq q_{2}, q_{1} + q_{2} \neq N$ 
cases respectively. The restrictions that apply in writing the interaction terms are mentioned in the subscript of the parenthesis. For example, $\Phi_1^m \Phi_2^n$ term 
can be written for $q_{1} \neq q_{2}, q_{1} + q_{2} \neq N$ case, only when $\dfrac{mq_1+nq_2}{N}\in \mathbb{N}$.

In the second part of \cref{tab:zn}, we have shown examples of $\mathcal{Z}_{2}$, $\mathcal{Z}_{3}$ and $\mathcal{Z}_{4}$ to write 
interaction terms between $\Phi_1,\Phi_2$ fields. For $\mathcal{Z}_{2}$, the 
charges are trivial and only caters to the possibility $q_1=q_2=1, q_1+q_2=2$; but for $\mathcal{Z}_{3}$ and $\mathcal{Z}_{4}$, other combination of charges as in 
eq\,.~\eqref{eq:class} are possible. In the table, we have omitted listing $q_{1} \neq q_{2}, q_{1} + q_{2} \neq N$ case for $\mathcal{Z}_{3}$ 
as it doesn't exist, and for $\mathcal{Z}_{4}$ as it can be stabilised by kinematic constraints without any fine tuning of the couplings. 
All the possibilities for $\mathcal{Z}_{2}$ and $\mathcal{Z}_{3}$ symmetries to contain two DM components will be discussed in this paper; for $\mathcal{Z}_{2}$ this turns out to be one combination, while there are six different 
combinations of choosing vanishingly small parameters for the case of $\mathcal{Z}_{3}$ to accommodate two DMs. 
Obviously, the number of possibilities increase with higher $N$. It is interesting to note that when we choose one or more couplings vanishingly small in the Lagrangian, 
they may indicate to further restrictions or symmetries imposed to the Lagrangian.

\begin{itemize}
\item For example, when $q_1\,=\,q_2, {\rm ~and }~ q_1+q_2\,=\, N$ as in the first case as in eq\,.~\eqref{eq:class}, 
imposing the constraints on relevant couplings to have two DM components leads to
$\mathcal{Z}_{N}\otimes\mathcal{Z}^{\prime}_{N}$ symmetric Lagrangian. This is {\em always} the case for $\mathcal{Z}_2$, given the only possible charge combination.
\item On the other hand, the second and third possibilities of eq\,.~\eqref{eq:class} eventually lead to the fourth scenario [$q_{1} \neq q_{2}, q_{1} + q_{2} \neq N$] of a higher symmetry group, after assuming relevant couplings to be vanishingly small to stop heavier DS particle decay. 
In such circumstances, $\mathcal{Z}_3$ and $\mathcal{Z}_4$ symmetric cases become 
$\mathcal{Z}_6$ and $\mathcal{Z}_8$ symmetric scenarios respectively. The possible charge combinations of the scalar fields pertaining to 
$\mathcal{Z}_3$ and $\mathcal{Z}_4$ symmetries for single DM case, and the final charge assignments under 
$\mathcal{Z}_6$ and $\mathcal{Z}_8$  symmetries having two DMs (after stabilisation of the heavy DS particle) are listed below. We will elaborate on $\mathcal{Z}_3$ 
later.
\begin{gather}
\mathcal{Z}_3~\left[\{q_1,q_2\}:~(1,1),~(2,2)\right]\supset \mathcal{Z}_6~\left[\{q_1,q_2\}:~(2,5),~(4,1)\right]\,,\\
\mathcal{Z}_3~\left[\{q_1,q_2\}:~(1,2),~(2,1)\right]\supset \mathcal{Z}_6~\left[\{q_1,q_2\}:~(2,1),~(4,5)\right]\,,\\
\mathcal{Z}_4 ~\left[\{q_1,q_2\}:~(1,1),~(3,3)\right]\supset \mathcal{Z}_8~\left[\{q_1,q_2\}:~(1,5),~(5,1),~(3,7),~(7,3)\right]\,,\\
\mathcal{Z}_4 ~\left[\{q_1,q_2\}:~(1,3),~(3,1)\right]\supset \mathcal{Z}_8~\left[\{q_1,q_2\}:~(1,3),~(3,1),~(5,7),~(7,5)\right]\,.
\end{gather}
\end{itemize}

\section{Two scalar DMs under single $\mathcal{Z}_2$ symmetry}
\label{sec:z2}
$\mathcal{Z}_2$ symmetric model has been studied widely in several contexts with real scalar DM $(\phi_i)$, odd ($\phi_i\to -\phi_i$) under $\mathcal{Z}_2$ 
\cite{McDonald:1993ex, Burgess:2000yq, Guo:2010hq, Profumo:2010kp, Djouadi:2012zc, Cline:2013gha, Feng:2014vea, Duerr:2015mva, Han:2015hda, GAMBIT:2017gge}. 
We will have a brief discussion here on the possibility of getting two DM components, where two scalar fields $\phi_1$ and $\phi_2$ are transforming under the same 
$\mathcal{Z}_2$ symmetry.
The corresponding Lagrangian density is,
\bea
\rm\mathcal{L}=\mathcal{L}_{\rm SM}+\dfrac{1}{2}\partial_{\mu}\phi_1\partial^{\mu}\phi_1+\dfrac{1}{2}\partial_{\mu}\phi_2\partial^{\mu}\phi_2-V(\phi_1,\phi_2,H)\,,
\label{eq:z2-lagrangian}
\eea
where, 
\begin{align}
\rm\nonumber V(\phi_1,\phi_2,H)&=\frac{1}{2}m_{\phi_1}^2\phi_1^2+\frac{1}{4!}\lambda_{\phi_1}\phi_1^4+\frac{1}{2}m_{\phi_2}^2\phi_2^2+\frac{1}{4!}\lambda_{\phi_2}\phi_2^4+\frac{1}{2}\lambda_{\phi_1{\rm H}}\phi_1^2({\rm H^{\dagger}H}-\frac{1}{2}v^2)\\\nonumber&+\frac{1}{2}\lambda_{\phi_2{\rm H}}\phi_2^2({\rm H^{\dagger}H}-\frac{1}{2}v^2)+\frac{1}{4}\lambda_{\phi_1\phi_2}\phi_1^2\phi_2^2+\frac{1}{3!}\lambda_{122}\phi_1\phi_2^3+\frac{1}{3!}\lambda_{112}\phi_1^3\phi_2\\&
+\lambda_{\phi_1\phi_2\rm H}\phi_1\phi_2 ({\rm H^{\dagger}H}-\frac{1}{2}v^2)\,.
\label{eq:z2-model}
\end{align}
Here, we have written the Lagrangian in physical basis where the kinetic and mass terms are diagonalized.
The model parameters obey constraints from unitarity, perturbativity and vacuum 
stability \cite{Yaguna:2008hd, Goudelis:2009zz, Gonderinger:2009jp, Guo:2010hq, Bhattacharya:2016ysw}.
By default, the model is as in eq\,.~\eqref{eq:z2-model}, represents a single component DM scenario, with lighter of $\phi_1$ or $\phi_2$ serving as DM;
where the heavier DS particle decays to DM and provides co-annihilation channels for the DM to freeze out. In order to make the 
heavier DS particle stable, we need to analyse the interaction vertices corresponding to the possible decay channels; a detailed calculation is furnished in the \cref{apndx:z2-decay}. Note for example, the presence of $\phi_1^3\phi_2$ term ($\phi_2^3\phi_1$) can lead $\phi_1$($\phi_{2}$) to decay into three $\phi_2$ ($\phi_1$). 
To stop this decay, one can choose a kinematical constraint like $3 m_{\rm DM}>m_{\rm DS}$. However, in the presence of Higgs portal couplings, two-body decay 
$(\phi_2 \to h ~\phi_1{\rm~or~}\phi_1 \to h ~\phi_2)$ shown in fig\,.~\ref{fig:z2-decay} (to on-shell Higgs) and three-body decays (via off-shell Higgs) are possible.
Notably, for non-degenerate $m_{\phi_1,\phi_2}$ masses, the decay of the heavier scalar to di-photon or di-gluon\footnote{However, two gluons eventually form bound states that are not massless. In that case, it will be subjected to the specific phase space constraints availability.} final states via a one-loop diagram mediated by an off-shell Higgs is kinematically allowed. 
So, one can choose a sufficiently small portal coupling ($\lambda_{\phi_1\phi_2H}$), so that the life-time of the heavy scalar becomes larger than the age of the universe. However, one loop decay via $\phi_1^3\phi_2$ and $\phi_2^3\phi_1$ term is still possible, see \cref{fig:z2-decay}.
To stop them, $\lambda_{112}$, and $\lambda_{122}$ couplings have to be sufficiently small as well, see \cref{tab:z2-benchmark} for the order of magnitude. 

Interestingly, when we block these couplings to stabilise the heavier DS particle, we essentially get rid of all terms having combinations of $\phi_1\phi_2$, then the Lagrangian eventually reduces to $\mathcal{Z}_{2} \otimes \mathcal{Z}^{\prime}_{2}$ symmetric one, well studied in different contexts, like in WIMP-WIMP 
scenario \cite{Bhattacharya:2016ysw}, WIMP-pFIMP scenario \cite{Bhattacharya:2022dco} etc. The only subtle difference between a pure 
$\mathcal{Z}_{2} \otimes \mathcal{Z}^{\prime}_{2}$ symmetric scenario to that of single $\mathcal{Z}_2$ with two DM is that, for the latter, we are getting one stable DM and one LLP, 
while under two different $\mathcal{Z}_2$ symmetry, we get two stable DMs, although the phenomenology is identical in both the cases.
\section{Two scalar DMs under single $\mathcal{Z}_3$ symmetry}
\label{sec:z3}
In this section, we will discuss two complex scalar fields transforming under single $\mathcal{Z}_3$ symmetry 
and its phenomenology in single and two-component DM scenarios. We will also discuss the comparison with two-component DM scenario
in $\mathcal{Z}_3\otimes\mathcal{Z}^{\prime}_{3}$ set up \cite{Bhattacharya:2017fid}.
\subsection{Model}
\label{sec:model-z3}
\begin{table}[htb!]
\begin{center}
\begin{tabular}{|c|c|}\hline
\rowcolor{magenta!15}{\bf DS Fields} &$\mathcal{Z}_3$\\
\rowcolor{cyan!15}  Complex scalar $\chi_1$&$\omega/\omega^2$\\
\rowcolor{lime!15}   Complex scalar $\chi_2$&$\omega^2/\omega$\\\hline
\end{tabular}
\end{center}
\caption{Model particle contains and their transformation way, with $\omega=e^{i2\pi/3}$.}
\label{tab:z3_transfrm}
\end{table}
\noindent 
The SM extended dark sector Lagrangian containing two complex scalar fields that transform differently under $\mathcal{Z}_3$ symmetry (see \cref{tab:z3_transfrm}) is written as \cite{Belanger:2014bga, Belanger:2012zr, Athron:2018ipf, Hektor:2019ote, Kannike:2019mzk, Ko:2020gdg, Bhattacharya:2017fid, Liu:2023kil},
\bea
\mathcal{L}\,=\,\mathcal{L}_{\rm SM}+|\partial_{\mu}\chi_1|^2+|\partial_{\mu}\chi_2|^2
-V(\chi_1,\chi_2,{\rm H})\,,
\label{eq:z3-model}
\eea
where,
\bea
\begin{split}
 V(\chi_1,\chi_2,{\rm H})
=&\sum_{i=1,2}\left(m_{\chi_i}^2|\chi_i|^2+\lambda_{\chi_i}|\chi_i|^4+\lambda_{iH}|\chi_i|^2({\rm H^{\dagger}H}-\dfrac{v^2}{2})\right)
\\+&\lambda_{12}|\chi_1|^2|\chi_2|^2+\dfrac{1}{2}\lambda_{12H}(\chi_1\chi_2+h.c.)({\rm H^{\dagger}H}-\dfrac{v^2}{2})+\dfrac{1}{2}\lambda_{12}^{\prime}(\chi_1^2\chi_2^2+h.c.)
\\+&\dfrac{1}{2}\biggl[\mu_{1}\chi_1^{*}\chi_2^2+\mu_{2}\chi_1^2\chi_2^{*}+\mu_{3}\chi_1^3+\mu_{4}\chi_2^3
+h.c.\biggr]
\\+&\dfrac{1}{2}\left[\lambda_{3}\chi_1\chi_2|\chi_1|^2+\lambda_{4}\chi_1\chi_2|\chi_2|^2 + h.c. \right]\,.
\end{split}
\label{eq:model-potentialz3}
\eea
In eq\,.~\eqref{eq:z3-model}, we may have the kinetic mixing terms and mass mixing terms, which can be removed through an appropriate field rotation and redefinition, as discussed in detail in \Cref{app:kinetic-mixing}. Here, we assume that eq\,.~\eqref{eq:model-potentialz3} is written in the physical basis obtained after two successive rotations, in which the kinetic and mass mixing matrices are diagonalized. We must also emphasize that the diagonalization does not alter the interaction terms.
For $\lambda_{H},\lambda_{i}>0$ and $\mu_H^2<0$, the Higgs field acquires non-zero vacuum expectation value (VEV) ($v$), with ${\rm H}\,=\,\left(0~~(h+v)/\sqrt{2}\right)^{\rm T}$, and results in electroweak symmetry breaking (EWSB). 
Here, we do not assume any VEV for the complex fields, and all couplings in the scalar potential are taken to be real in order to ensure CP invariance.
The theoretical and experimental constraints on the model parameters are as follows:

\paragraph{$\bullet~$Unitarity\\}
The unitarity bounds, derived in the infinite scattering energy limit and considering only the quartic terms, are given by \cite{Kahlhoefer:2015bea, Hektor:2019ote, Goodsell:2018tti},
\begin{gather}
|\lambda_H|\leq 4\pi\,,\ \
|\lambda_{12}-2\lambda_{12}^{\prime}|\leq 8\pi\,,\\
|\lambda_{1H}+\lambda_{2H}\pm \sqrt{(\lambda_{1H}-\lambda_{2H})^2+\lambda_{12H}^2}|\leq 16\pi\,,
\end{gather}
while for the others, please follow \cref{appndx:unitarity_bounds}.
\paragraph{$\bullet~$Perturbativity\\}

To ensure the validity of perturbation theory, loop corrections to the couplings should be smaller than their tree-level values. The perturbative bound for the model is given by \cite{Lerner:2009xg},
\bea
|\lambda_{iH}|\leq 4\pi,~|\lambda_{i}|<\pi\,.
\eea

\paragraph{$\bullet~$Vacuum stability\\}
The necessary conditions required to stabilise the potential are,
\bea
\lambda_H>0,~\lambda_{i}>0,~\lambda_{iH}+2\sqrt{\lambda_{i}\lambda_{H}}>0\,.
\eea
The maximal allowed value of the cubic parameters $\mu_{3,4}$ is approximately equal to $\mu_{3}=2\sqrt{\lambda_{\chi_1}}m_{\chi_1}$, $\mu_{4}=2\sqrt{\lambda_{\chi_2}}m_{\chi_2}$ \cite{Hektor:2019ote,Kannike:2012pe,Kannike:2016fmd,Cai:2015tam}. However, we take the cubic coupling up to twice the lightest DM mass for simplicity.
\paragraph{$\bullet~$Higgs invisible decay width\\}

The search for invisible decays with the ATLAS detector \cite{ATLAS:2023tkt}, using $139~\mathrm{fb}^{-1}$ of proton–proton collision data at a centre-of-mass energy of $\sqrt{s} = 13~\mathrm{TeV}$ at the LHC, combined with results obtained at $\sqrt{s} = 7~\mathrm{TeV}$ and $\sqrt{s} = 8~\mathrm{TeV}$, sets an observed (expected) upper limit of $0.107~(0.077)$ on the Higgs boson branching ratio to invisible final states at the $95\%$ confidence level. A combined analysis of $\sqrt{s} = 7$, $8$, and $13~\mathrm{TeV}$ proton–proton collision data at the CMS detector \cite{CMS:2023sdw} sets an observed (expected) $95\%$ confidence level upper limit of $0.15~(0.08)$ on $\mathcal{B}(h \to \mathrm{inv})$.

\subsection{Single component DM}
\label{sec:z3-single}
In the potential given by eq\,.~\eqref{eq:model-potentialz3}, we have written all possible terms are respecting the $\mathcal{Z}_3$ symmetry. Note that
the term $\chi_1 \chi_{2}$ can be written since both $\chi_{1}$ and $\chi_{2}$ rotate under the same $\mathcal{Z}_{3}$, which is not present in 
$\mathcal{Z}_3\otimes\mathcal{Z}^{\prime}_{3}$ \cite{Bhattacharya:2017fid}. But, the presence of this term gives us a non-diagonal mass term
unless we choose $\mu^2_{12}=-\lambda_{12H}v^2/2$.

\begin{figure}
\centering
\subfloat[]{\includegraphics[width=0.475\linewidth]{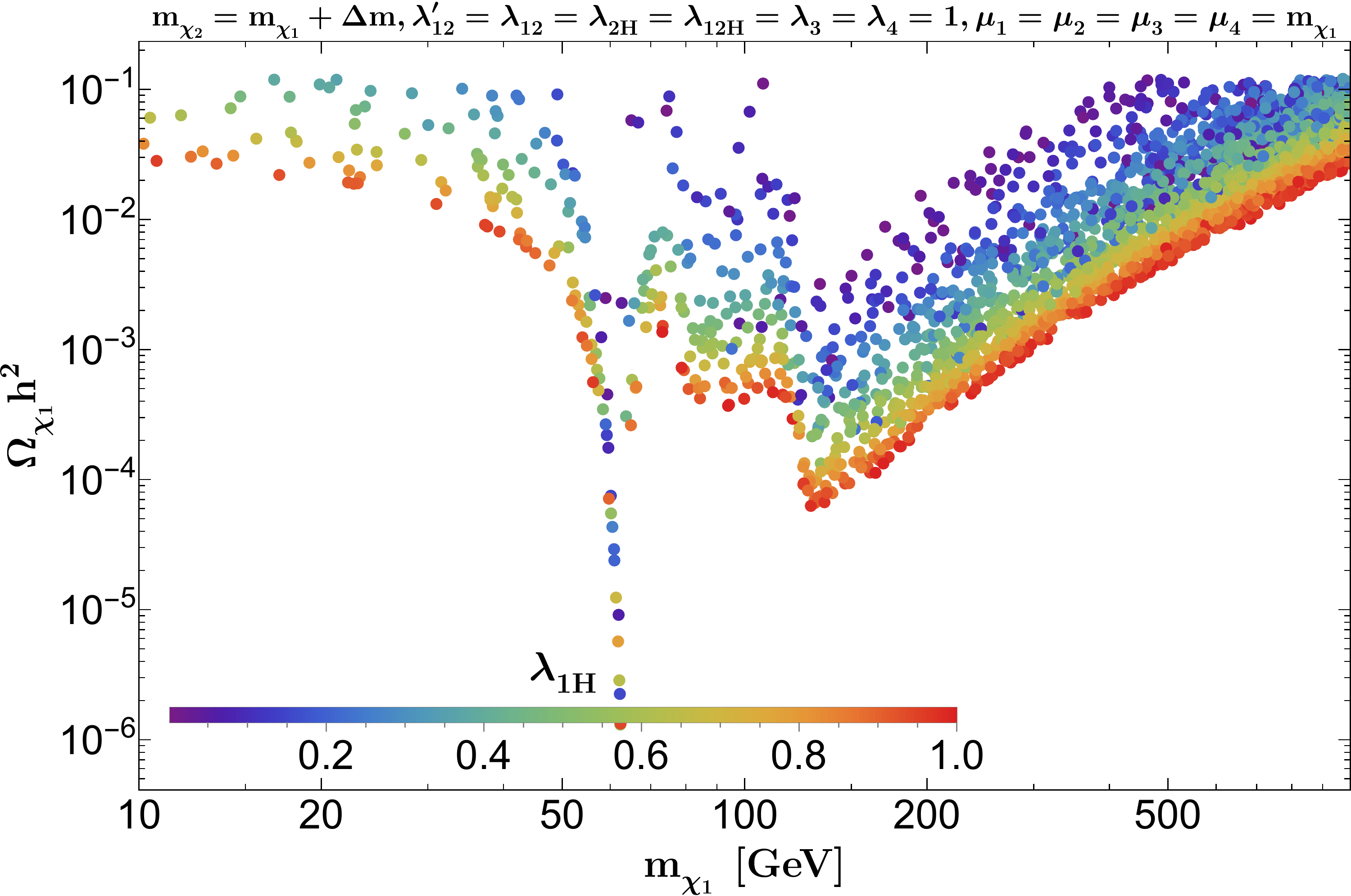}\label{fig:z3-single-relic}}\quad
\subfloat[]{\includegraphics[width=0.475\linewidth]{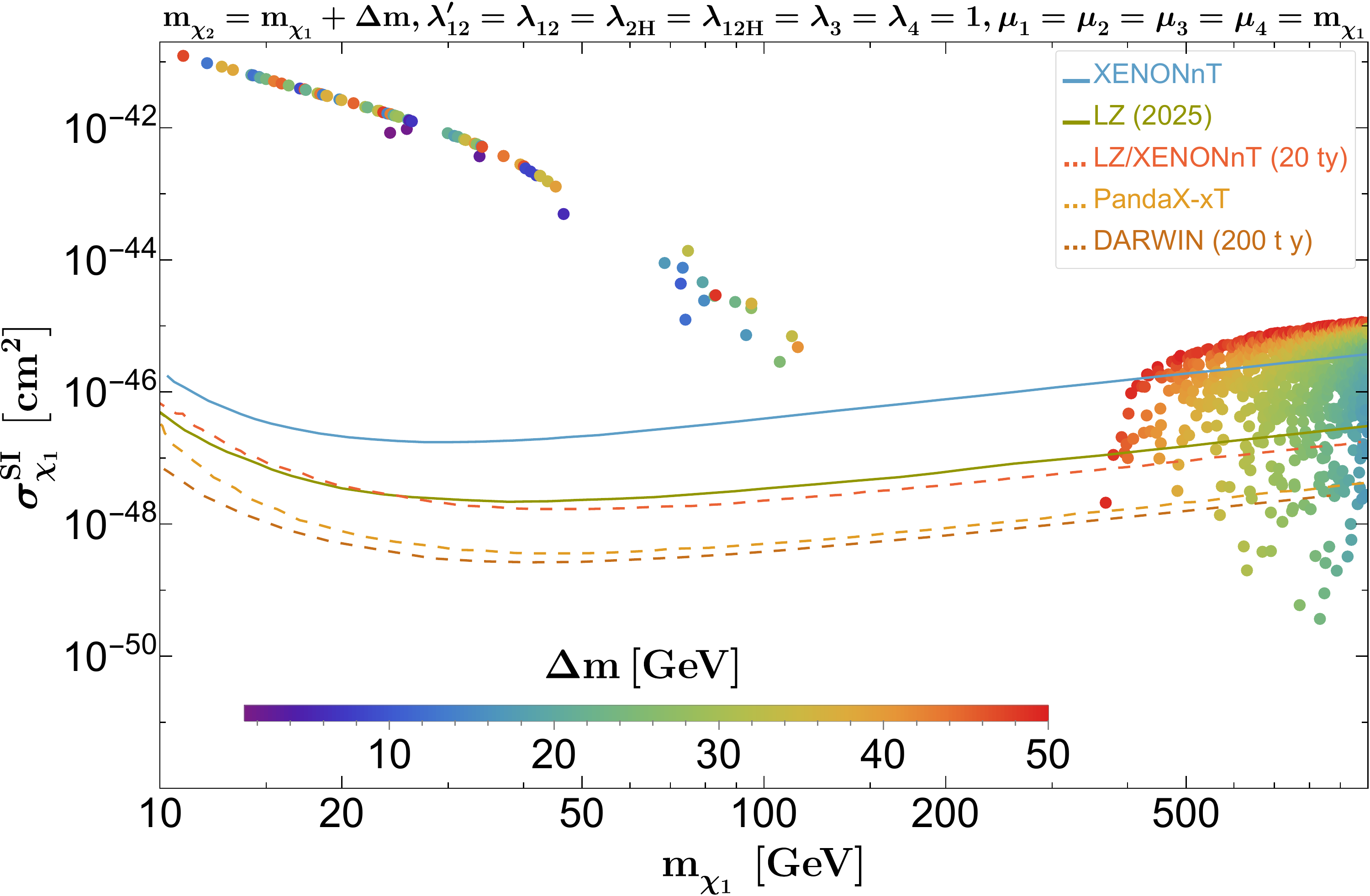}\label{fig:z3-single-dd}}

\subfloat[]{\includegraphics[width=0.475\linewidth]{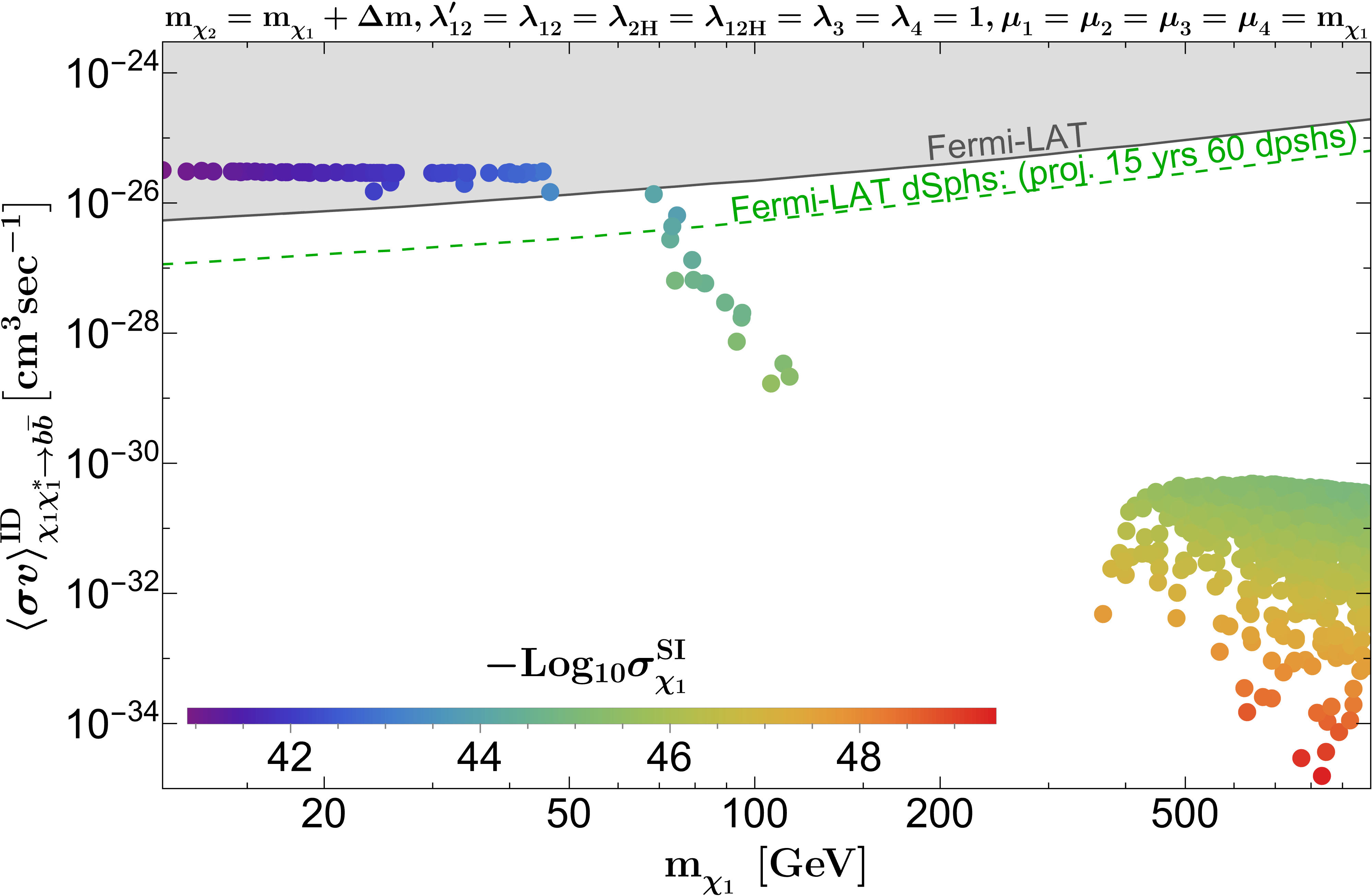}\label{fig:z3-single-id}}\quad
\subfloat[]{\includegraphics[width=0.475\linewidth]{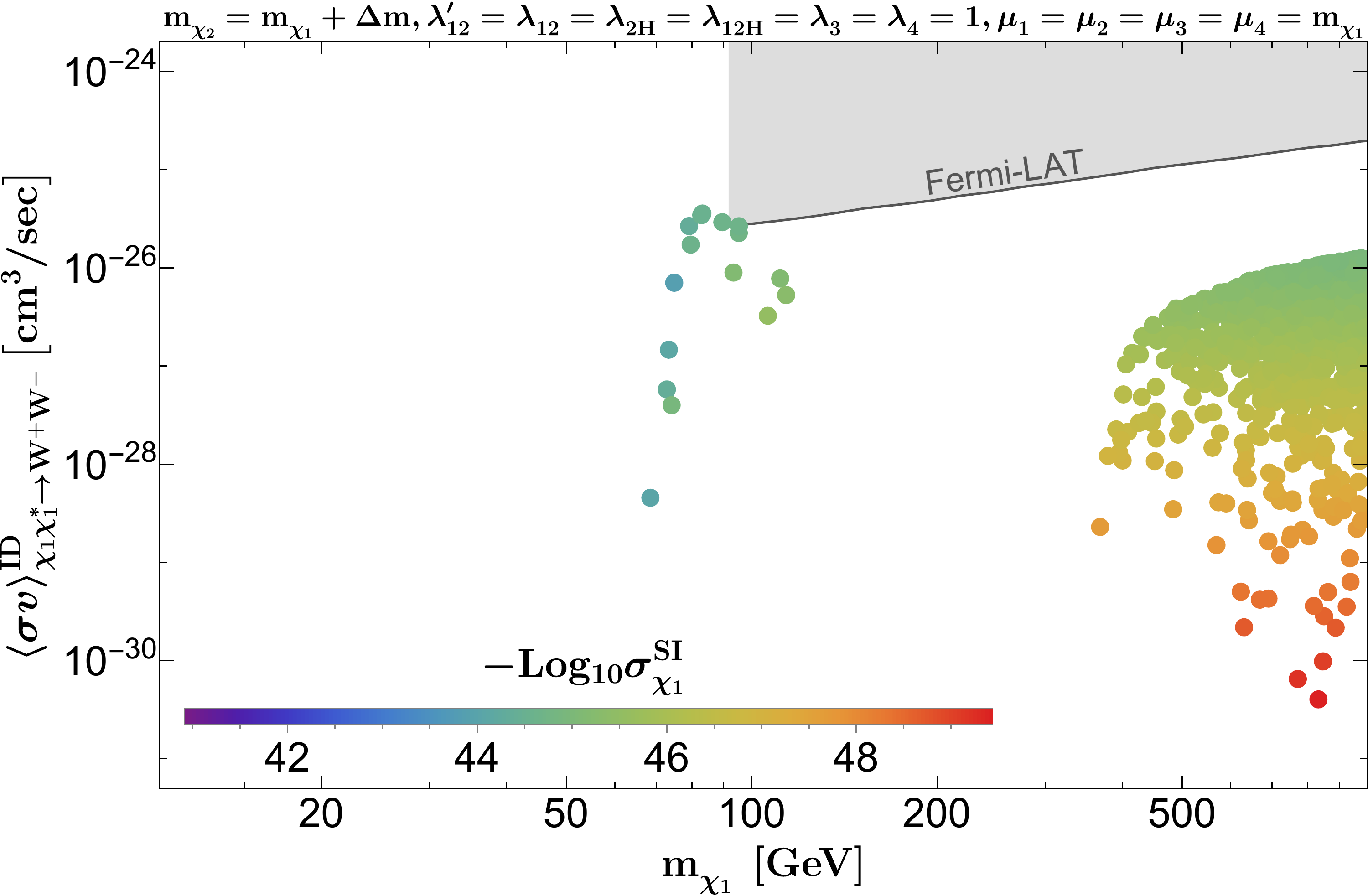}\label{fig:z3-single-id-WW}} \\
\subfloat[]{\includegraphics[width=0.475\linewidth]{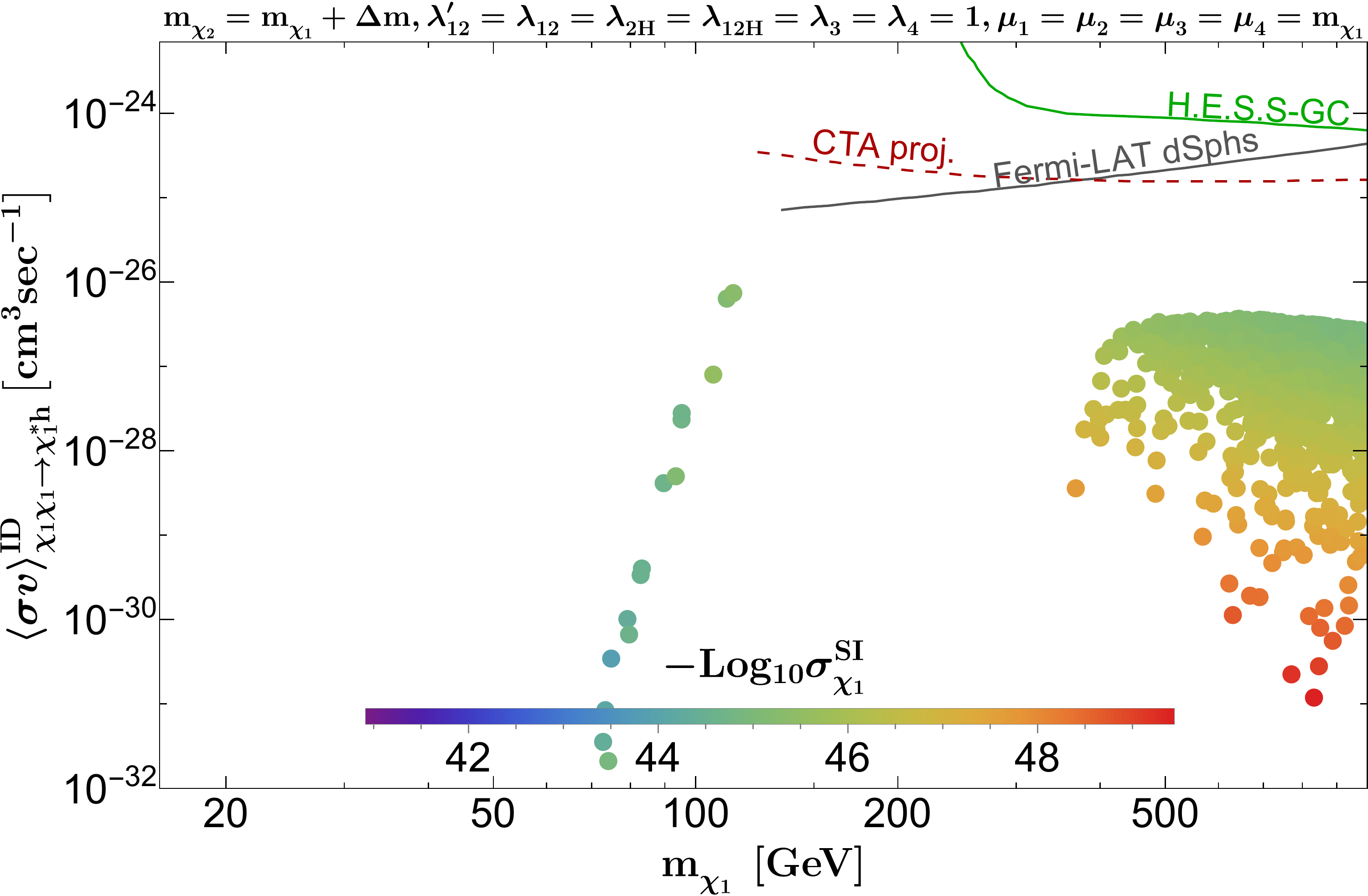}\label{fig:z3-single-id-semi}} 
\caption{Relic density allowed parameter space of the single component DM model with two scalars transforming under the same $\mathcal{Z}_3$ symmetry, 
confronting direct and indirect search limits. \Cref{fig:z3-single-relic} shows relic under abundant $(\Omega_{\chi_1}h^2\leq 0.1212)$ parameter space in 
$m_{\chi_1}-\Omega_{\chi_1}h^2$ plane where the rainbow color bar shows the variation of the Higgs portal coupling ($\lambda_{1H}$) with DM ($\chi_1$) mass.
\Cref{fig:z3-single-dd} shows relic allowed parameter space in $m_{\chi_1}-\sigma_{\chi_1}^{\rm SI}$ plane and the rainbow color bar here shows DM mass splitting 
with the co-annihilating partner $\chi_2$, i.e, $\Delta m=m_{\chi_2}-m_{\chi_1}$. The light blue and olive green colored lines represent the XENONnT and LZ-2022 bound.
\Cref{fig:z3-single-id} (\ref{fig:z3-single-id-WW}) shows the relic density allowed parameter space in $m_{\chi_1}-\langle\sigma v\rangle_{\chi_1\chi_1^*\to b\overline{b}\, (W^{+}W^{-})}$  plane and rainbow color bar shows the variation of $\sigma_{\chi_1}^{\rm SI}$ in unit of $\rm cm^2$. Thick grey and dashed green lines portray Fermi-LAT observation and projected limit\cite{Fermi-LAT:2015att, MAGIC:2016xys, Fermi-LAT:2016afa}, respectively.
\Cref{fig:z3-single-id-semi} shows the relic density allowed region in $m_{\chi_1}-\langle\sigma v\rangle_{\chi_1\chi_1\to \chi_1^*h}$ plane, and the rainbow color bar represents the spin-independent inelastic scattering cross-section ($\sigma_{\chi_1}^{\rm SI}$) variation. The thick green, thick grey, and dashed blue lines represent the 
upper bound of the DM semi-annihilation from H.E.S.S, Fermi-LAT and CTA, respectively.}
\label{fig:z3-single}
\end{figure}

By default, the Lagrangian describes a complex scalar singlet DM, stable under $\mathcal{Z}_3$ symmetry, which has been studied in different contexts 
\cite{Kang:2017mkl, Kannike:2019mzk}. Here we recap this scenario with the trending bounds on DM available from direct and indirect searches. 
Here, we have taken $\chi_1$ as our DM, but $\chi_2$ is equally possible, and DM phenomenology remains unaltered. The relic density of DM 
is governed by its annihilation channels. Additionally, we can have co-annihilation and semi-annihilation contribution in presence of heavy DS particle. 
This new degrees of freedom gives more allowed parameter space compared to the only one $\mathcal{Z}_3$ symmetric DM scenario \cite{Belanger:2012zr, Hektor:2019ote}.
As we focus on GeV scale DM, $n\to 2$ $(n>2)$ depletion processes are always subdominant to $2\to 2$ annihilation process.
The heavy DS particle decays to DM via off-shell or on-shell Higgs in tree or loop level, due to sizeable portal couplings: 
$(\lambda_{12H},~\lambda_3,~\mu_2)$. The key free parameters that govern DM analysis are,
\bea
\{m_{\chi_i}, ~ \lambda_{j},~\mu_{j},~\lambda_{iH},~\lambda_{12H},~ \lambda_{12},~ \lambda^{\prime}_{12}\}\,.
\label{eq:model-free}
\eea 
where $i=1,2$ and $j=1,...,4$ and all of these are taken as real parameters. Mass kinematics and couplings are the key parameters for the heavier DS decay to DM, 
and to get only one stable DM.
\paragraph{$\bullet$~BEQ and Relic density\\}
The Boltzmann equation (BEQ) in this case is relevant to the total yield, $Y_{\chi}=\sum\limits_{i}Y_i=Y_{\chi_1}+Y_{\chi^{*}_1}+Y_{\chi_2}+Y_{\chi^{*}_2}$ including the 
complex conjugate fields and heavier dark sector fields as the heavier ones decay to DM, 
and is given by \cite{Bhattacharya:2022vxm, Bhattacharya:2017fid, Yaguna:2019cvp, Belanger:2020hyh, Belanger:2022esk, Yaguna:2021vhb},
\bea
\dfrac{dY_{\chi}}{dx}=-\dfrac{s}{x~H(x)}\left[\langle\sigma v\rangle_{\rm SM}^{\rm eff}(Y_{\chi}^{2}-Y_{\chi}^{eq^2})+\frac{1}{2}\langle\sigma v\rangle_{\rm semi}^{\rm eff}(Y_{\chi}^{2}-Y_{\chi}Y_{\chi}^{eq})\right]\,;
\label{eq:z3-beq}
\eea
where,
$$\langle\sigma v\rangle_{\rm SM}^{\rm eff}=\sum\limits_{i,j}\langle\sigma v\rangle_{i~j\to\rm SM~SM}\dfrac{n_{i}^{\rm eq}n_{j}^{\rm eq}}{\left(n_{\chi}^{\rm eq}\right)^2}, ~~\langle\sigma v\rangle_{\rm semi}^{\rm eff}=\sum\limits_{i,j,k}\langle\sigma v\rangle_{i~j\to~k~{\rm SM}}\dfrac{n_{i}^{\rm eq}n_{j}^{\rm eq}}{\left(n_{\chi}^{\rm eq}\right)^2},$$ 
and 
$$n^{\rm eq}_{\chi}=\sum\limits_{i}n_{i}^{\rm eq}=n_{\chi_1}^{eq}+n_{\chi^{*}_1}^{eq}+n_{\chi_2}^{eq}+n_{\chi^{*}_2}^{eq}.$$
We can also express the effective semi-annihilation cross-section as, 
\bea
\langle\sigma v\rangle_{\rm SM}^{\rm eff}=\left[\sum_ig_im_i^2K_2\left(\frac{m_i}{T}\right)\right]^{-2}\sum_{i,j} \langle\sigma v\rangle_{ij\to \rm SM~ SM}~g_ig_j~m_i^2m_j^2K_2\left(\frac{m_i}{T}\right)K_2\left(\frac{m_j}{T}\right)\,,
\eea
where $g_i$ denotes internal degrees of freedom and $K_2$ denotes Bessel function of second kind. 
We have numerically solved the BEQ, eq\,.~\eqref{eq:z3-beq}, and verified it using micrOMEGAs \cite{Belanger:2018ccd} after importing the $\mathcal{Z}_3$ symmetric model 
using FeynRules \cite{Christensen:2008py}. The solutions of the BEQ and relic density allowed parameter space in terms of relevant parameters as well as direct and 
indirect search constraints are shown in \cref{fig:z3-single}. 

In \cref{fig:z3-single-relic}, we show the relic under abundance in terms of DM mass and $\Omega h^2$. 
The rainbow color bar shows the variation of $\lambda_{1H}$. With the presence of Higgs-mediated $s-$channel diagrams, 
Higgs resonance drop is observed at $m_{\chi_1}\sim m_h/2$. The presence of a semi-annihilation process, $\chi_i\chi_j\to\chi_k h$ provides another dip near 
$m_{\chi_1}\sim m_h$. DM relic density decreases with larger portal and cubic couplings through annihilation, co-annihilation and semi-annihilation. 
When we move away from Higgs mass regime, the relic density decreases due to less (semi) annihilation contribution and requires larger $\lambda_{1H}$ coupling. 
The shape of the parameter space is otherwise very typical to single component scalar DM model connected via Higgs portal. Semi-annihilation and co-annihilation contribution 
moves the whole spectrum downwards towards smaller relic density and allows smaller values of $\lambda_{1H}$.

\paragraph{$\bullet$~Direct detection limit from WIMP-nucleon inelastic scattering\\}
The relic density allowed parameter space is further constrained by non-observation of DM in direct search, resulting in a  
lower bound on DM nucleon scattering cross section from direct detection experiments like, XENONnT \cite{XENON:2023cxc} and LUX-ZEPLIN \cite{LZ:2022lsv} etc,. 
The self-annihilation and semi-annihilation of this $\mathcal{Z}_3$ WIMP are also constrained by the Fermi-LAT \cite{Fermi-LAT:2015att}, 
CTA \cite{Silverwood:2014yza}, and H.E.S.S. \cite{HESS:2016mib} data.

The DM-nuclei scattering is governed mainly by the Higgs portal interaction term $\chi_1\chi_1^*({\rm H^{\dagger}H})$. For complex scalar DM, 
the spin-independent DM-nucleon inelastic scattering cross-section is an observed quantity in direct detection experiments,
\bea
\sigma_{\chi_1}^{\rm SI}\,=\,\frac{\mu_n^2m_n^2}{4\pi v^2m_{\chi_1}^2}\frac{f_n^2}{m_h^4}\lambda_{h\chi_1\chi_1^*}^2\,,
\label{eq:z3-dd-single}
\eea
where $\mu_n=\frac{m_nm_{\chi_1}}{m_n+m_{\chi_1}}$, $\lambda_{h\chi_1\chi_1^*}=\lambda_{1H}v$, 
and other parameters have usual meaning. Eq\,.~\eqref{eq:z3-dd-single} show that the spin-independent scattering cross-section depends on the DM mass 
$(m_{\chi_1})$ and Higgs portal coupling $(\lambda_{1H})$. So, to satisfy direct search bound, we wish to decrease $\lambda_{1H}$ and adjust other parameters to satisfy 
relic density, where co-annihilation and semi-annihilation play a major role. 

In fig\,.~\ref{fig:z3-single-dd}, we have shown the relic density allowed parameter space in 
$m_{\chi_1}-\sigma_{\chi_1}^{\rm SI}$ plane and color bar represents the mass splitting between $\chi_1$ and $\chi_2$, which characterises co-annihilation contribution to relic density. Parameters kept fixed are mentioned in the figure inset. 

We see that no relic density allowed points are seen at resonance region, or in the semi-annihilation region, they correspond to under abundant region.
This is simply due to the range of $\lambda_{1H}$ chosen for the scan, as it requires much smaller values to satisfy relic, 
due to enhancement in effective annihilation cross-section in these regions. If the DM mass is adequately large and away 
from the Higgs mass region, then semi-annihilation is small and we need smaller $\Delta m$ which enhances co-annihilation. 
In fig\,.~\ref{fig:z3-single-dd}, some region of the parameter space above the Higgs mass regime is excluded by the recent XENONnT and LZ-2022 bound 
represented by light blue and olive green lines, respectively. The three dashed lines correspond to the projected limit spin-independent DM direct detection 
cross-section. The parameter space illustrated here is particular for this benchmark, and one can get more relic density-allowed regions depending on the values of the 
parameters kept fixed.
\paragraph{$\bullet$~Indirect detection limit on the self-annihilation of WIMP\\}
The limits on DM self-annihilation impose another constraint on the relic and DD-allowed parameter space of the model. Here, we are using the indirect search 
limit on DM self-annihilation from Fermi collaboration, six years of observation using 15 dwarf spheroidal galaxies (dSphs) \cite{Fermi-LAT:2015att} and a projected 
sensitivity for 45 dSphs and 15 years of observation \cite{Fermi-LAT:2016afa}. The thick grey and dashed green line in \cref{fig:z3-single-id} 
represents the observed and projected limit from Fermi-LAT\cite{Fermi-LAT:2015att}. At the same time, the grey-shaded region is excluded by the observed Fermi-LAT limit on 
DM self-annihilation into bottom pairs.

The indirect detection limit on the thermal average DM $(\chi_1)$ self-annihilation to the bottom pair is evaluated at the DM freeze-out point, $\rm T^{FO}_{\chi_1}\sim m_{\chi_1}/25$, which depends on the DM mass $(m_{\chi_1})$ and Higgs portal coupling $(\lambda_{1H})$. The thermal average cross-section, $\langle\sigma v \rangle_{\chi_1\chi_1^*\to b\overline{b}}$, is nearly constant below $m_h/2$ region, but, maximum near Higgs resonance. Near Higgs mass, semi-annihilation helps decrease the $\lambda_{1H}$ coupling.
The DM mass region, near and below the Higgs resonance, is mostly excluded for the specific choice of the benchmark point and allowed above the resonance region. To see the corresponding DD cross section, see the rainbow color bar in \cref{fig:z3-single-id}.
\paragraph{$\bullet$~Indirect detection limit on the semi-annihilation of WIMP\\}
Recently, the gamma-ray observation from Fermi-LAT\cite{Fermi-LAT:2015att}, H.E.S.S \cite{HESS:2016mib} telescope, and also CTA \cite{Silverwood:2014yza} provides a conservative projected limit on the DM ($\chi_1$) semi-annihilation. The presence of cubic interaction in $\mathcal{Z}_3$ symmetric WIMP provides semi-annihilation channel via $\langle\sigma v\rangle_{\chi_1\chi_1\to \chi_1^*h}$ 
and is evaluated at the freeze-out point of $\chi_1$ having s-wave contribution of $\chi_1\chi_1^*\to\chi_1 h$. This cross-section depends on $m_{\chi_1},~m_{\chi_2},~\lambda_{1H},~\lambda_{12H},~\mu_2,$ and $~\mu_3$. In fig\,.~\ref{fig:z3-single-id-semi}, we show the relic density allowed parameter space $m_{\chi_1}-\langle\sigma v\rangle^{\rm ID}_{\chi_1\chi_1\to \chi_1^*h}$ plane. The color bar 
shows variation with respect to the DD cross-section. We see that for the chosen values of the parameters, relic density allowed points lie below the existing limits. Like other 
plots, the absence of points near the Higgs resonance and the Higgs mass can be seen, mainly due to the range of parameters chosen for the scan. However, the DM is well allowed up to 
TeV, due to co-annihilation and self-annihilation processes through the presence of additional complex scalar $\chi_2$, transforming under the same symmetry \cite{Belanger:2012zr}.
.
\subsection{Two component DM }
Eq\,.~\eqref{eq:z3-model} describes the extension of the SM containing two complex scalar fields that transform differently under a single $ \mathcal{Z}_3$ symmetry with the charge charge combinations of the fields $q_{1} \neq q_{2}$ and $q_{1}+q_{2} =N$. Interaction terms depend on their transformation charges; see \cref{tab:z3} for details. For the charge combinations $q_{1} = q_{2}$ and $q_{1} + q_{2} \neq N$, the interaction terms can be found from table~\ref{tab:z3_q1_neq_q2}. Such interaction terms between $\chi_1$ and $\chi_2$, like $\chi_1^2\chi_2^*$ or $\chi_2^2\chi_1^*$, etc., open up the decay of the heavier component to the lighter one(s).
A suitable choice of mass hierarchy, $2\,m_{\rm lighter}>m_{\rm higher}$, can stop such decay.
Further, both fields are connected with the visible sector via Higgs portal interactions. The one connecting both the fields, like $\chi_1\chi_2|H|^2$ leads to the decay of the heavier component to the lighter one plus the SM at tree level. A mass hierarchy like $m_{\chi_2}>m_{\chi_1}+m_h$ can stop such on-shell decay; however, off-shell Higgs decay to di-photon or di-gluon is always permitted for non-degenerate masses of the dark sector particles.
Considering sufficiently small portal coupling ($\lambda_{12H}$) associated with this term can stop on-shell or off-shell 
decay of the heavier particle to SM at tree level. 
However, one-loop and two-loop decay terms are still possible in the presence of  self-interaction terms between 
$\chi_1$ and $\chi_2$.
Hence, even after neglecting $\lambda_{12H}$, one needs to 
choose very small $\lambda_3^{}$, $\lambda_4^{}$, and $\mu_2$ to restrict one-loop decay, although the choice of $\mu_2$ isn't unique. However, after sacrificing all of these parameters, the two-loop decay (see \cref{sec:decay-z3}) is still possible.
To restrict the two-loop decay, different other parameters can be chosen to be small, giving rise to different kinds of scenarios, as shown in \cref{tab:z3}. 
The red colored terms in \cref{tab:z3} can be sacrificed to obtain the second long lived DM component.
In the \cref{tab:z3-benchmark} of \cref{sec:decay-z3}, we have also given an estimate of smallness of these couplings, $\lambda_{3},~\lambda_{4},~\lambda_{12H},~\mu_2\lesssim 10^{-15}$, 
so that the decay lifetime is larger than the age of the universe. Note that for $\mathcal{Z}_{3}$ model, 
the possibility of $q_1\neq q_2, q_1+q_2 \neq N$ do not arise. Therefore, one has to sacrifice some couplings to make both components stable. We also provide, for reference, the terms that are absent in the $q_{1} = q_{2}$ scenario (table~\ref{tab:z3_q1_neq_q2}) (marked in red color), though we will not analyze this case in detail. 
\begin{table}[htb!]
\centering
\scalebox{0.75}{
\begin{tabular}{|c |c |c |c |}\hline
\cellcolor{gray!30}Scenarios&\cellcolor{gray!30}Interaction terms of two DMs: $\chi_1$ and $\chi_2$ under $\mathcal{Z}_3$ symmetry \\ \hline\hline
\rowcolor{gray!20}\multicolumn{2}{|c|}{\text{\bf$q_1=1,~q_2=2~{\rm or}~q_1=2,~q_2=1$}}\\\hline
\multirow{2}{*}{A}&\multirow{2}{*}{${\rm \left| \chi _1\right|^2H^{\dagger}H,~\left| \chi _2\right|^2H^{\dagger}H},~\chi_1^3,~{\color{black}\chi_2^3},~|\chi_1|^4,~|\chi_2|^4,~{\color{red}\chi_1\chi_2\rm H^{\dagger}H},~{\color{red}\chi_1^2\chi_2^2},~{\color{black}|\chi_1\chi_2|^2},~{\color{red}\chi_2^2\chi_1^*},~{\color{red}\chi_1^2\chi_2^*},~{\color{red}\chi_1\chi_2(|\chi_1|^2+|\chi_2|^2)}$}\\& \\\hline
\multirow{2}{*}{B}&\multirow{2}{*}{${\rm \left| \chi _1\right|^2H^{\dagger}H,~{\color{black}\left| \chi _2\right|^2H^{\dagger}H}},~\chi_1^3,~{\color{red}\chi_2^3},~|\chi_1|^4,~|\chi_2|^4,~{\color{red}\chi_1\chi_2\rm H^{\dagger}H},~{\color{black}\chi_1^2\chi_2^2},~{\color{black}|\chi_1\chi_2|^2},~{\color{black}\chi_2^2\chi_1^*},~{\color{red}\chi_1^2\chi_2^*},~{\color{red}\chi_1\chi_2(|\chi_1|^2+|\chi_2|^2)}$}\\& \\\hline
\multirow{2}{*}{C}&\multirow{2}{*}{${\rm \left| \chi _1\right|^2H^{\dagger}H,~{\color{black}\left| \chi _2\right|^2H^{\dagger}H}},~{\color{red}\chi_1^3},~{\color{black}\chi_2^3},~|\chi_1|^4,~|\chi_2|^4,~{\color{red}\chi_1\chi_2\rm H^{\dagger}H},~{\color{black}\chi_1^2\chi_2^2},~{\color{black}|\chi_1\chi_2|^2},~{\color{red}\chi_2^2\chi_1^*},~{\color{black}\chi_1^2\chi_2^*},~{\color{red}\chi_1\chi_2(|\chi_1|^2+|\chi_2|^2)}$}\\& \\\hline
\end{tabular}}
\caption{Terms having two dark sector scalar fields oppositely transforming under discrete symmetry $\mathcal{Z}_3$. Red colour terms are the minimum number of terms to be sacrificed to stop the tree and loop-level decays of the heavier DS particle.}
\label{tab:z3}
\end{table}

\begin{table}[htb!]
\centering
\scalebox{0.75}{
\begin{tabular}{|c |c |c |c |}\hline
\cellcolor{gray!30}Scenarios&\cellcolor{gray!30}Interaction terms of two DMs: $\chi_1$ and $\chi_2$ under $\mathcal{Z}_3$ symmetry \\ \hline\hline
\rowcolor{gray!20}\multicolumn{2}{|c|}{\text{$\bf q_1=q_2=1,~2$}}\\\hline
\multirow{2}{*}{D}&\multirow{2}{*}{${\rm \left| \chi _1\right|^2H^{\dagger}H,~\left| \chi _2\right|^2H^{\dagger}H},~\chi_1^3,~{\color{black}\chi_2^3},~|\chi_1|^4,~|\chi_2|^4,~{\color{red}\chi_1\chi_2^*\rm H^{\dagger}H},~{\color{red}(\chi_1\chi_2^*)^2},~{\color{black}|\chi_1\chi_2|^2},~{\color{red}\chi_2^2\chi_1},~{\color{red}\chi_1^2\chi_2},~{\color{red}\chi_1\chi_2^*(|\chi_1|^2+|\chi_2|^2)}$}\\& \\\hline
\multirow{2}{*}{E}&\multirow{2}{*}{${\rm \left| \chi _1\right|^2H^{\dagger}H,~{\color{black}\left| \chi _2\right|^2H^{\dagger}H}},~\chi_1^3,~{\color{red}\chi_2^3},~|\chi_1|^4,~|\chi_2|^4,~{\color{red}\chi_1\chi_2^*\rm H^{\dagger}H},~{\color{black}(\chi_1\chi_2^*)^2},~{\color{black}|\chi_1\chi_2|^2},~{\color{black}\chi_2^2\chi_1},~{\color{red}\chi_1^2\chi_2},~{\color{red}\chi_1\chi_2^*(|\chi_1|^2+|\chi_2|^2)}$}\\& \\\hline
\multirow{2}{*}{F}&\multirow{2}{*}{${\rm \left| \chi _1\right|^2H^{\dagger}H,~{\color{black}\left| \chi _2\right|^2H^{\dagger}H}},~{\color{red}\chi_1^3},~{\color{black}\chi_2^3},~|\chi_1|^4,~|\chi_2|^4,~{\color{red}\chi_1\chi_2^*\rm H^{\dagger}H},~{\color{black}(\chi_1\chi_2^*)^2},~{\color{black}|\chi_1\chi_2|^2},~{\color{red}\chi_2^2\chi_1},~{\color{black}\chi_1^2\chi_2},~{\color{red}\chi_1\chi_2^*(|\chi_1|^2+|\chi_2|^2)}$}\\& \\\hline
\end{tabular}}
\caption{Interaction terms between two DS scalar fields, $\chi_{1}$ and $\chi_{2}$, carrying identical $\mathcal{Z}_3$ charges. The highlighted red terms represent the minimal set that must be eliminated to prevent both tree-level and loop-level decays of the heavier DS particle.}
\label{tab:z3_q1_neq_q2}
\end{table}

Depending on the choice of the couplings required to stabilize the heavier DM component, six possibilities emerge as shown in \cref{tab:z3} $(q_1\neq q_2)$ and \cref{tab:z3_q1_neq_q2} $(q_1=q_2)$. 
The absence of red-coloured terms indicates the presence of a new discrete symmetry, as mentioned before.
For example, scenario A and D correspond to $\mathcal{Z}_{3} \otimes \mathcal{Z}^{\prime}_{3}$ scenario.
Similarly, scenario B and C turns out to be $\mathcal{Z}_6$ symmetry with $q_1=4,~q_2=5$ or $q_1=2,~q_2=1$ and $q_1=5,~q_2=4$ or $q_1=1,~q_2=2$, respectively.
Scenario E and F falls under $\mathcal{Z}_6$ with $q_1=4,~q_2=1$ or $q_1=2,~q_2=5$ and $q_1=1,~q_2=4$ or $q_1=5,~q_2=2$, respectively. These observations are summarised in table \ref{tab:z3new}.
Let us recall that when we can have $q_1\neq q_2, q_1+q_2 \neq N$, we can stabilise both the components kinematically. 

\begin{table}[htb!]
\centering
\scalebox{1.0}{
\begin{tabular}{|>{\columncolor{gray!10}}c |>{\columncolor{gray!10}}c |>{\columncolor{gray!5}}c |>{\columncolor{gray!5}}c |>{\columncolor{gray!5}} c|>{\columncolor{gray!5}} c|}\hline
\multicolumn{2}{|>{\columncolor{gray!10}}c|}{DS particles and charges} & \multicolumn{3}{>{\columncolor{gray!5}}c|}{DM particles symmetry and charges} \\ 
\hline
symmetry and charges & Scenarios & Resulting Symmetry & $q_{1}$ & $q_{2}$ \\ \hline
&  &  & 1 & 1,2 \\ \cline{4-5}
& \multirow{-2}{*}{A} & \multirow{-2}{*}{$\mathcal{Z}_{3} \otimes \mathcal{Z}_{3}'$} & 2 & 1,2 \\ \cline{2-5} 
 & &   & 4 & 5 \\ \cline{4-5}
&  \multirow{-2}{*}{B} & \multirow{-2}{*}{$\mathcal{Z}_{6} $} & 2 & 1 \\ \cline{2-5} 
& &  & 5 & 4 \\ \cline{4-5}
\multirow{-6}{*}{$\mathcal{Z}_{3}: (q_{1},q_{2}) = (1,2) $ or $(2,1)$}& \multirow{-2}{*}{C}  &  \multirow{-2}{*}{$\mathcal{Z}_{6} $} & 1 & 2 \\ 
\hline
\hline
&  &  & 1 & 1,2 \\\cline{4-5}
& \multirow{-2}{*}{D} &\multirow{-2}{*}{$\mathcal{Z}_{3} \otimes \mathcal{Z}_{3}'$}  & 2 & 1,2 \\ \cline{2-5} 
&  &   & 4 & 1 \\ \cline{4-5}
& \multirow{-2}{*}{E} &\multirow{-2}{*}{$\mathcal{Z}_{6} $}  & 2 & 5 \\ \cline{2-5} 
& &  & 1 & 4 \\ \cline{4-5}
\multirow{-6}{*}{$\mathcal{Z}_{3}: (q_{1},q_{2}) = (1,1) $ or $(2,2)$ }& \multirow{-2}{*}{F}  & \multirow{-2}{*}{$\mathcal{Z}_{6} $}  & 5 & 2 \\ \hline
\end{tabular}}
\caption{The resultant symmetries after imposing stabilization conditions on DS particles from \(\mathcal{Z}_{3}\) symmetric Lagrangian.}
\label{tab:z3new}
\end{table}

Further, if we want one of the components to be pFIMP, we have to choose the corresponding portal coupling feeble, but DM-DM interaction sizeable. Now the scenarios 
that arise here, having two DM components, are not all independent.  
In the following, we elaborate upon scenarios A and B; scenarios C, D, E, and F are somewhat similar to them. 
\subsubsection{Scenario-A}
\label{subsec:A}
From \cref{tab:z3}, we see that this particular scenario arises via tiny couplings associated with $\chi_1\chi_2\rm H^{\dagger}H,~\chi_1^2\chi_2^2,~\chi_2^2\chi_1^*,~\chi_1^2\chi_2^*,~\chi_1\chi_2|\chi_1|^2,~\chi_1\chi_2|\chi_2|^2$ interaction terms, to give rise to one LLP and one stable DM.
This is then equivalent to $\mathcal{Z}_3\otimes\mathcal{Z}^{\prime}_{3}$ scenario \cite{Bhattacharya:2017fid}. But, in the two WIMP case, limited parameter space is 
left for future direct detection. However, WIMP-pFIMP possibility enhances the allowed parameter space, as discussed below.

\begin{figure}[htb!]
\centering
\subfloat[]{\begin{tikzpicture}
\begin{feynman}
\vertex(a);
\vertex[left=1cm and 1cm  of a] (a1){\(\chi_1\)};
\vertex[right=1cm and 1cm  of a] (a2){\(\chi_1\)}; 
\vertex[below=1.5cm of a] (b); 
\vertex[left=1cm and 1cm of b] (c1){\(N\)};
\vertex[right=1cm and 1cm of b] (c2){\(N\)};
\diagram* {
(a1) -- [ line width=0.25mm,charged scalar, arrow size=0.7pt,style=blue] (a) -- [ line width=0.25mm,charged scalar, arrow size=0.7pt,style=black] (a2),
(a) -- [ line width=0.25mm,scalar, edge label={\(\color{black}{h}\)}, style=red] (b) ,
(c1) --[ line width=0.25mm,fermion, arrow size=0.7pt,style=blue] (b)  --[ line width=0.25mm,fermion, arrow size=0.7pt,style=black] (c2)};
\end{feynman}
\end{tikzpicture}\label{fig:z3-wimp_dd}}\quad
\subfloat[]{\begin{tikzpicture}
\begin{feynman}
\vertex (a);
\vertex[above left=0.5cm and 0.5cm of a] (a1){\(\chi_1\)};
\vertex[below left=0.5cm and 0.5cm of a] (a2){\(\chi_1\)}; 
\vertex[right=1cm of a] (b); 
\vertex[above right=0.5cm and 0.5cm of b] (b1){\(\rm b\)};
\vertex[below right=0.5cm and 0.5cm of b] (b2){\(\rm b\)}; 
\diagram*{
(a1) -- [ line width=0.25mm,charged scalar, arrow size=0.7pt, style=blue] (a) -- [ line width=0.25mm,charged scalar, arrow size=0.7pt, style=blue] (a2),
(b1) -- [ line width=0.25mm, fermion, arrow size=0.7pt, style=black] (b) -- [ line width=0.25mm, fermion, arrow size=0.7pt, style=black] (b2), 
(b) -- [ line width=0.25mm, scalar, arrow size=0.7pt, edge label={\(\rm \color{black}{h}\)},style=red] (a)};
\end{feynman}
\end{tikzpicture}\label{fig:z3-wimp_id}}\quad
\subfloat[]{\begin{tikzpicture}
\begin{feynman}
\vertex (a);
\vertex[above left=0.5cm and 0.5cm of a] (a1){\(\chi_1\)};
\vertex[below left=0.5cm and 0.5cm of a] (a2){\(\chi_1\)}; 
\vertex[right=1cm of a] (b); 
\vertex[above right=0.5cm and 0.5cm of b] (b1){\(\rm h\)};
\vertex[below right=0.5cm and 0.5cm of b] (b2){\(\rm \chi_1\)}; 
\diagram* {
(a1) -- [ line width=0.25mm,charged scalar, arrow size=0.7pt, style=blue] (a),
(a2) -- [ line width=0.25mm,charged scalar, arrow size=0.7pt, style=blue] (a), 
(b) -- [ line width=0.25mm,charged scalar, arrow size=0.7pt, edge label={\(\rm \color{black}{\chi_1}\)},style=red] (a),
(b2) -- [ line width=0.25mm,charged scalar, arrow size=0.7pt, style=black] (b),
(b1)-- [ line width=0.25mm, scalar, arrow size=0.7pt, style=black] (b), 
};
\end{feynman}
\end{tikzpicture}\label{fig:z3-wimp_semi-id1}}\quad
\subfloat[]{\begin{tikzpicture}
\begin{feynman}
\vertex (a);
\vertex[left=1cm and 1cm  of a] (a1){\(\chi_1\)};
\vertex[right=1cm and 1cm  of a] (a2){\(\rm h\)}; 
\vertex[below=1.5cm of a] (b); 
\vertex[left=1cm and 1cm of b] (c1){\(\chi_1\)};
\vertex[right=1cm and 1cm of b] (c2){\(\chi_1\)};
\diagram* {
(a1) -- [ line width=0.25mm,charged scalar, arrow size=0.7pt,style=blue] (a),
(a) -- [ line width=0.25mm, scalar, arrow size=0.7pt,style=black] (a2),
(a) -- [ line width=0.25mm,charged scalar,  arrow size=0.7pt,edge label={\(\color{black}{\chi_1}\)}, style=red] (b) ,
(c1) --[ line width=0.25mm,charged scalar , arrow size=0.7pt,style=blue] (b),
(c2)  --[ line width=0.25mm,charged scalar, arrow size=0.7pt,style=black] (b)};
\end{feynman}
\end{tikzpicture}\label{fig:z3-wimp_semi-id2}}

\subfloat[]{\begin{tikzpicture}
\begin{feynman}
\vertex (a);
\vertex[left=1cm and 1cm  of a] (a1){\(\chi_1\)};
\vertex[ right=1cm and 1cm  of a] (a2){\(\rm h\)}; 
\vertex[below=1.5cm of a] (b); 
\vertex[ left=1cm and 1cm of b] (c1){\(\chi_1\)};
\vertex[ right=1cm and 1cm of b] (c2){\(\chi_1\)};
\diagram* {
(a1) -- [ line width=0.25mm,charged scalar, arrow size=0.7pt,style=blue] (a),
(c2) -- [ line width=0.25mm,charged scalar, arrow size=0.7pt,style=black] (a),
(b) -- [ line width=0.25mm,charged scalar,  arrow size=0.7pt,edge label={\(\color{black}{\chi_1}\)}, style=red] (a) ,
(c1) -- [ line width=0.25mm,charged scalar , arrow size=0.7pt,style=blue] (b),
(b)  -- [ line width=0.25mm, scalar, arrow size=0.7pt,style=black] (a2)};
\end{feynman}
\end{tikzpicture}\label{fig:z3-wimp_semi-id3}}\quad
\subfloat[]{\begin{tikzpicture}
\begin{feynman}
\vertex (a);
\vertex[left=1cm and 1cm of a] (a1){\(\chi_2\)};
\vertex[right=1cm and 1cm of a] (a2){\(\chi_2\)}; 
\vertex[below=1.5cm of a] (b); 
\vertex[below=0.75cm of a] (c); 
\vertex[left=1cm and 1cm of b] (c1){\(N\)};
\vertex[right=1cm and 1cm of b] (c2){\(N\)};
\diagram* {
(a1) -- [ line width=0.25mm,charged scalar, arrow size=0.7pt, style=blue] (a),
(a) -- [ line width=0.25mm,charged scalar, arrow size=0.7pt, style=black] (a2),
(c) -- [ line width=0.25mm, scalar, edge label={\(\color{black}{\rm h}\)}, style=red] (b),
(c) -- [ line width=0.25mm, charged scalar, half left, arrow size=0.7pt, edge label={\(\color{black}{\chi_1}\)}, style=gray] (a),
(a) -- [ line width=0.25mm, charged scalar, half left, arrow size=0.7pt, edge label={\(\color{black}{\chi_1}\)}, style=gray] (c) ,
(c1) -- [ line width=0.25mm, fermion, arrow size=0.7pt, style=blue] (b),
(b)  --[ line width=0.25mm, fermion, arrow size=0.7pt] (c2)};
\node at (a)[circle,fill,style=gray,inner sep=1pt]{};
\node at (b)[circle,fill,style=gray,inner sep=1pt]{};
\node at (c)[circle,fill,style=gray,inner sep=1pt]{};
\end{feynman}
\end{tikzpicture}\label{feyn:pfimp-dd}}\quad
\subfloat[]{\begin{tikzpicture}
\begin{feynman}
\vertex (a);
\vertex[above left=0.75cm and 1cm of a] (a1){\(\chi_2\)};
\vertex[below left=0.75cm and 1cm of a] (a2){\(\chi_2\)}; 
\vertex[right=0.75cm of a] (b); 
\vertex[right=1.5cm of a] (c); 
\vertex[above right=0.75cm and 1cm of c] (c1){\(\rm b\)};
\vertex[below right=0.75cm and 1cm of c] (c2){\(\rm b\)};
\diagram* {
(a1) -- [ line width=0.25mm,charged scalar, arrow size=0.7pt, style=blue] (a) -- [ line width=0.25mm,charged scalar, arrow size=0.7pt, style=blue] (a2),
(c) -- [ line width=0.25mm, scalar, edge label={\(\color{black}{\rm h}\)}, style=red] (b),
(b) -- [ line width=0.25mm, charged scalar, half left, arrow size=0.7pt, edge label={\(\color{black}{\chi_1}\)}, style=gray] (a),
(a) -- [ line width=0.25mm, charged scalar, half left, arrow size=0.7pt, edge label={\(\color{black}{\chi_1}\)}, style=gray] (b) ,
(c1) --[ line width=0.25mm, fermion, arrow size=0.7pt,style=black] (c) --[ line width=0.25mm, fermion, arrow size=0.7pt,style=black] (c2)};
\node at (a)[circle,fill,style=gray,inner sep=1pt]{};
\node at (b)[circle,fill,style=gray,inner sep=1pt]{};
\node at (c)[circle,fill,style=gray,inner sep=1pt]{};
\end{feynman}
\end{tikzpicture}\label{feyn:pfimp-id}}

\subfloat[]{\begin{tikzpicture}
\begin{feynman}
\vertex (a);
\vertex[above left=0.75cm and 1cm of a] (a1){\(\chi_2\)};
\vertex[below left=0.75cm and 1cm of a] (a2){\(\chi_2\)}; 
\vertex[right=1cm of a] (b); 
\vertex[below right=0.5cm and 0.5cm of b] (c); 
\vertex[above right=0.75cm and 1cm of b] (c1){\(\rm \chi_2\)};
\vertex[below right=0.75cm and 1cm of b] (c2){\(\rm h\)};
\diagram*{
(a1) -- [ line width=0.25mm,charged scalar, arrow size=0.7pt, style=blue] (a),
(a2) -- [ line width=0.25mm,charged scalar, arrow size=0.7pt, style=blue] (a),
(b) -- [ line width=0.25mm,charged scalar, arrow size=0.7pt, edge label={\(\color{black}{\rm \chi_2}\)}, style=red] (a),
(b) -- [ line width=0.25mm,charged scalar, half left, arrow size=0.7pt, edge label={\(\color{black}{\chi_1}\)}, style=gray] (c),
(c) -- [ line width=0.25mm, charged scalar, half left, arrow size=0.7pt, edge label={\(\color{black}{\chi_1}\)}, style=gray] (b),
(c1) --[ line width=0.25mm,charged scalar, arrow size=0.7pt,style=black] (b),
(c) -- [ line width=0.25mm, scalar, arrow size=0.7pt,style=black] (c2)};
\node at (a)[circle,fill,style=gray,inner sep=1pt]{};
\node at (b)[circle,fill,style=gray,inner sep=1pt]{};
\node at (c)[circle,fill,style=gray,inner sep=1pt]{};
\end{feynman}
\end{tikzpicture}\label{fig:pfimp-semiid-1}}\quad
\subfloat[]{\begin{tikzpicture}
\begin{feynman}
\vertex (a);
\vertex[left=1cm of a] (a1){\(\chi_2\)};
\vertex[right=1cm of a] (a2){\(\chi_2\)}; 
\vertex[below=1.5cm of a] (b); 
\vertex[right=0.75cm of b] (c); 
\vertex[left=1cm of b] (c1){\(\rm \chi_2\)};
\vertex[right=0.5cm of c] (c2){\(\rm h\)};
\diagram*{
(a1) -- [ line width=0.25mm,charged scalar, arrow size=0.7pt, style=blue] (a),
(a2) -- [ line width=0.25mm,charged scalar, arrow size=0.7pt, style=black] (a),
(b) -- [ line width=0.25mm,charged scalar, arrow size=0.7pt, edge label={\(\color{black}{\rm \chi_2}\)}, style=red] (a),
(b) -- [ line width=0.25mm,charged scalar, half left, arrow size=0.7pt, edge label={\(\color{black}{\chi_1}\)}, style=gray] (c),
(c) -- [ line width=0.25mm, charged scalar, half left, arrow size=0.7pt, edge label={\(\color{black}{\chi_1}\)}, style=gray] (b),
(c1) --[ line width=0.25mm,charged scalar, arrow size=0.7pt,style=blue] (b),
(c) -- [ line width=0.25mm, scalar, arrow size=0.7pt,style=black] (c2)};
\node at (a)[circle,fill,style=gray,inner sep=1pt]{};
\node at (b)[circle,fill,style=gray,inner sep=1pt]{};
\node at (c)[circle,fill,style=gray,inner sep=1pt]{};
\end{feynman}
\end{tikzpicture}\label{fig:pfimp-semiid-2}}\quad
\subfloat[]{\begin{tikzpicture}
\begin{feynman}
\vertex (a);
\vertex[left=1cm of a] (a1){\(\chi_2\)};
\vertex[right=1cm of a] (a2){\(\chi_2\)}; 
\vertex[below=1.5cm of a] (b); 
\vertex[right=0.75cm of b] (c); 
\vertex[left=1cm of b] (c1){\(\rm \chi_2\)};
\vertex[right=0.5cm of c] (c2){\(\rm h\)};
\diagram*{
(c1) -- [ line width=0.25mm,charged scalar, arrow size=0.7pt, style=blue] (a),
(a2) -- [ line width=0.25mm,charged scalar, arrow size=0.7pt, style=black] (a),
(b) -- [ line width=0.25mm,charged scalar, arrow size=0.7pt, edge label={\(\color{black}{\rm \chi_2}\)}, style=red] (a),
(b) -- [ line width=0.25mm,charged scalar, half left, arrow size=0.7pt, edge label={\(\color{black}{\chi_1}\)}, style=gray] (c),
(c) -- [ line width=0.25mm, charged scalar, half left, arrow size=0.7pt, edge label={\(\color{black}{\chi_1}\)}, style=gray] (b) ,
(a1) --[ line width=0.25mm,charged scalar, arrow size=0.7pt,style=blue] (b),
(c) -- [ line width=0.25mm, scalar, arrow size=0.7pt,style=black] (c2)};
\node at (a)[circle,fill,style=gray,inner sep=1pt]{};
\node at (b)[circle,fill,style=gray,inner sep=1pt]{};
\node at (c)[circle,fill,style=gray,inner sep=1pt]{};
\end{feynman}
\end{tikzpicture}\label{fig:pfimp-semiid-3}}
\caption{Top: Figs\,.~\ref{fig:z3-wimp_dd} and \ref{fig:z3-wimp_id} represent Feynman diagrams corresponding to the direct and indirect detection of WIMP ($\chi_1$) 
respectively. \Cref{fig:z3-wimp_semi-id1,fig:z3-wimp_semi-id2,fig:z3-wimp_semi-id3} correspond to WIMP semi-annihilation processes contributing to its indirect search.
Bottom: Figs\,.~\ref{feyn:pfimp-dd} shows pFIMP direct search, and \cref{feyn:pfimp-id,fig:pfimp-semiid-1,fig:pfimp-semiid-2,fig:pfimp-semiid-3} represent processes that 
contribute to indirect detection of pFIMP ($\chi_2$).}
\label{fig:z3-wimp_dd-id}
\end{figure}
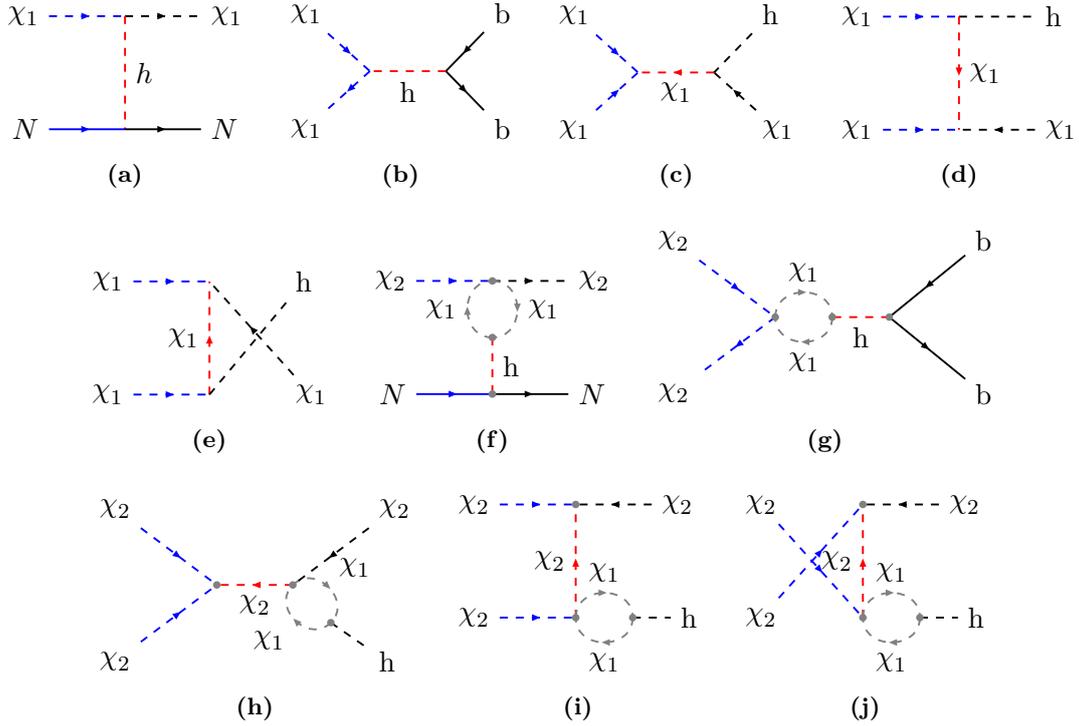

To make the LLP ($\chi_2$) pFIMP, we further need $\lambda_{2H}$ to be very tiny, see \cref{sec:decay-z3} to know the range of couplings.
Then the available parameters, that could be utilised in WIMP-pFIMP phenomenology, are,
\bea
\{m_{\chi_1},~ m_{\chi_2}, ~ \lambda_{\chi_1},~\lambda_{\chi_2},~\lambda_{1H} ,~ \lambda_{12},~\mu_3,{\rm ~and ~} \mu_4\}\,.
\label{eq:z3-free}
\eea 
\paragraph{$\bullet$~Relic density\\}
The total relic abundance for this WIMP-pFIMP set-up, comes from $\chi_1$ as WIMP and $\chi_2$ as pFIMP, to stipulate to,
\bea
{\rm\Omega_{DM}h^2}=2.744\times 10^8 \sum\limits_{i=1,2}m_{\chi_i}Y_{i}[x_{\infty}]\,,
\label{eq:cbeq-relic}
\eea
where $Y_i=n_i/s$, $n_i$ is number density, and $s$ is the entropy density. The DM yield $(Y_i)$ is calculated by solving coupled BEQ for $\chi_1$ and $\chi_2$ 
numerically using micrOmega. The self-annihilation, semi-annihilation, conversion and semi-conversion are the main number-changing processes that contribute 
to the DM relic. It is worthy recalling that after stabilisation of the heavy component, there is no co-annihilation contribution. 
\paragraph{$\bullet$~Direct detection limits on WIMP and pFIMP\\}
The WIMP is weakly coupled to the visible sector via the Higgs portal interaction and can scatter with the detector nuclei. The effective WIMP-nucleon inelastic 
scattering cross section at zero transfer momentum limit $(q_h^2=t\to 0)$ turns out to be, see fig\,.~\ref{fig:z3-wimp_dd},
\bea
\sigma_{\chi_1}^{\rm eff}\,=\,\frac{\Omega_{\chi_1}h^2}{\Omega_{\chi_1}h^2+\Omega_{\chi_2}h^2}\frac{\mu_n^2m_n^2}{4\pi v^2m_{\chi_1}^2}\frac{f_n^2}{m_h^4}|\lambda_{h\chi_1\chi_1^*}|^2\,.
\label{eq:wimp-z3-sigma-ddA}
\eea
In the above, $\mu_n=\frac{m_nm_{\chi_1}}{m_n+m_{\chi_1}}$ where $m_n$ is the nucleon mass, $f_n=\frac{2}{9}+\frac{7}{9}\displaystyle\sum_{u,d,s}f_{T_q}^n$ with $f_{T_u}^{p(n)}=0.018~(0.013),~f_{T_d}^{p(n)}=0.027~(0.040),$ and $f_{T_s}^{p(n)}=0.037~(0.037)$ \cite{Abe:2015rja}.

The direct detection of pFIMP is possible only by the WIMP loop-mediated penguin and vertex correction diagrams, see fig\,.~\ref{feyn:pfimp-dd}. 
The effective spin-independent pFIMP-nucleon $(\chi_2-N)$ inelastic scattering cross-section at zero transfer momentum $(q_h^2=t\to 0)$ limit is given by,
\bea
\sigma_{\chi_2}^{\rm eff}\,=\,\frac{\Omega_{\chi_2}h^2}{\Omega_{\chi_1}h^2+\Omega_{\chi_2}h^2}\frac{\mu_n^2m_n^2}{4\pi v^2m_{\chi_2}^2}\frac{f_n^2}{m_h^4}|\Gamma_{h\chi_2\chi_2^*}^{\rm total}|^2_{t\to 0}\,,
\label{eq:pfimp-z3-sigma-ddA}
\eea
where, $\mu_n=\frac{m_nm_{\chi_2}}{m_n+m_{\chi_2}}$ and $\Gamma_{h\chi_2\chi_2^*}^{\rm total}=\Gamma_{h\chi_2\chi_2^*}^{\ref{feyn:pfimp-dd}}$. One can easily calculate,
\begin{gather}
\Gamma_{h\chi_2\chi_2^*}^{\ref{feyn:pfimp-dd}}\left(q_h^2\right)=  -i\lambda^{\rm relic}_{h\chi_2\chi_2^*}-i\frac{\lambda_{\chi_1\chi_1^*\chi_2\chi_2^*}\lambda_{h\chi_1\chi_1^*}}{16\pi^2}\bigintsss\limits_0^1 dx\ln\left[\frac{m_{\chi_1}^2-x(1-x)4m_{\chi_2}^2}{m_{\chi_1}^2-x(1-x)q_h^2}\right]\,,
\end{gather}
where, $q_h$ is transfer momentum associated with Higgs, $\lambda_{\chi_1\chi_1^*\chi_2\chi_2^*}=-i\lambda_{12}{\rm,~ and}~\lambda_{h\chi_1\chi_1^*}=-i\lambda_{1H}v.$
Using eqs\,.~\ref{eq:wimp-z3-sigma-ddA} and \ref{eq:pfimp-z3-sigma-ddA}, we impose the presently available direct detection constraint from 
LUX-ZEPLIN, XENONnT, and PandaX-xT experiments on the relic density allowed parameter space of the model.
\begin{figure}[htb!]
\centering
\subfloat[]{\includegraphics[width=0.475\linewidth]{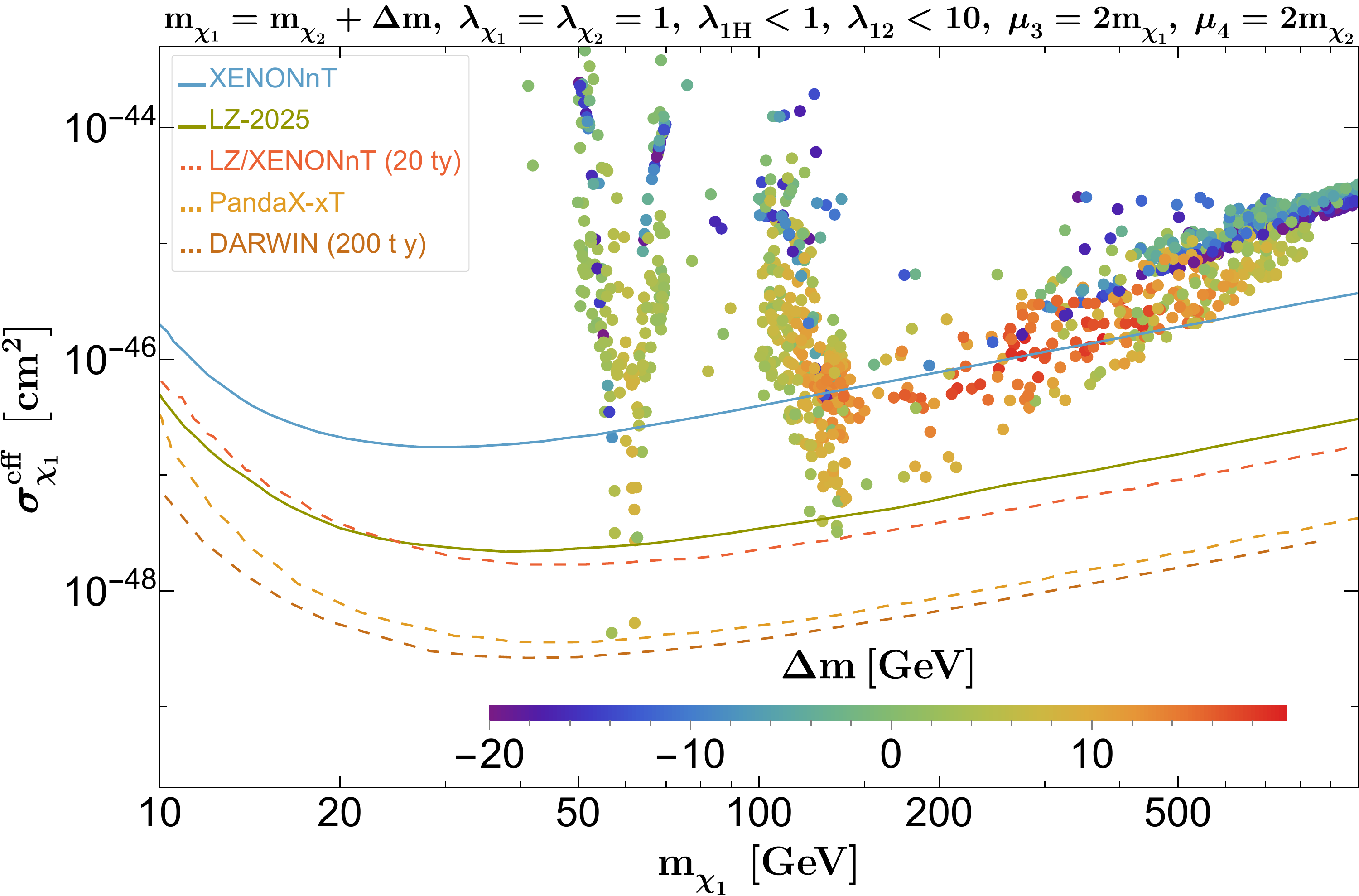}\label{fig:z3-twoscalar_chi1-dd1}}~~
\subfloat[]{\includegraphics[width=0.475\linewidth]{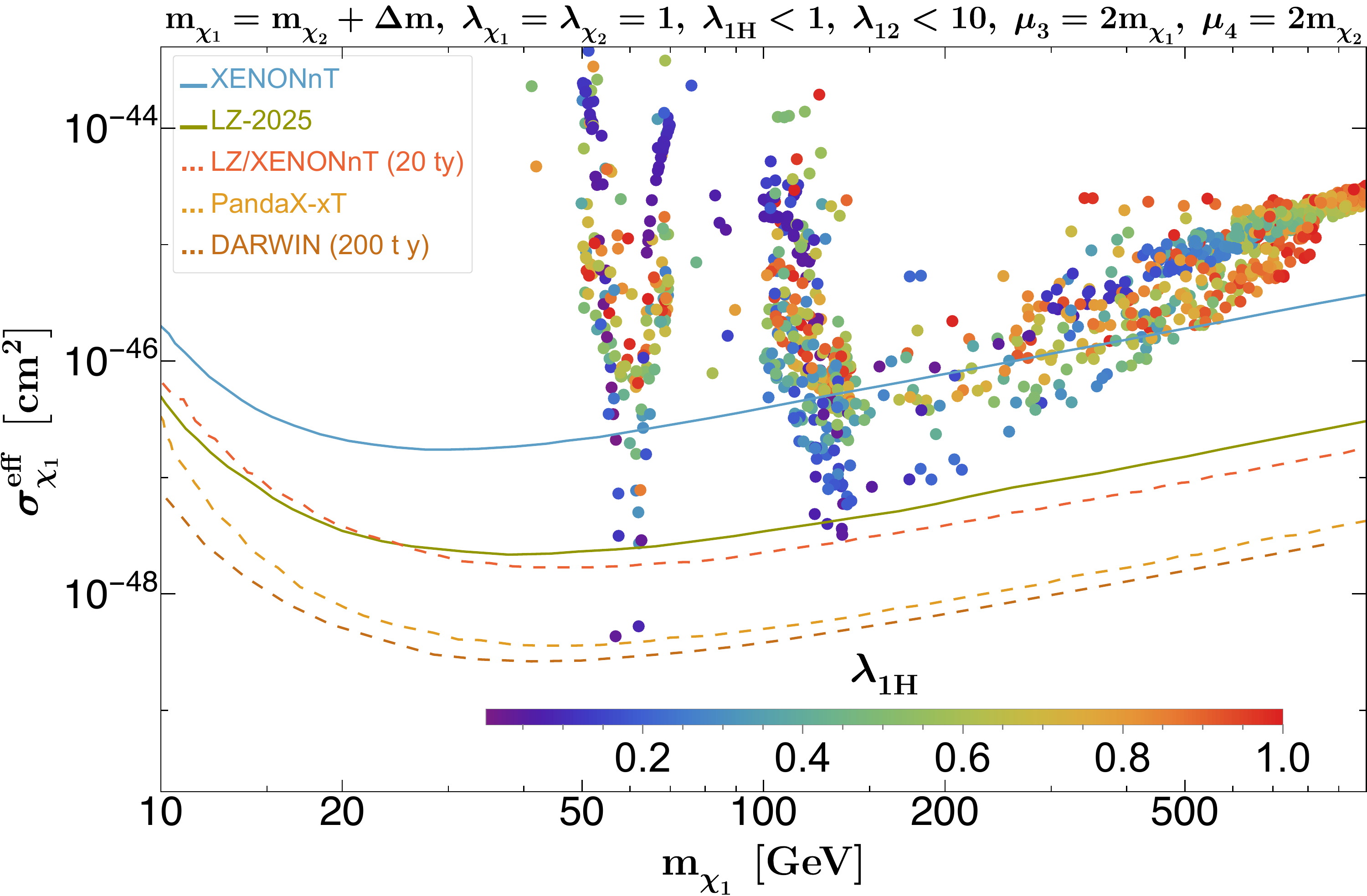}\label{fig:z3-twoscalar_chi1-dd2}}

\subfloat[]{\includegraphics[width=0.475\linewidth]{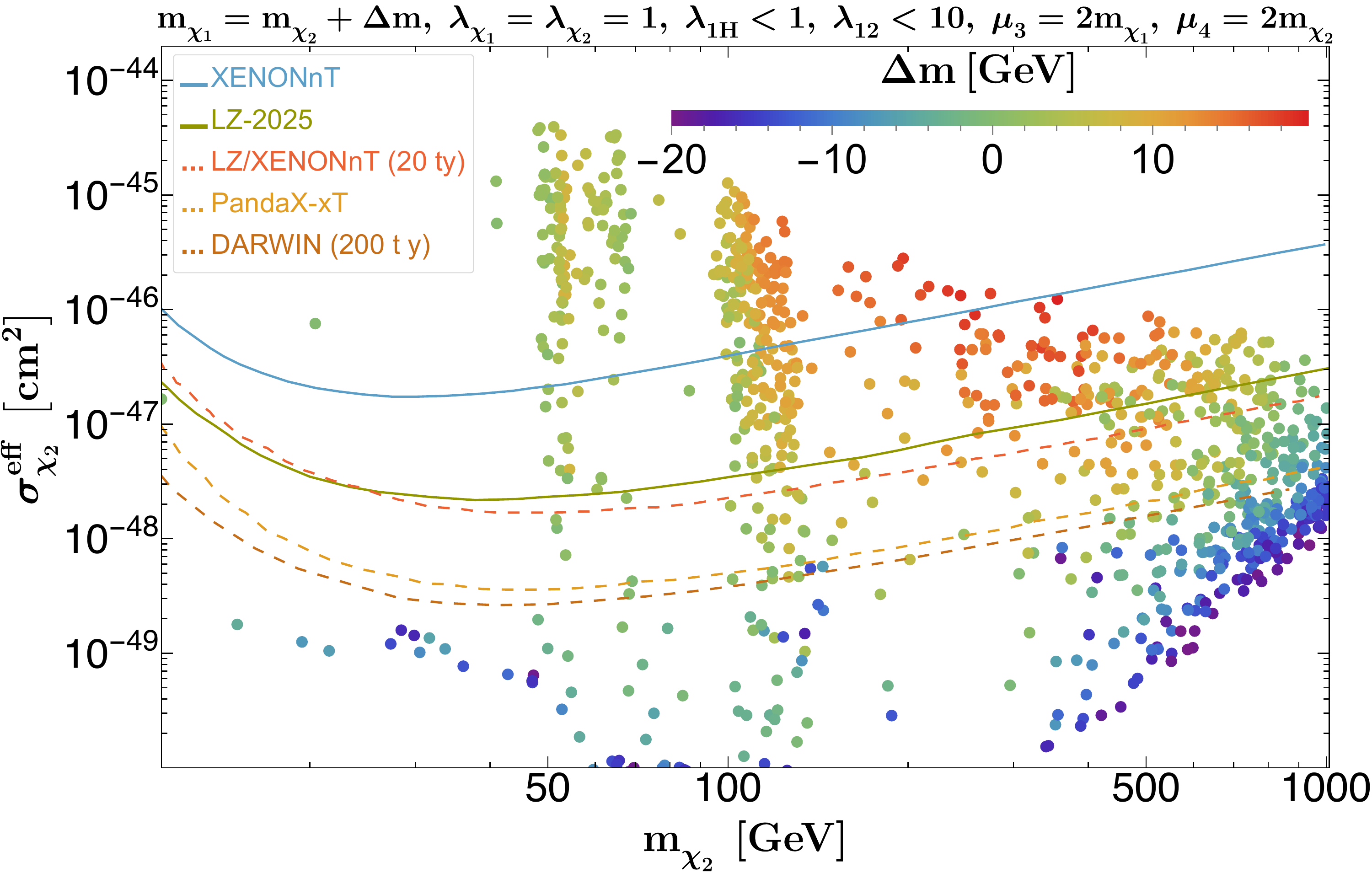}\label{fig:z3-twoscalar_chi2-dd1}}~~
\subfloat[]{\includegraphics[width=0.475\linewidth]{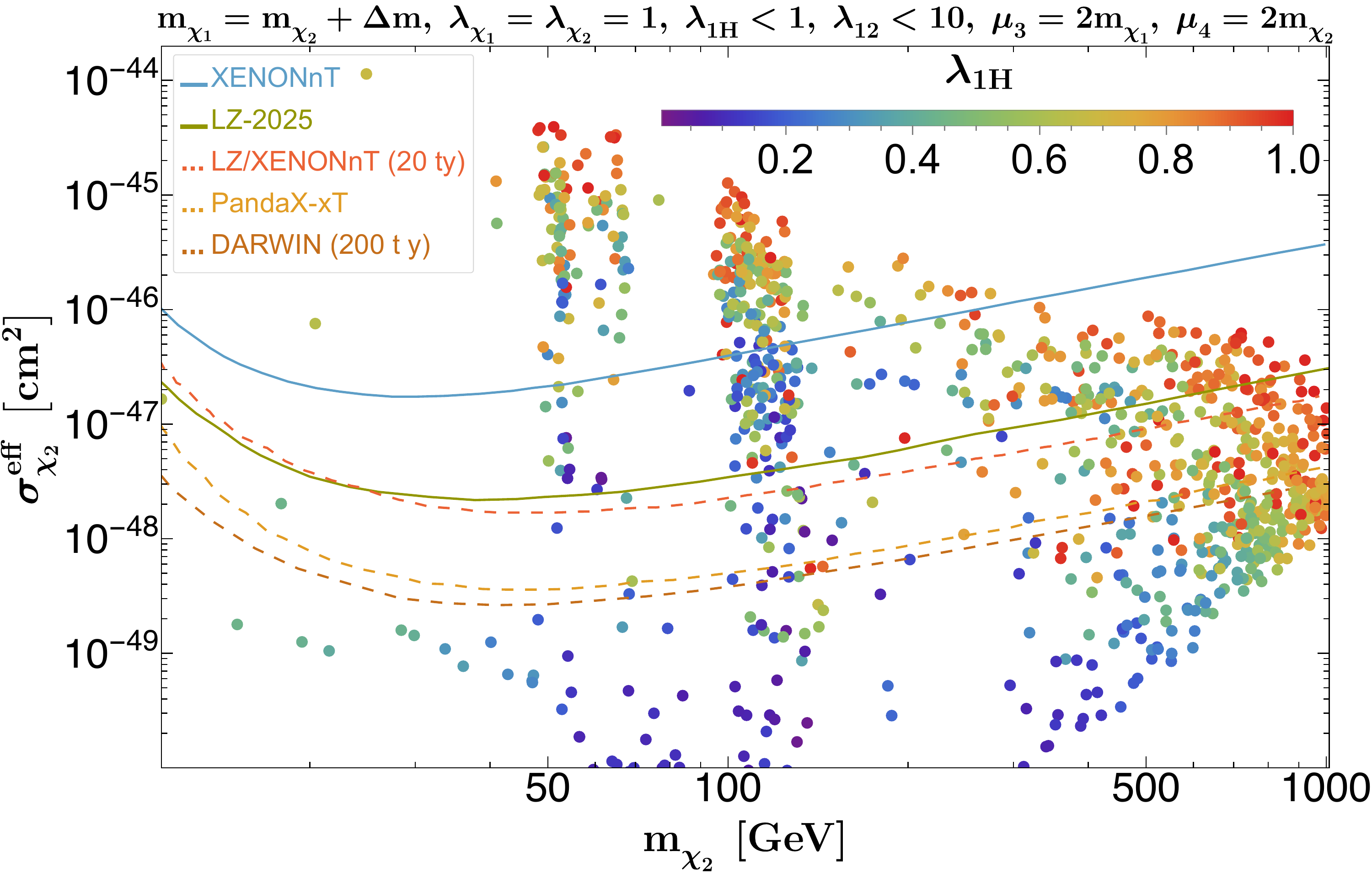}\label{fig:z3-twoscalar_chi2-dd2}}
\caption{In figs\,.~\ref{fig:z3-twoscalar_chi1-dd1}, \ref{fig:z3-twoscalar_chi1-dd2}, \ref{fig:z3-twoscalar_chi2-dd1}, and \ref{fig:z3-twoscalar_chi2-dd2}, we have shown the relic 
density allowed parameter space of the two component WIMP-pFIMP set up, on $m_{\chi_1}-\sigma_{\chi_1}^{\rm eff}$ (top panel) and 
$m_{\chi_2}-\sigma_{\chi_2}^{\rm eff}$ planes (bottom panel). We have taken $\lambda_{12}^{\prime}=\lambda_{3}=\lambda_4=10^{-20}$ and 
$\mu_1=\mu_2=10^{-15}$ GeV to stabilise $\chi_2$, and choose $\lambda_{2H}=10^{-12}$ to make $\chi_2$ a pFIMP. Other parameters kept fixed or varied are shown in the figure heading and inset.}
\label{fig:dd-A}
\end{figure}
\Cref{fig:z3-twoscalar_chi1-dd1,fig:z3-twoscalar_chi1-dd2}, shows the relic density allowed parameter space in
$m_{\chi_1}-\sigma_{\chi_1}^{\rm eff}$ plane (corresponding to WIMP) while the rainbow colour bar represents the DM mass difference and $\lambda_{1H}^{}$ respectively.
Here, we have taken $\mu_3=2m_{\chi_1}$ and $\mu_4=2m_{\chi_2}$, which are most useful parameters for DM semi-annihilation. The Higgs resonance ($m_{\chi_1}\sim m_{h}/2$) 
and semi-annihilation ($m_{\chi_1}\sim m_{h}$) of WIMP help it to acquire under abundance, while the total relic is adjusted by the pFIMP, and relax the 
$\lambda_{1H}$ to come under the present DD bound. Beyond the Higgs resonance or semi-annihilation regime, cross-section decreases and relic density increases, which is adjusted by enhancing the portal coupling $\lambda_{1H}^{}$, respecting DD bound.

\Cref{fig:z3-twoscalar_chi2-dd1,fig:z3-twoscalar_chi2-dd2} represent DM relic density allowed parameter space in $m_{\chi_2}-\sigma_{\chi_2}^{\rm eff}$ plane 
corresponding to pFIMP. The spin-independent scattering cross-section, $\sigma_{\chi_2}^{\rm eff}$ depends on the total loop amplitude 
($\Gamma_{h\chi_2\chi_2^*}^{\rm total}$), which is function of $\lambda_{12},~\lambda_{1H},~{\rm and}~\mu_1$ couplings, and they
decide which process in \cref{feyn:pfimp-dd} dominantly contribute to pFIMP direct search cross-section. Notably, $\sigma_{\chi_2}^{\rm eff}$ 
also depends on the mass splitting $\Delta m$ and pFIMP relic density $\Omega_{\chi_2}h^2$, which is small for large conversion rate. 
The different thick (dashed) coloured lines correspond to existing (projected) bounds from the different experiments mentioned in the figure inset.
Essentially the part of relic density allowed parameter space with $\lambda_{1H} \gtrsim 0.5$ is under conflict with DD limits.  

\paragraph{$\bullet$~Indirect detection limits on WIMP and pFIMP\\}
The limit on DM self-annihilation into $b\overline{b},~W^+W^-,~ZZ$ and $t\overline{t}$ is obtained from the data of Fermi collaboration from 6 years of observation of 15 dwarf 
spheroidal galaxies (dSphs) \cite{Fermi-LAT:2015att}. They also provide projected sensitivity for 45 dSphs of 16 years of observation \cite{Fermi-LAT:2016afa}.
Various gamma-ray observations from Fermi-LAT \cite{Fermi-LAT:2015att}, H.E.S.S. \cite{HESS:2016mib} and Cherenkov Telescope Array (CTA) 
\cite{Silverwood:2014yza} put a bound on DM semi-annihilation $\chi_i~\chi_i\to \chi_i^{*}~h$. We will calculate both the self and semi-annihilation rates in this two 
component WIMP-pFIMP set up and apply the bounds to find allowed parameter space that can be probed further.

The effective WIMP ($\chi_1$) self and semi-annihilation cross-section are given by,
\bea
\langle\sigma v\rangle_{\chi_1\chi_1^*\to b\overline{b}}^{\rm eff}=\left(\dfrac{\Omega_{\chi_1}h^2}{\Omega_{\chi_1}h^2+\Omega_{\chi_2}h^2}\right)^2 \langle\sigma v\rangle_{\chi_1\chi_1^*\to b\overline{b}}\,.
\label{eq:z3-wimp-self}
\eea
and
\bea
\langle\sigma v\rangle_{\chi_1\chi_1\to\chi_1^* h}^{\rm eff}=\left(\dfrac{\Omega_{\chi_1}h^2}{\Omega_{\chi_1}h^2+\Omega_{\chi_2}h^2}\right) \langle\sigma v\rangle_{\chi_1\chi_1\to\chi_1^* h}\,.
\label{eq:z3-wimp-semi}
\eea

\begin{figure}[htb!]
\centering
\subfloat[]{\includegraphics[width=0.475\linewidth]{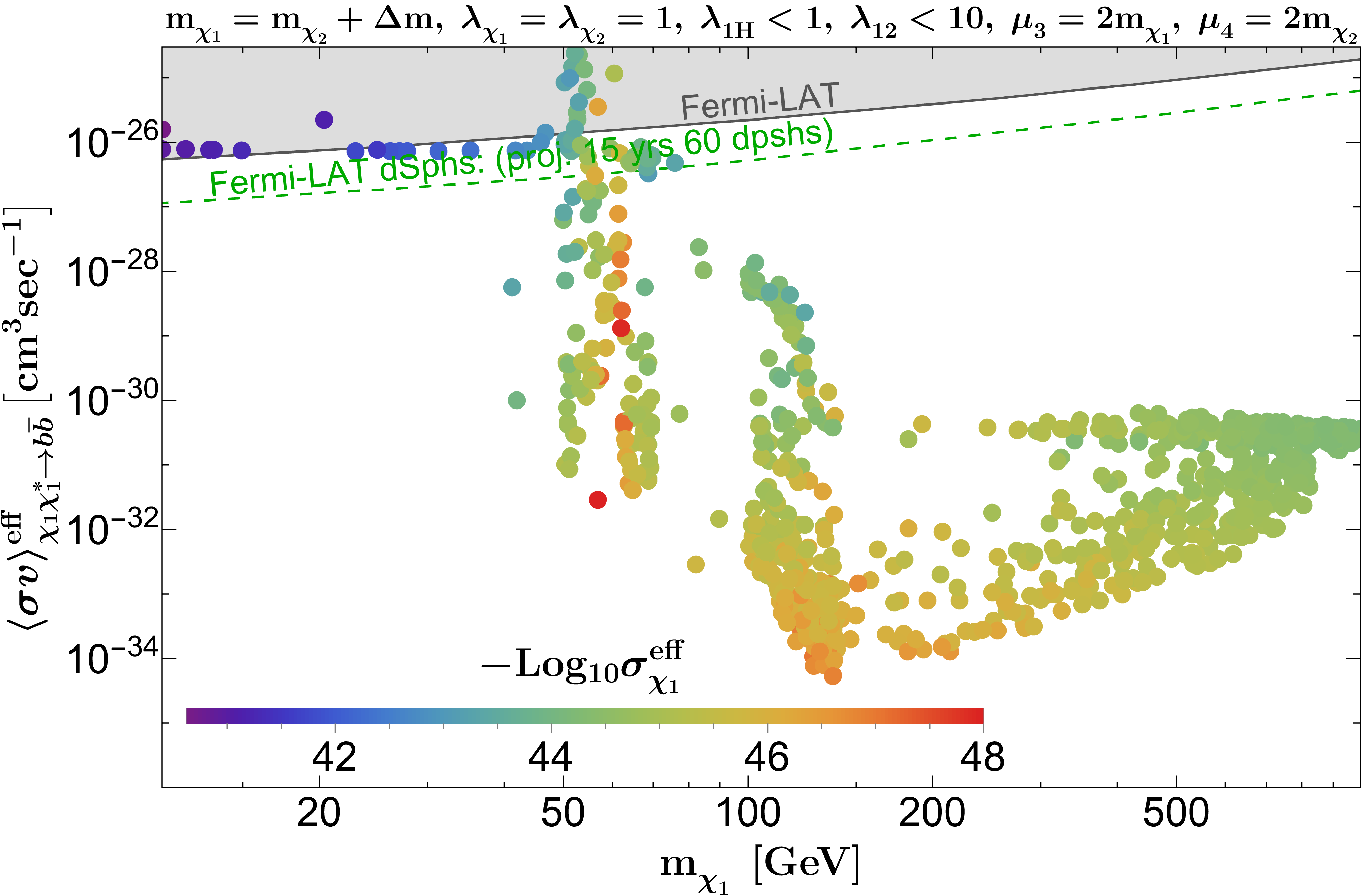}\label{fig:z3-twoscalar_chi1-idA}}~~
\subfloat[]{\includegraphics[width=0.5\linewidth]{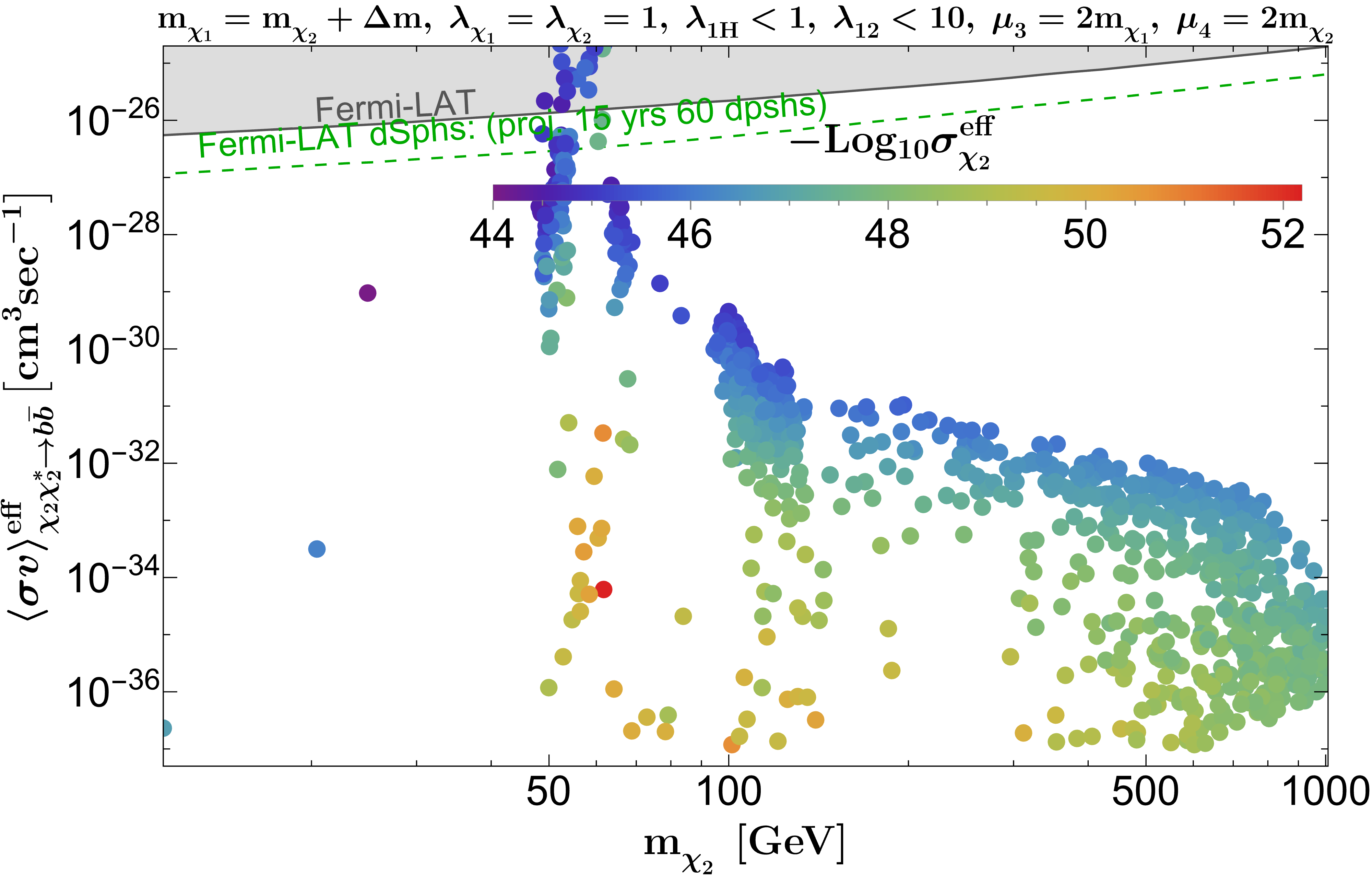}\label{fig:z3-twoscalar_chi2-idA}}

\subfloat[]{\includegraphics[width=0.475\linewidth]{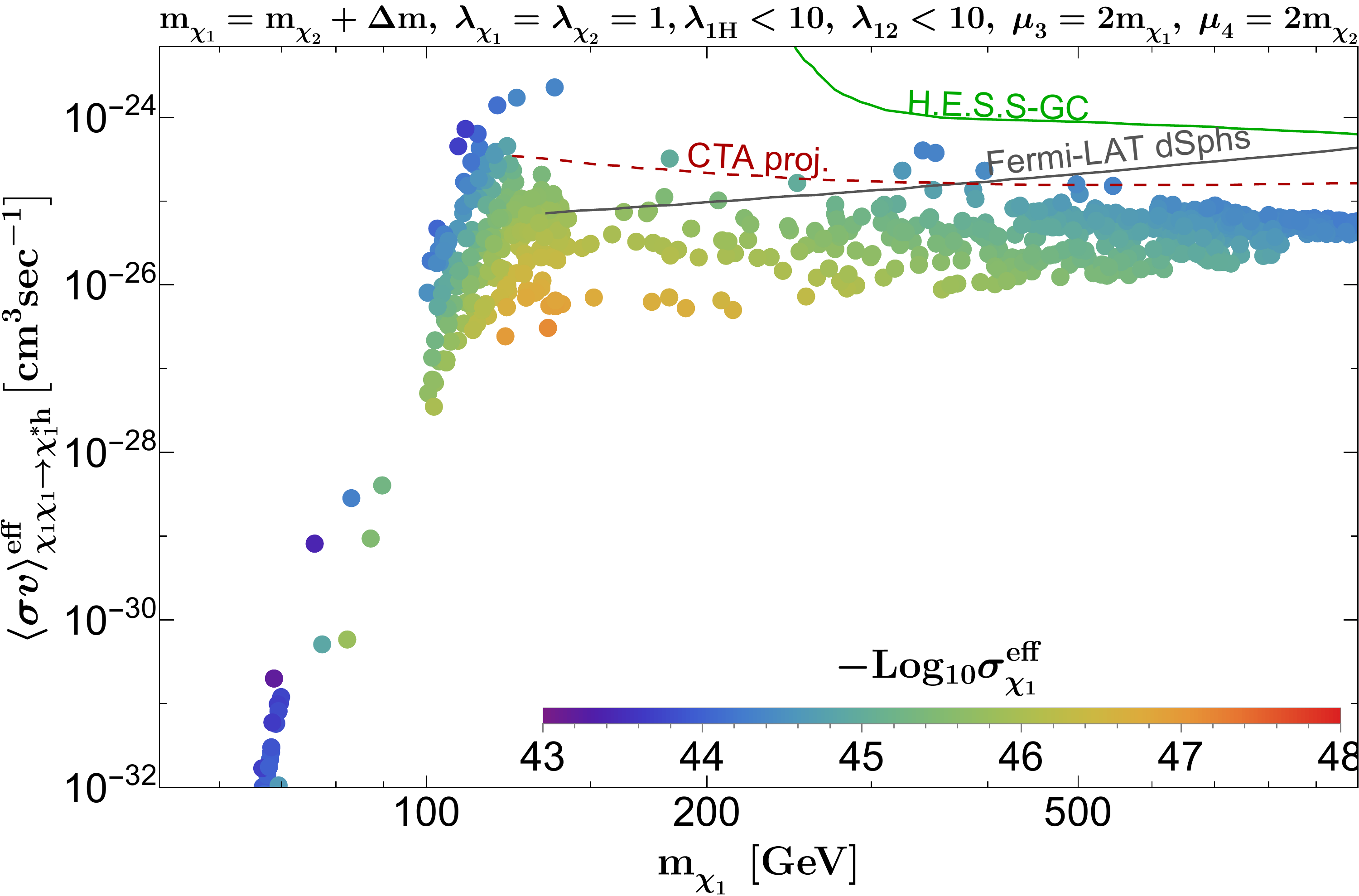}\label{fig:z3-twoscalar_chi1-semiidA}}~~
\subfloat[]{\includegraphics[width=0.5\linewidth]{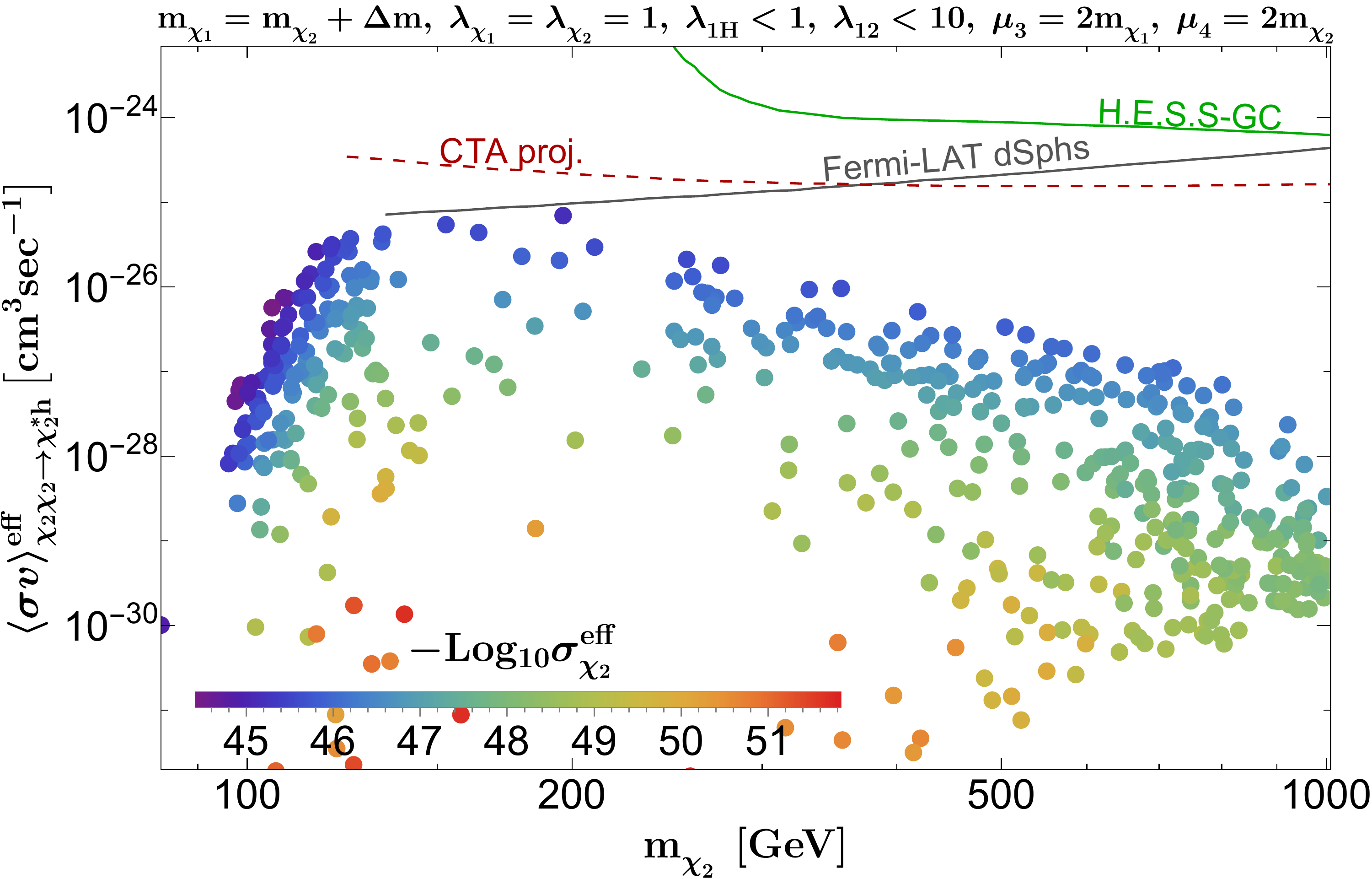}\label{fig:z3-twoscalar_chi2-semiidA}}

\subfloat[]{\includegraphics[width=0.5\linewidth]{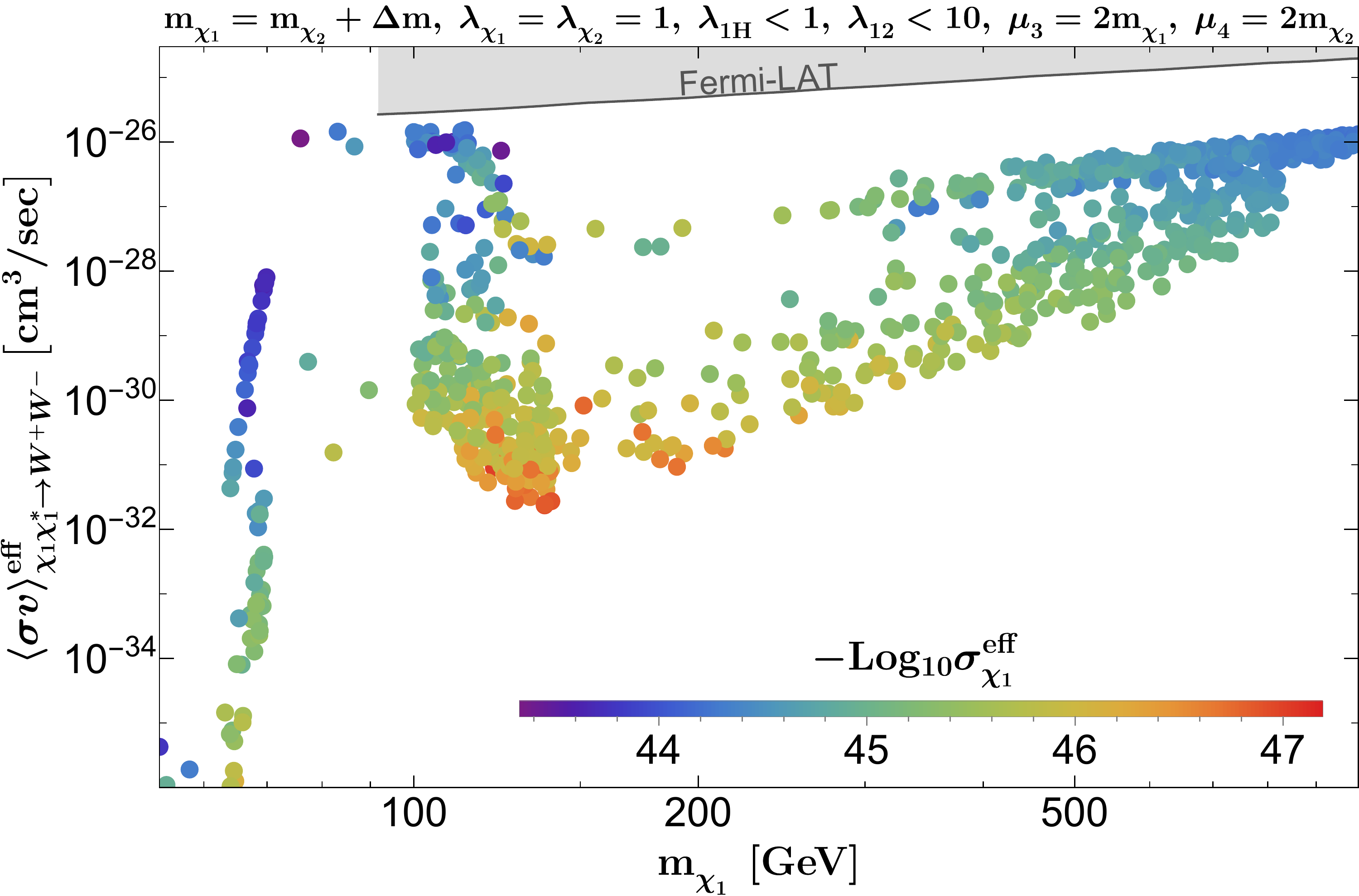}\label{fig:z3-twoscalar_chi1-idA_WW}}~~
\subfloat[]{\includegraphics[width=0.5\linewidth]{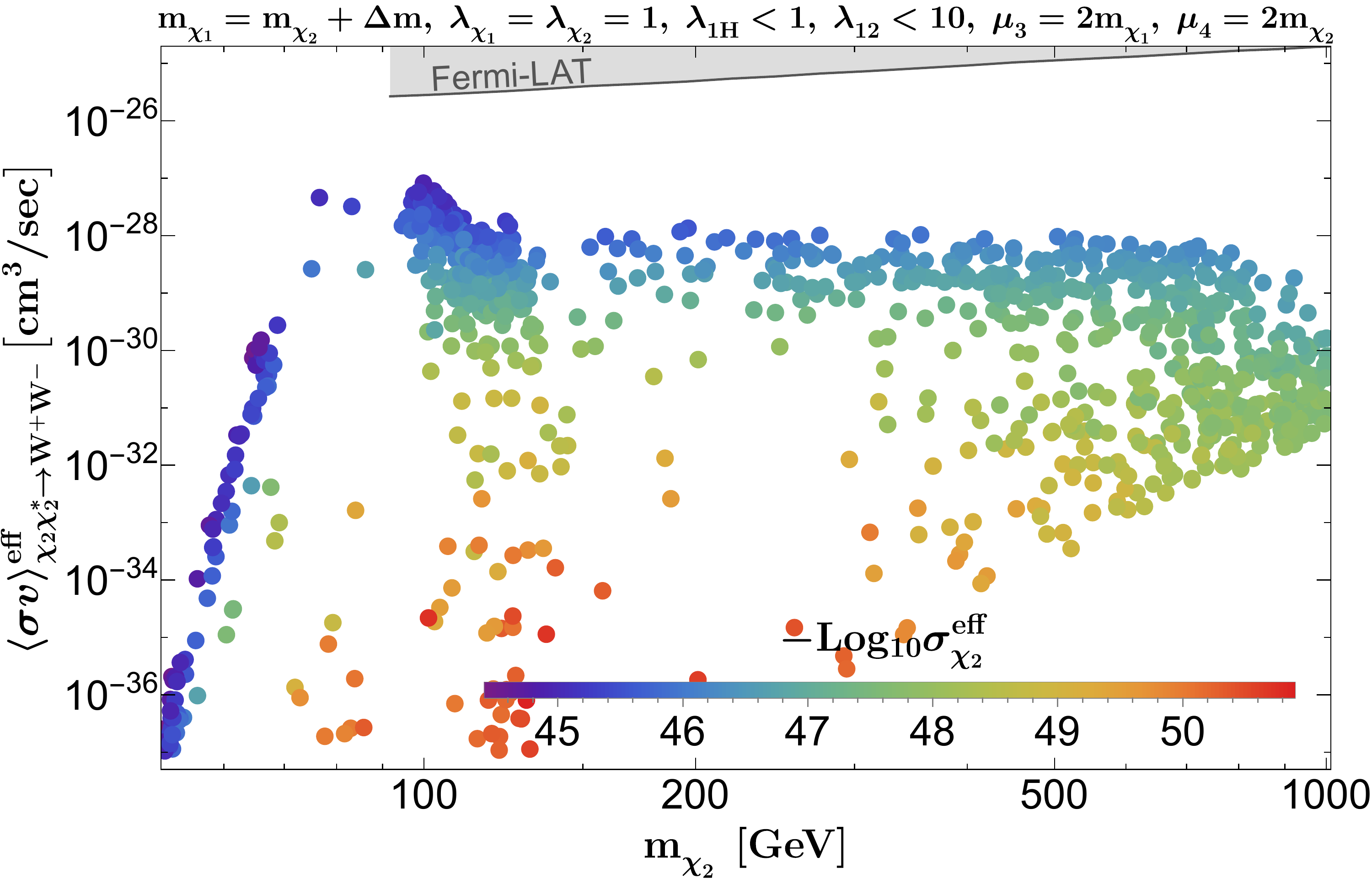}\label{fig:z3-twoscalar_chi2-idA_WW}}
\caption{\Cref{fig:z3-twoscalar_chi1-idA} (\ref{fig:z3-twoscalar_chi1-idA_WW}), and \cref{fig:z3-twoscalar_chi2-idA} (\ref{fig:z3-twoscalar_chi2-idA_WW}) show the relic allowed parameter space $m_{\chi_1}-\langle \sigma v\rangle^{\rm eff}_{\chi_1\chi_1^{*}\to b\overline{b}~(W^+W^-)}$ and $m_{\chi_2}-\langle \sigma v\rangle^{\rm eff}_{\chi_2\chi_2^{*}\to b\overline{b}~(W^+W^-)}$ plane, respectively for WIMP and pFIMP while \cref{fig:z3-twoscalar_chi1-semiidA,fig:z3-twoscalar_chi2-semiidA} corresponds to WIMP and pFIMP semi-annihilation. We have taken $\lambda_{12}^{\prime}=\lambda_{3}=\lambda_{4}=10^{-20}$ and $\mu_1=\mu_2=10^{-15}$ GeV to stabilise the heavier DM and $\lambda_{2H}=10^{-12}$ to become $\chi_2$ as pFIMP.}
\label{fig:id-A}
\end{figure}

In eqs\,.~\eqref{eq:z3-wimp-self} and \eqref{eq:z3-wimp-semi}, the thermal average of self and semi-annihilation cross-sections are evaluated 
at the WIMP freeze-out point, $\rm T_{\chi_1}^{FO}\sim m_{\chi_1}/25$ and plotted in $m_{\chi_1}-\langle \sigma v\rangle^{\rm eff}_{\chi_1\chi_1^{*}\to b\overline{b}}$ 
plane in figs\,.~\ref{fig:z3-twoscalar_chi1-idA} and \ref{fig:z3-twoscalar_chi1-semiidA}, respectively.
The rainbow colour bar shows the spin-independent direct detection cross-section of pFIMP in $\rm cm^2 $ unit. The effective self and semi-annihilation 
cross-section of WIMP depend on $\lambda_{1H}$, WIMP mass and also on the effective WIMP contribution to the total DM relic. Fermi-LAT observation and projection data 
are shown by thick grey and dashed green lines, respectively. We see that most of the parameter space probed here obey the limit.

pFIMP $\chi_2$ connects with the visible sector via WIMP loop-mediated interaction, as shown in \cref{feyn:pfimp-id,fig:pfimp-semiid-1,fig:pfimp-semiid-2,fig:pfimp-semiid-3}. 
These diagrams are subdominant to the tree-level processes. The effective self-annihilation cross-section of pFIMP ($\chi_2$) is given by,
\bea
\langle\sigma v\rangle_{\chi_2\chi_2^*\to b\overline{b}}^{\rm eff}=\left(\dfrac{\Omega_{\chi_2}h^2}{\Omega_{\chi_1}h^2+\Omega_{\chi_2}h^2}\right)^2 \langle\sigma v\rangle_{\chi_2\chi_2^*\to b\overline{b}}\,,
\label{eq:z3-pfimp-self}
\eea
and the pFIMP semi-annihilation cross-section is given by,
\bea
\langle\sigma v\rangle_{\chi_2\chi_2\to\chi_2^* h}^{\rm eff}=\left(\dfrac{\Omega_{\chi_2}h^2}{\Omega_{\chi_1}h^2+\Omega_{\chi_2}h^2}\right) \langle\sigma v\rangle_{\chi_2\chi_2\to\chi_2^* h}\,.
\label{eq:z3-pfimp-semi}
\eea
The dominant self-annihilation of $\chi_2$ to SM pair comes from diagram \cref{feyn:pfimp-id}, while others are suppressed due to tiny couplings assumed for the stability of 
$\chi_2$. We also need to remove the divergence contribution from the vertex correction diagram \cref{feyn:pfimp-id}. We take $q_h^2=4m_{\chi_2}^2$ as our renormalisation scale \cite{Bhattacharya:2022dco}. Then, we have calculated the thermal average annihilation cross-section at the pFIMP freeze-out point $\rm T_{\chi_2}^{\rm FO}\sim m_{\chi_2}/25$.

\Cref{fig:z3-twoscalar_chi2-idA,fig:z3-twoscalar_chi2-semiidA} show the relic density allowed parameter space in 
$m_{\chi_2}-\langle \sigma v\rangle^{\rm eff}_{\chi_2\chi_2^{*}\to b\overline{b}}$ and $m_{\chi_2}-\langle \sigma v\rangle^{\rm eff}_{\chi_2\chi_2\to \chi_2^*h}$ planes, respectively.
The pFIMP effective semi-annihilation cross-section is written in eq\,.~\eqref{eq:z3-pfimp-semi} and is proportional to the $\lambda_{1H}$, $\lambda_{12}$, and $\mu_4$ couplings.
Among these three, $\lambda_{1H}$ and $\lambda_{12}$ contribute to the pFIMP DD cross-section. With larger $\lambda_{1H}$, 
both the self-annihilation and direct detection cross-section of pFIMP increases, which is also reflected in \cref{fig:z3-twoscalar_chi2-idA}.
Finally, we constrain the parameter space by imposing the Fermi-LAT observation and projection bound in 
$m_{\chi_2}-\langle\sigma v\rangle_{\chi_2\chi_2^*\to b\overline{b}}^{\rm eff}$ plane and the colour bar represents the variation of effective pFIMP 
direct detection cross-section. The pFIMP semi-annihilation is also mediated via WIMP loop, see \cref{fig:pfimp-semiid-1,fig:pfimp-semiid-2,fig:pfimp-semiid-3}.
We have used the bound on DM semi-annihilation from H.E.S.S (green), Fermi-LAT (grey), and CTA (dashed red) in \cref{fig:z3-twoscalar_chi2-semiidA}. 
These constraints exclude some regions, which are also excluded by the presently available DD bounds. 

In summary, the two component DM scenario of type A after stabilising the heavier component by adjusting coupling parameters behaves like 
$\mathcal{Z}_3\otimes\mathcal{Z}^{\prime}_{3}$ scenario. Additionally, when the heavier DM component behaves like pFIMP having tiny portal interaction, 
the under abundant parameter space of WIMP is adequately utilised by the second component and enhance the allowed parameter space. 
Having the second component as pFIMP reduces the DD and ID bounds as they are primarily governed by loop mediated interactions and allows 
one to exploit larger parameter space compared to the WIMP-WIMP case \cite{Bhattacharya:2017fid}. Both DMs are allowed within $\rm GeV~to~ TeV$ 
range, where semi-annihilation and conversion play crucial roles for yielding correct DM relic density. 
\subsubsection{Scenario-B}
\label{subsec:B}
In scenario B (\cref{tab:z3}), the absence of $\chi_2^3$ helps in the stabilisation of the heavier particle. This parameter 
doesn't have any significance in WIMP-pFIMP set up. But the presence of $\chi_1^2\chi_2^2$ and $\chi_2^2\chi_1^*$ terms 
has an important role in the semi-annihilation of $\chi_2$ for the WIMP-WIMP scenario.
They open up many conversion and semi-conversion channels, shown in fig\,.~\ref{fig:z3-feynman3}, which are absent in scenario A or WIMP-WIMP set up under 
$\mathcal{Z}_3\otimes\mathcal{Z}^{\prime}_{3}$ model. We will do a similar kind of analysis to understand the benefits, such as the scenario A.
We study both (I) WIMP$-$WIMP $(\lambda_{2H}\,\neq\, 0)$, and (II) WIMP-pFIMP $(\lambda_{2H}\,\to\, 0)$ cases below.
\paragraph{$\bullet$~WIMP-WIMP\\}
In case of a real scalar singlet WIMP as a single component DM, some breathing space is left near the Higgs resonance region, or in the high mass regime, 
$\gtrsim 1$ TeV, but the direct detection sensitivity is less for the latter. In the two-component real scalar singlet WIMP scenario, the 
choices could be: (I) one is near Higgs resonance, and another in the higher mass regime, $m_{\rm DM}\gtrsim 1$ TeV, or
(II) both DM masses are far above the Higgs resonance. In scenario B, the presence $\chi_1^3$ term opens up semi-annihilation channels, which brings 
more relic and DD-allowed parameter space.

However, in two-component complex scalar WIMP scenario as in here, $\chi_2$ semi-annihilation is inefficient due to the feeble $\chi_2^3$, but opens the door for 
conversion, semi-conversion channels via $\chi_1^2\chi_2^2$ and $\chi_2^2\chi_1^*$ terms.
Along with the standard DM-DM conversion via $|\chi_1|^2|\chi_2|^2$ term, we then have $\chi_1$ semi-annihilation ($\chi_1\chi_1\to\chi_1^{*}h$) due to 
$\chi_1^3$ and $\chi_2$ mediated DM-DM conversion \cref{feyn:conv-1,feyn:conv-4}, and semi-conversion ($\chi_2\chi_2\to\chi_1^{*}h$) \cref{feyn:conv-6,feyn:conv-7} in
due to the presence of $\chi^2\chi_1^*$ term, all of which help in acquiring correct relic, without perturbing DD or ID searches.
The $\chi_1-\chi_2$ conversion can also be possible via Higgs mediation  and $\chi_2$ mediation along with the four-point scattering.
So, the Higgs resonance effect is not only limited to DM annihilation to SM particles but also to DM-DM conversion.
The s-channel semi-conversion of $\chi_2\chi_2\to\chi_1^{*} h$ is more effective near the Higgs mass regime.
These processes help to relax the Higgs portal coupling, but keep adequate depletion via conversion processes and save the DM components from DD constraints.
\begin{figure}[htb!]
\centering
\subfloat[]{\includegraphics[width=0.475\linewidth]{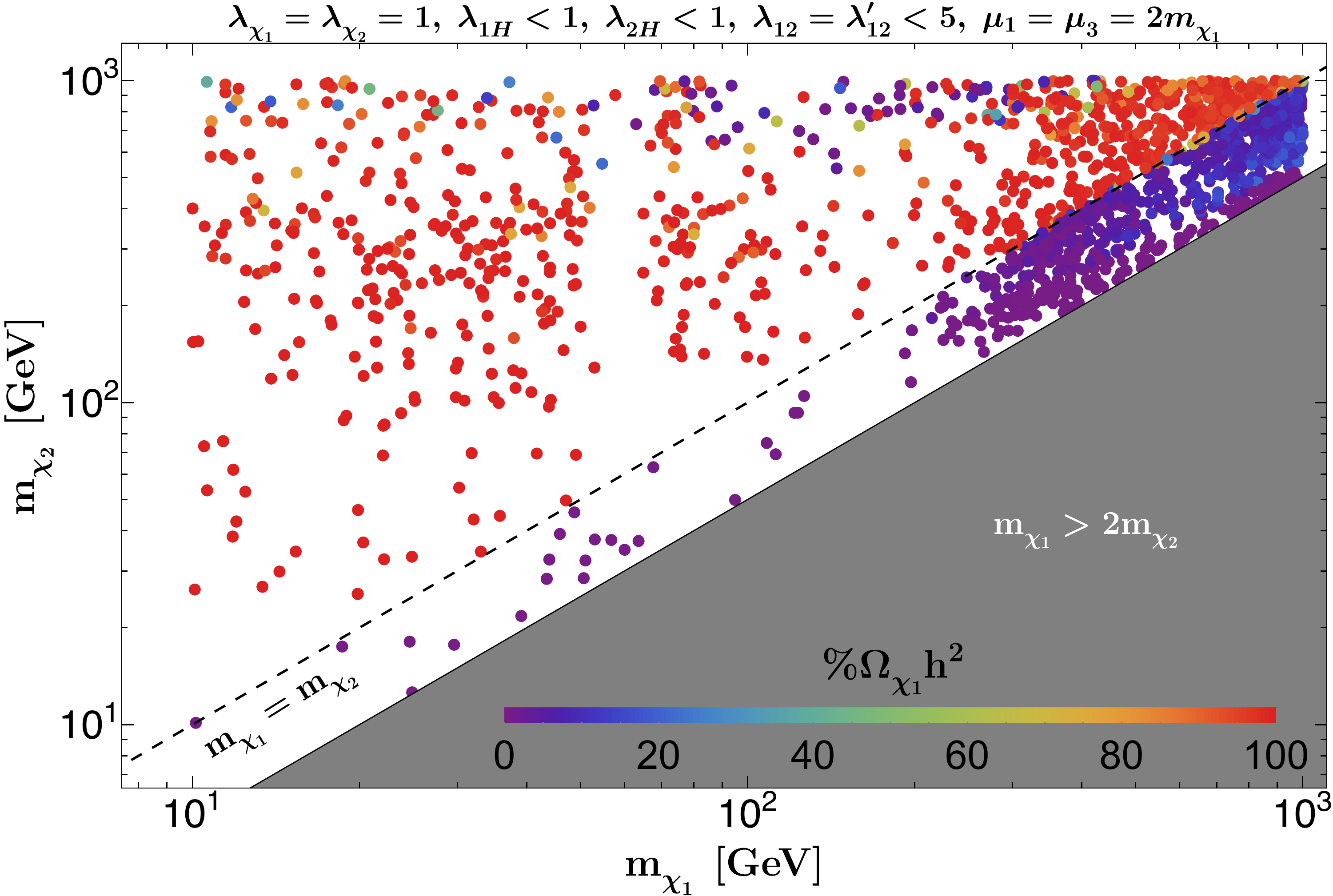}\label{fig:z3-2wimp_mchi1-mchi2p}}~~
\subfloat[]{\includegraphics[width=0.475\linewidth]{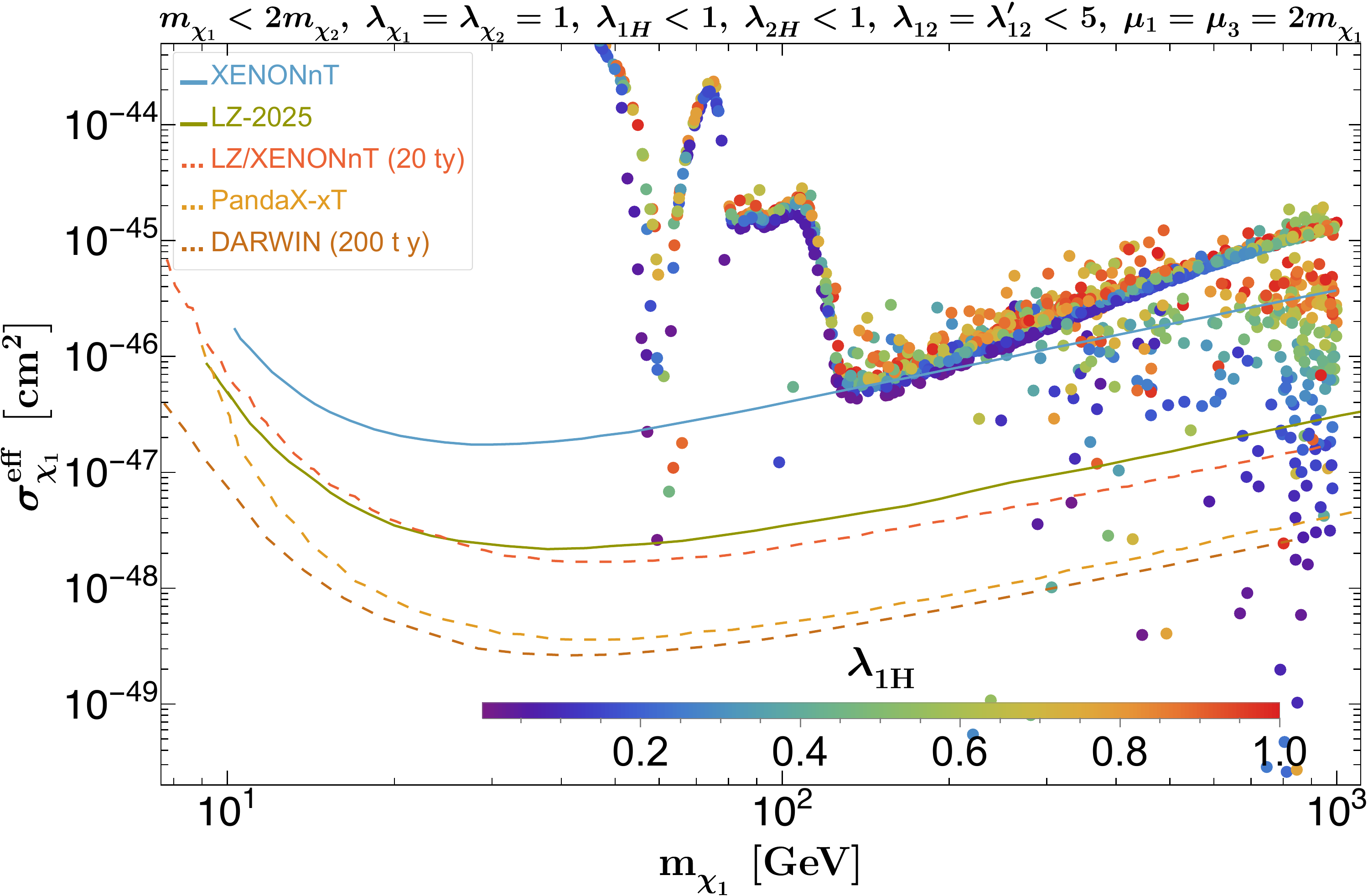}\label{fig:2wimp-m1-dd1-l1h}}

\subfloat[]{\includegraphics[width=0.475\linewidth]{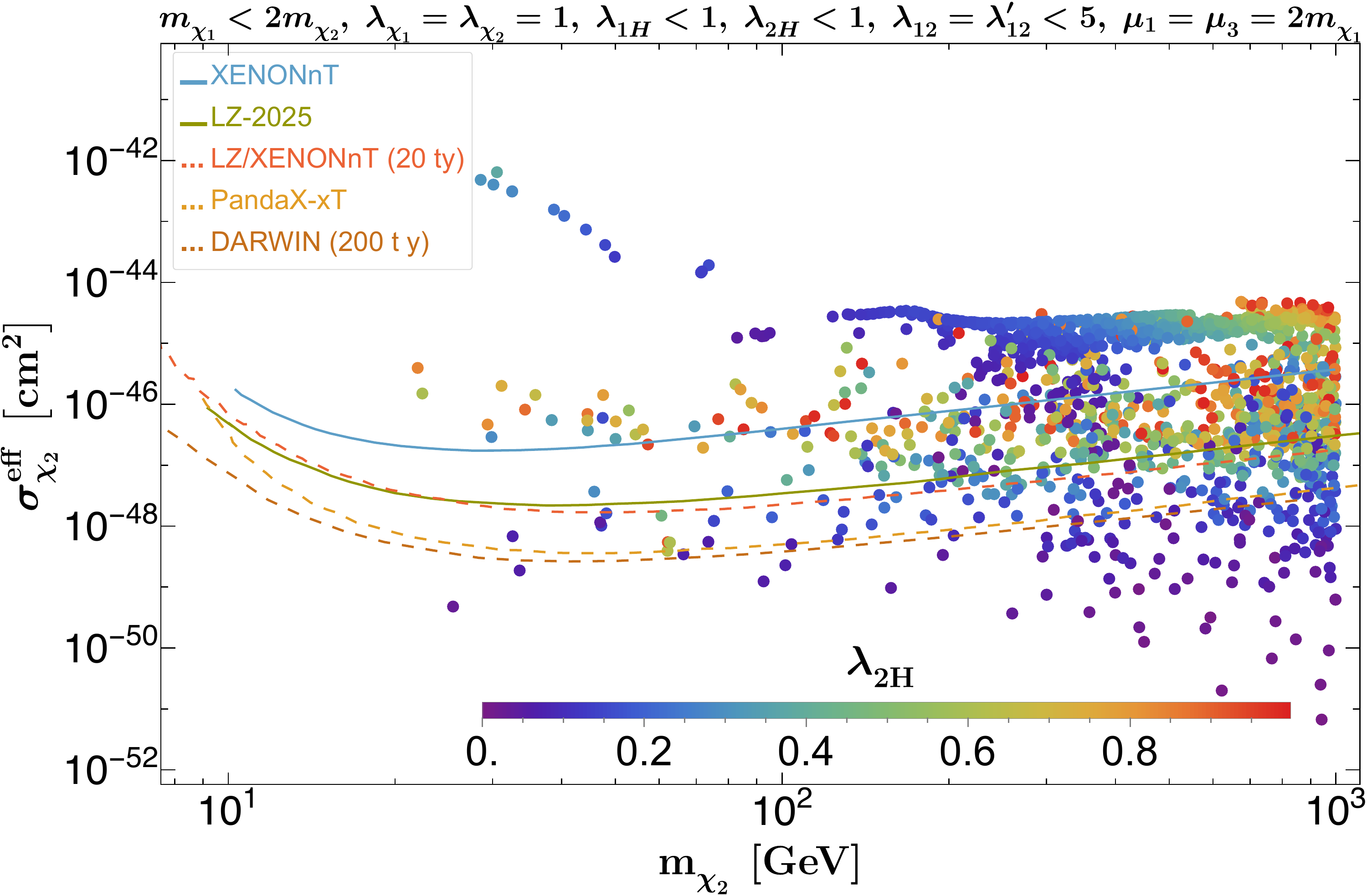}\label{fig:2wimp-m2-dd2-l2h}}~~
\subfloat[]{\includegraphics[width=0.475\linewidth]{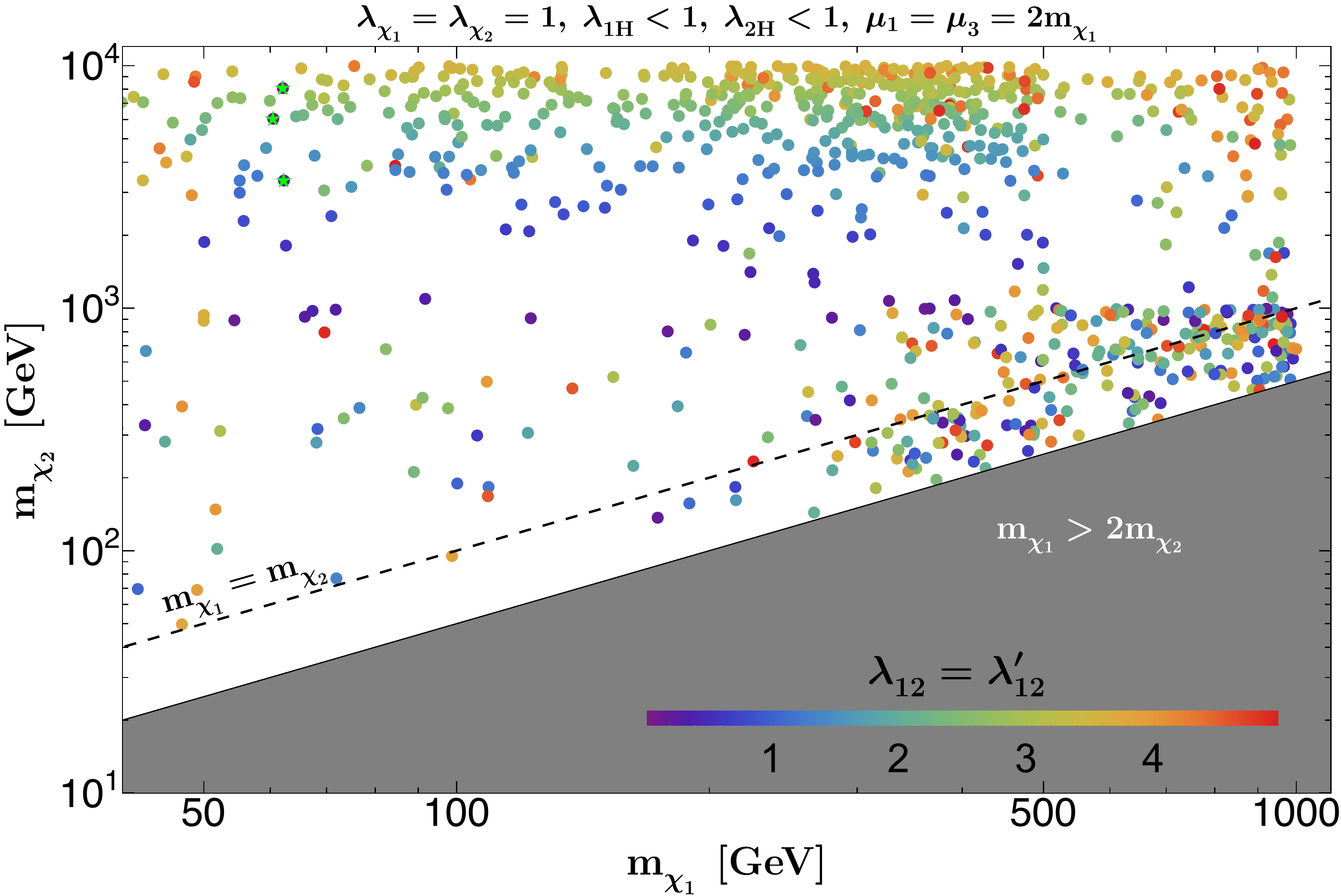}\label{fig:2wimp-m1-m2-l12}}
\caption{\Cref{fig:2wimp-m1-dd1-l1h,fig:2wimp-m2-dd2-l2h} shows the relic density allowed parameter space for two-component WIMP in category B (\cref{tab:z3}) in
$m_{\chi_1}-\sigma_{\chi_1}^{\rm eff}$, and $m_{\chi_2}-\sigma_{\chi_2}^{\rm eff}$ planes respectively.
\Cref{fig:z3-2wimp_mchi1-mchi2p,fig:2wimp-m1-m2-l12} show the relic allowed parameter space in $m_{\chi_1}-m_{\chi_2}$ plane, where rainbow color depicts variation in 
$\Omega_{\chi_1} h^2$ and $\lambda_{12}$ respectively. The couplings ($\mu_2,\mu_4,\lambda_3,\lambda_4$) are taken adequately small ($\sim 10^{-20}$) to stabilise the 
heavier DM component. The grey-shaded region is excluded from the heavier particle stability. In fig\,.~\ref{fig:2wimp-m1-m2-l12} green star points depict the case where both DM components obey the stringent limit on spin-independent DM-nucleon scattering cross section from LZ-2025.}
\label{fig:2wimp-ddB}
\end{figure}
In \cref{fig:2wimp-ddB}, we have shown the relic density allowed parameter space and incorporated possible direct detection bounds. \Cref{fig:z3-2wimp_mchi1-mchi2p} 
represents the relic allowed parameter space in $m_{\chi_1}-m_{\chi_2}$ plane, while the rainbow colour bar shows the variation of percentage contribution 
of $\chi_1$ in the total relic. If $m_{\chi_1}>m_{\chi_2}$, then $\chi_1$ relic is subdominant due to conversion of $\chi_1$ to $\chi_2$ decrease the $\chi_1$ number 
density but $\chi_2$ number density is enhanced simultaneously. Around the Higgs resonance and semi-annihilation region, the conversion effect could be ignored, 
and equal contribution arises; see green colour points in \cref{fig:z3-2wimp_mchi1-mchi2p}.
\Cref{fig:2wimp-m1-dd1-l1h,fig:2wimp-m2-dd2-l2h} represents the relic allowed point in $m_{\chi_1}-\sigma_{\chi_1}^{\rm eff}$ and $m_{\chi_2}-\sigma_{\chi_2}^{\rm eff}$ 
plane, respectively. In \cref{fig:2wimp-m1-dd1-l1h}, we see two kinks, $m_{\chi_1}$ around the Higgs resonance and Higgs-mass region, where semi-annihilation process is 
more effective because of the presence of $\chi_1^3$ term. Beyond the Higgs mass, semi-annihilation and also self-annihilation cross-section decreases, 
which can be adjusted by $\lambda_{1H}$, and $\lambda_{12}$, but $\lambda_{1H}$ is strongly restricted by DD.
On the contrary, as the trilinear coupling of $\chi_2$ is very tiny to keep it stable, there is no kink due to semi-annihilation near Higgs mass in 
\cref{fig:2wimp-m2-dd2-l2h}. So, with larger mass, the decrease in cross-section is adjusted by the enhancement of $\lambda_{2H}$. Although 
semi-annihilation of $\chi_2$ is not active in absence of $\chi_2^3$ term, but semi-conversion channels are there due to $\chi_2^2\chi_1^*$ terms, 
which help to relax the $\lambda_{2H}$ coupling. In \cref{fig:2wimp-m1-m2-l12}, we show the effect of conversion in $m_{\chi_1}-m_{\chi_2}$ plane. 

The main outcome of this scenario is that we are getting relic and DD-allowed points near the Higgs resonance, Higgs mass, 
and also some above the Higgs mass, represented by the green stars in \cref{fig:2wimp-ddB}. Indirect search limits are less stringent than DD and do not alter 
the allowed parameter space. 
\paragraph{$\bullet$~WIMP-pFIMP\\}
In scenario-B (\cref{tab:z3}), we have identified $\chi_2$ as pFIMP by choosing $\lambda_{2H}\sim 10^{-12}$, so that it doesn't have a direct SM connection, however sizeable 
$\lambda_{12}$ keeps it in thermal bath via interaction with $\chi_1$.
\begin{figure}[htb!]
\centering
\subfloat[]{\begin{tikzpicture}
\begin{feynman}
\vertex (a);
\vertex[above=1cm of a] (b); 
\vertex[above=1cm of b] (c); 
\vertex[left=0.5cm of c] (c11);
\vertex[left=0.75cm of c11] (c1){\(\chi_2\)}; 
\vertex[right=0.5cm of c] (c22);
\vertex[right=0.75cm of c22] (c2){\(\chi_2\)}; 
\vertex[left=1cm of a] (a1){\(N\)};
\vertex[right=1cm of a] (a2){\(N\)};
\diagram* {
(c1) -- [ line width=0.25mm,charged scalar, edge label={\(\color{black}{\rm}\)}, arrow size=0.7pt, style=blue] (c11),
(c22) -- [ line width=0.25mm, charged scalar, arrow size=0.7pt, edge label'={\(\color{black}{\chi_2}\)}, style=gray] (c11),
(c22)  -- [ line width=0.25mm, charged scalar, arrow size=0.7pt, style=black] (c2),
(c11) -- [ line width=0.25mm, charged scalar, arrow size=0.7pt, edge label'={\(\color{black}{\chi_1}\)}, style=gray] (b), 
(b) -- [ line width=0.25mm, charged scalar, arrow size=0.7pt, edge label'={\(\color{black}{\chi_1}\)}, style=gray] (c22),
(a) -- [ line width=0.25mm, scalar, arrow size=0.7pt, edge label={\(\color{black}{h}\)}, style=red] (b) ,
(a1) -- [ line width=0.25mm, fermion, arrow size=0.7pt, style=blue] (a)  --[ line width=0.25mm, fermion, arrow size=0.7pt] (a2)};
\node at (a)[circle,fill,style=gray,inner sep=1pt]{};
\node at (b)[circle,fill,style=gray,inner sep=1pt]{};
\node at (c11)[circle,fill,style=gray,inner sep=1pt]{};
\node at (c22)[circle,fill,style=gray,inner sep=1pt]{};
\end{feynman}
\end{tikzpicture}\label{feyn:pfimp-ddB1}}
\subfloat[]{\begin{tikzpicture}
\begin{feynman}
\vertex (a);
\vertex[left=1cm and 1cm of a] (a1){\(\chi_2\)};
\vertex[right=1cm and 1cm of a] (a2){\(\chi_2\)}; 
\vertex[below=2cm of a] (b); 
\vertex[below=1cm of a] (c); 
\vertex[left=1cm and 1cm of b] (c1){\(N\)};
\vertex[right=1cm and 1cm of b] (c2){\(N\)};
\diagram* {
(a1) -- [ line width=0.25mm,charged scalar, arrow size=0.7pt, style=blue] (a),
(a) -- [ line width=0.25mm,charged scalar, arrow size=0.7pt, style=black] (a2),
(c) -- [ line width=0.25mm, scalar, edge label={\(\color{black}{\rm h}\)}, style=red] (b),
(c) -- [ line width=0.25mm, charged scalar, half left, arrow size=0.7pt, edge label={\(\color{black}{\chi_1}\)}, style=gray] (a),
(a) -- [ line width=0.25mm, charged scalar, half left, arrow size=0.7pt, edge label={\(\color{black}{\chi_1}\)}, style=gray] (c) ,
(c1) -- [ line width=0.25mm, fermion, arrow size=0.7pt, style=blue] (b),
(b)  --[ line width=0.25mm, fermion, arrow size=0.7pt] (c2)};
\node at (a)[circle,fill,style=gray,inner sep=1pt]{};
\node at (b)[circle,fill,style=gray,inner sep=1pt]{};
\node at (c)[circle,fill,style=gray,inner sep=1pt]{};
\end{feynman}
\end{tikzpicture}\label{feyn:pfimp-ddB2}}
\subfloat[]{\begin{tikzpicture}
\begin{feynman}
\vertex (a);
\vertex[above left=0.5cm and 0.5cm of a] (a11);
\vertex[below left=0.5cm and 0.5cm of a] (a22); 
\vertex[above left=0.75cm and 1cm of a] (a1){\(\chi_2\)};
\vertex[below left=0.75cm and 1cm of a] (a2){\(\chi_2\)}; 
\vertex[right=1cm of a] (c); 
\vertex[above right=0.75cm and 1cm of c] (c1){\(\rm b\)};
\vertex[below right=0.75cm and 1cm of c] (c2){\(\rm b\)};
\diagram* {
(a11) -- [ line width=0.25mm,charged scalar, arrow size=0.7pt, edge label'={\(\color{black}{\chi_2}\)}, style=gray] (a22),
(a) -- [ line width=0.25mm,charged scalar, arrow size=0.7pt, edge label'={\(\color{black}{\chi_1}\)},  style=gray] (a11) -- [ line width=0.25mm,charged scalar, arrow size=0.7pt, style=blue] (a1),
(a22) -- [ line width=0.25mm,charged scalar, arrow size=0.7pt, edge label'={\(\color{black}{\chi_1}\)},  style=gray] (a),
(a2) -- [ line width=0.25mm,charged scalar, arrow size=0.7pt, style=blue] (a22),
(c) -- [ line width=0.25mm, scalar, arrow size=0.7pt, edge label={\(\color{black}{h}\)}, style=red] (a),
(c1) --[ line width=0.25mm, fermion, arrow size=0.7pt,style=black] (c) --[ line width=0.25mm, fermion, arrow size=0.7pt,style=black] (c2)};
\node at (a11)[circle,fill,style=gray,inner sep=1pt]{};
\node at (a22)[circle,fill,style=gray,inner sep=1pt]{};
\node at (a)[circle,fill,style=gray,inner sep=1pt]{};
\node at (c)[circle,fill,style=gray,inner sep=1pt]{};
\end{feynman}
\end{tikzpicture}\label{feyn:pfimp-idB1}}
\subfloat[]{\begin{tikzpicture}
\begin{feynman}
\vertex (a);
\vertex[above left=0.75cm and 1cm of a] (a1){\(\chi_2\)};
\vertex[below left=0.75cm and 1cm of a] (a2){\(\chi_2\)}; 
\vertex[right=0.75cm of a] (b); 
\vertex[right=1.5cm of a] (c); 
\vertex[above right=0.75cm and 1cm of c] (c1){\(\rm b\)};
\vertex[below right=0.75cm and 1cm of c] (c2){\(\rm b\)};
\diagram* {
(a1) -- [ line width=0.25mm,charged scalar, arrow size=0.7pt, style=blue] (a) -- [ line width=0.25mm,charged scalar, arrow size=0.7pt, style=blue] (a2),
(c) -- [ line width=0.25mm, scalar, edge label={\(\color{black}{\rm h}\)}, style=red] (b),
(b) -- [ line width=0.25mm, charged scalar, half left, arrow size=0.7pt, edge label={\(\color{black}{\chi_1}\)}, style=gray] (a),
(a) -- [ line width=0.25mm, charged scalar, half left, arrow size=0.7pt, edge label={\(\color{black}{\chi_1}\)}, style=gray] (b) ,
(c1) --[ line width=0.25mm, fermion, arrow size=0.7pt,style=black] (c) --[ line width=0.25mm, fermion, arrow size=0.7pt,style=black] (c2)};
\node at (a)[circle,fill,style=gray,inner sep=1pt]{};
\node at (b)[circle,fill,style=gray,inner sep=1pt]{};
\node at (c)[circle,fill,style=gray,inner sep=1pt]{};
\end{feynman}
\end{tikzpicture}\label{feyn:pfimp-idB2}}
\caption{\Cref{feyn:pfimp-ddB1,feyn:pfimp-ddB2}, and \cref{feyn:pfimp-idB1,feyn:pfimp-idB2} represent the Feynman diagrams related to the direct and 
indirect detection of pFIMP ($\chi_2$), respectively.}
\label{fig:pfimp-detectionB}
\end{figure}
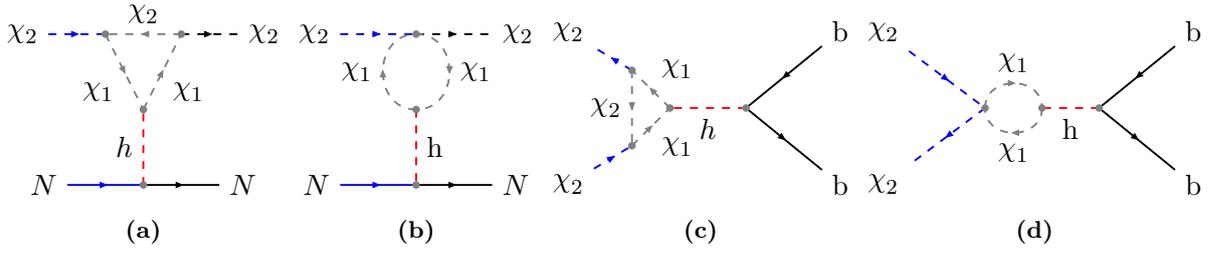
\begin{figure}[htb!]
\centering
\subfloat[]{\includegraphics[width=0.475\linewidth]{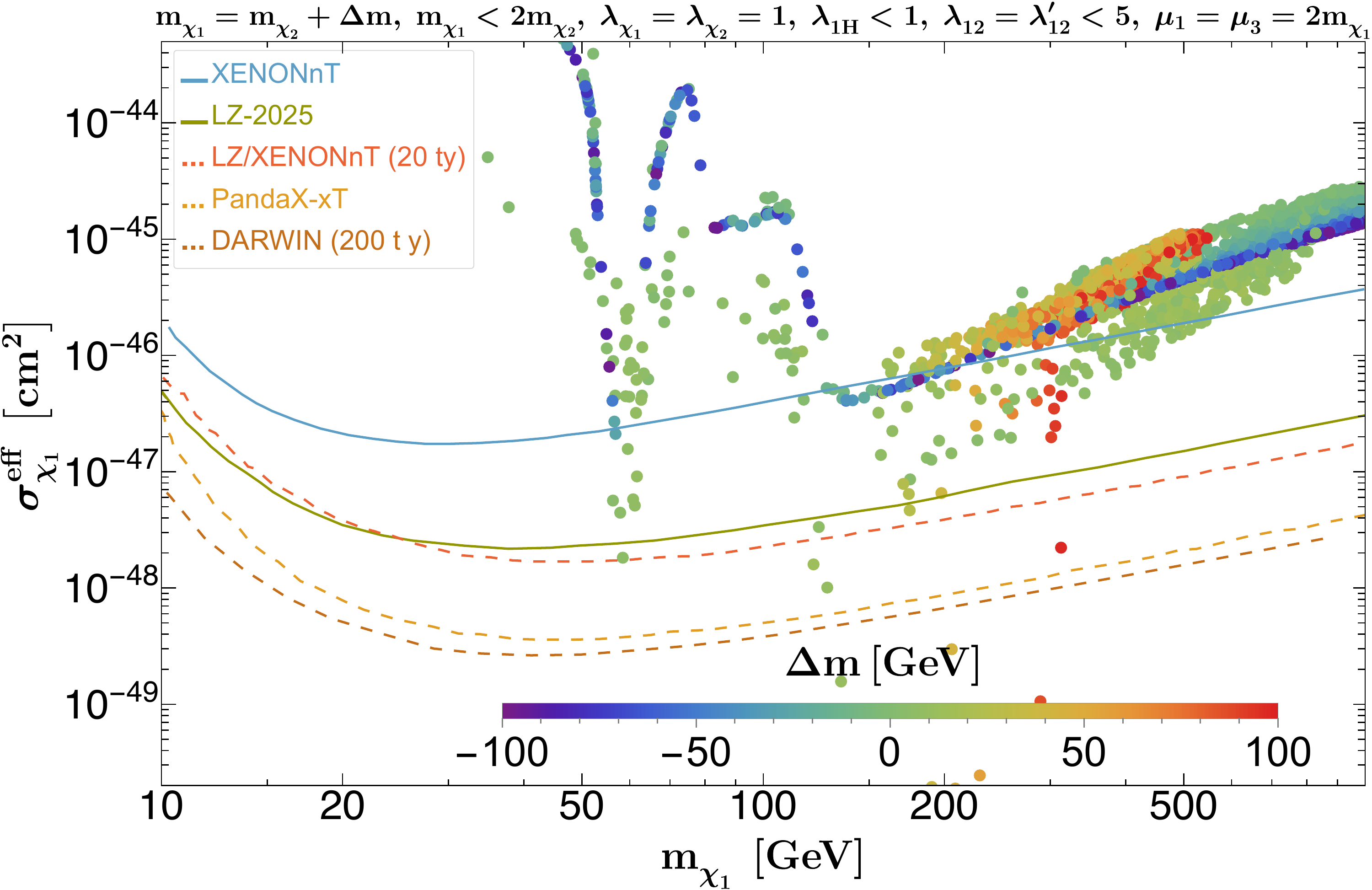}\label{fig:z3-twoscalar_chi1-dd1B}}~~
\subfloat[]{\includegraphics[width=0.475\linewidth]{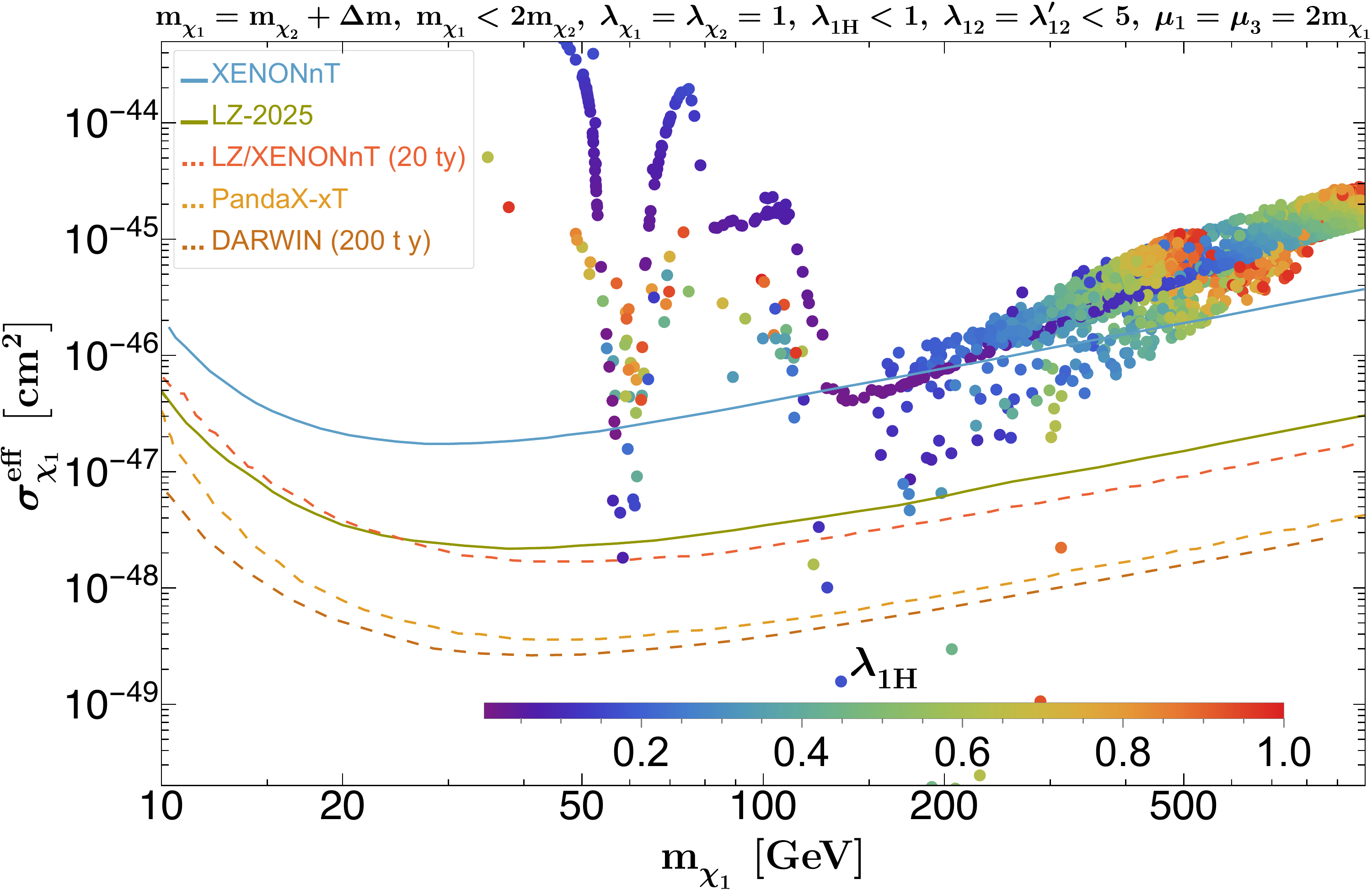}\label{fig:z3-twoscalar_chi1-dd2B}}

\subfloat[]{\includegraphics[width=0.475\linewidth]{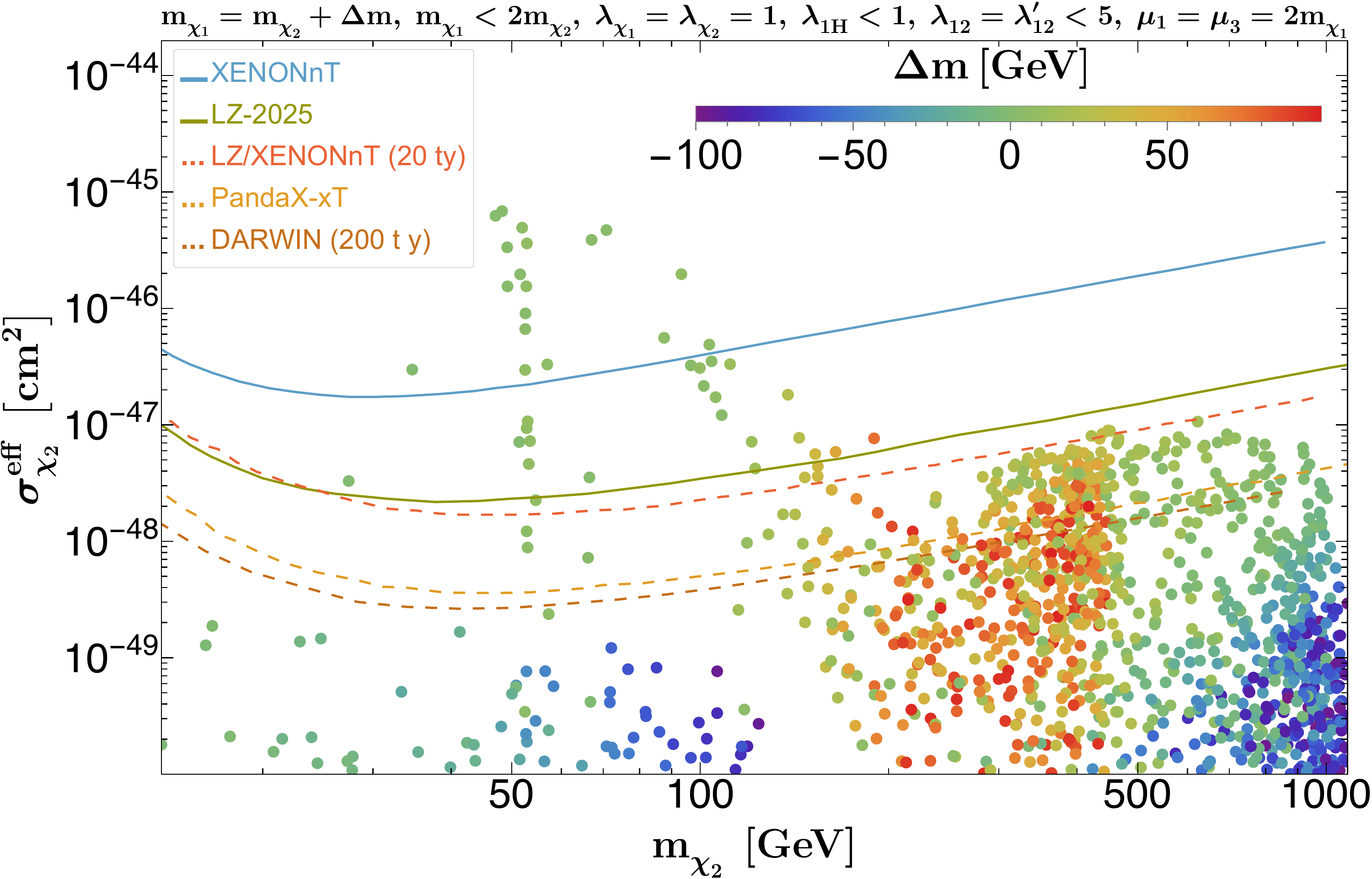}\label{fig:z3-twoscalar_chi2-dd1B}}~~
\subfloat[]{\includegraphics[width=0.475\linewidth]{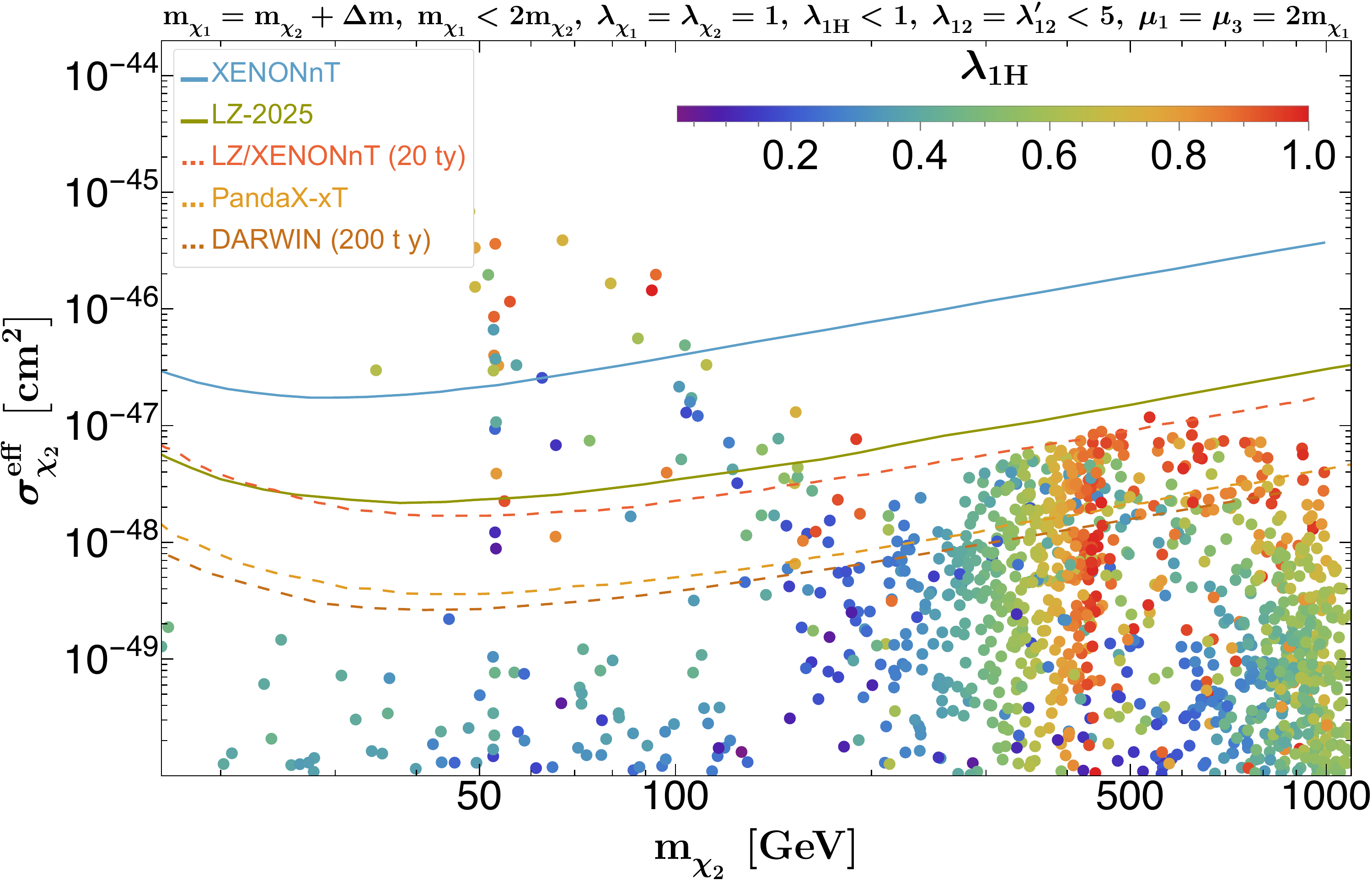}\label{fig:z3-twoscalar_chi2-dd2B}}
\caption{Figs\,.~\ref{fig:z3-twoscalar_chi1-dd1B}, \ref{fig:z3-twoscalar_chi1-dd2B}, \ref{fig:z3-twoscalar_chi2-dd1B}, and \ref{fig:z3-twoscalar_chi2-dd2B}  show the relic density allowed parameter space in $m_{\chi_1}-\sigma_{\chi_1}^{\rm eff}$ and $m_{\chi_2}-\sigma_{\chi_2}^{\rm eff}$ plane, respectively. 
The couplings ($\mu_2,\mu_4,\lambda_3,\lambda_4$) are taken adequately small to stabilise the heavier DM $\chi_2$ so that it falls in category B of two component DM (\cref{tab:z3}). The rainbow colour bar show the variation of $\Delta m$ (left) and $\lambda_{1H}$ (right). The thick and dashed lines correspond to the 
lower limits on the observed and projected DM-nucleon spin-independent DD cross-section respectively, while different colors refer to the experiments as mentioned in the 
figure inset.}
\label{fig:z3-ddB}
\end{figure}

\begin{figure}[htb!]
\centering
\subfloat[]{\includegraphics[width=0.475\linewidth]{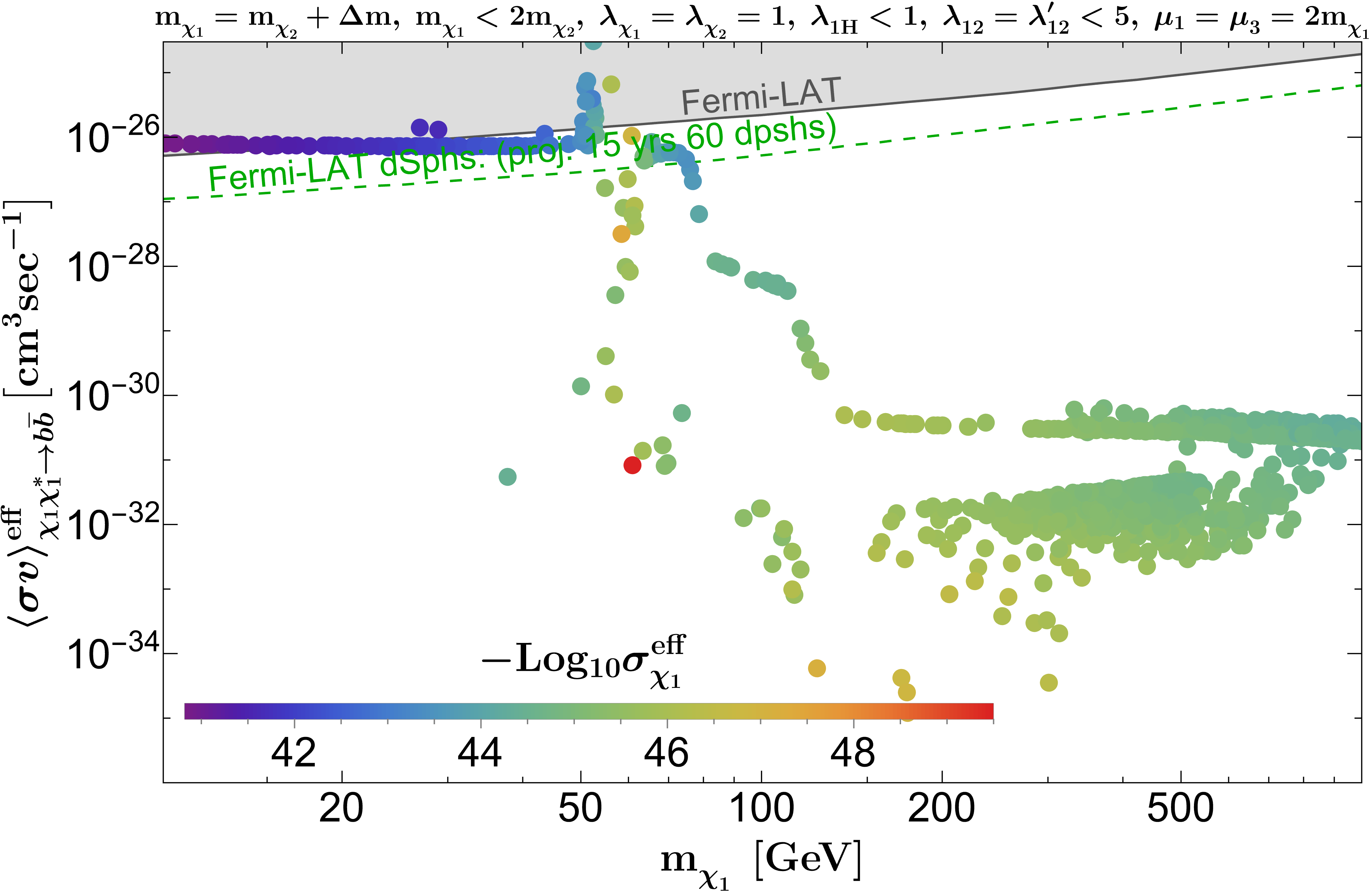}\label{fig:mchi1gmchi2_chi1idB}}~~
\subfloat[]{\includegraphics[width=0.475\linewidth]{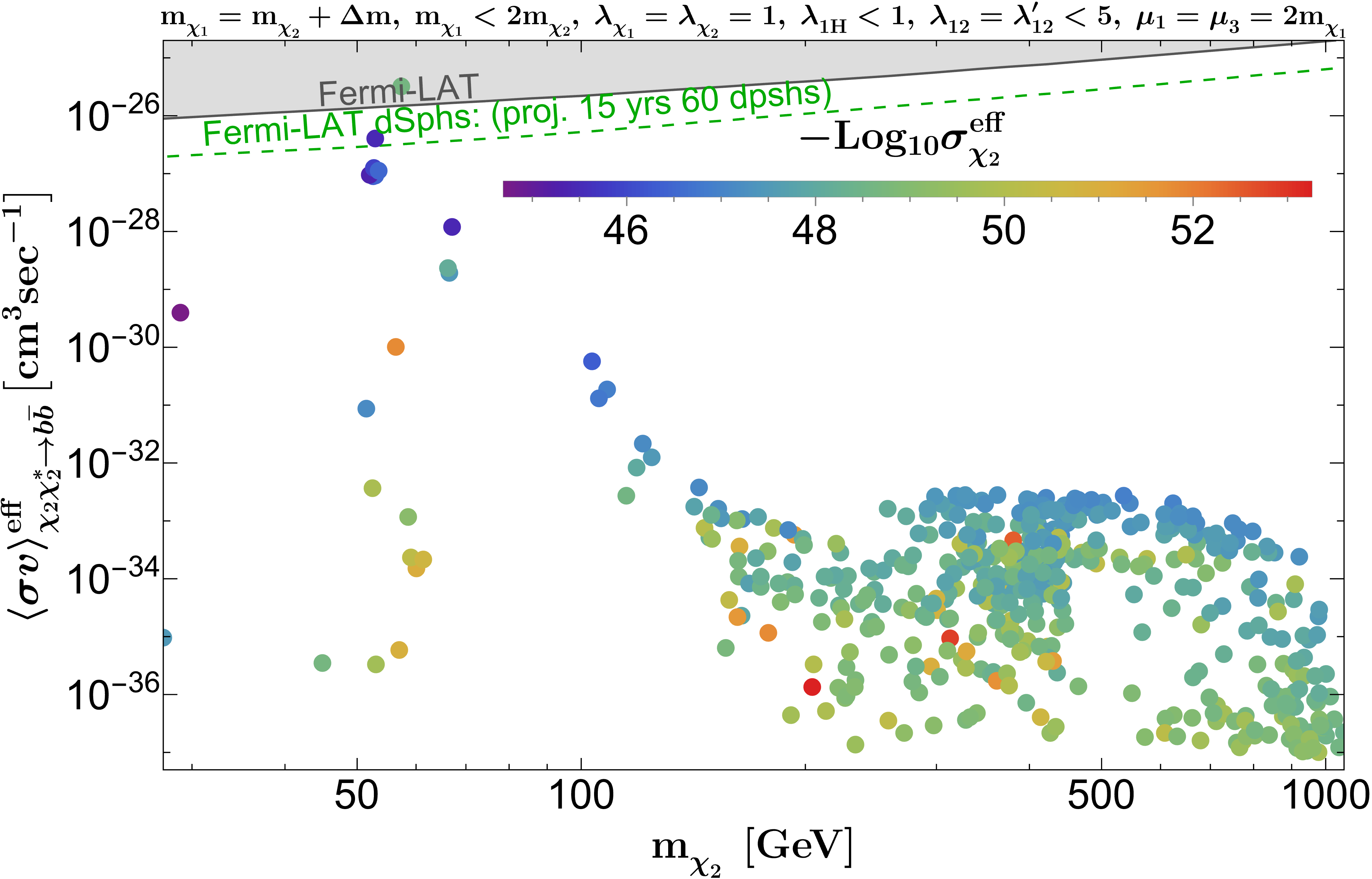}\label{fig:z3-twoscalar_chi2-idB}}

\subfloat[]{\includegraphics[width=0.475\linewidth]{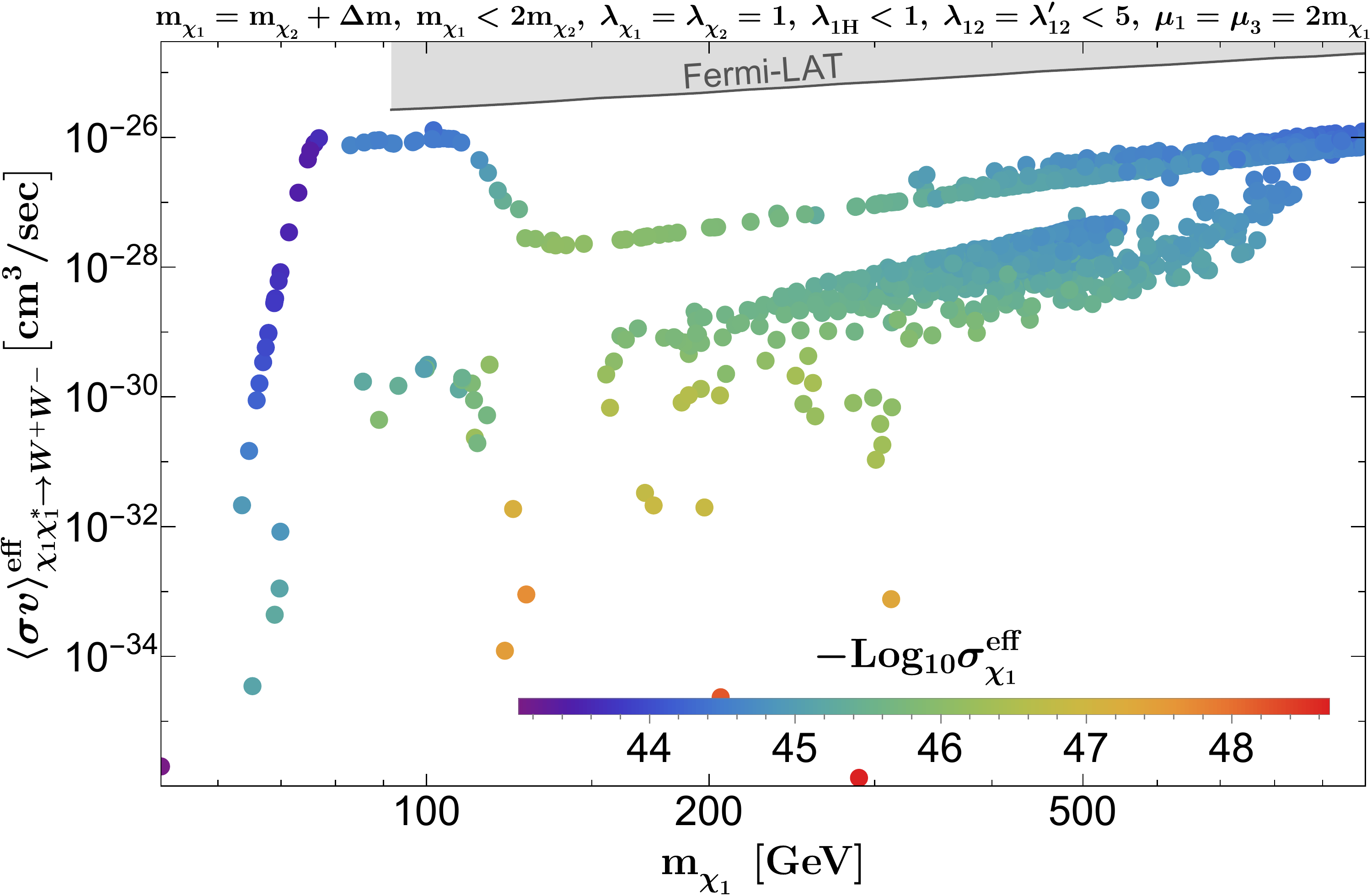}\label{fig:mchi1gmchi2_chi1idB_WW}}~~
\subfloat[]{\includegraphics[width=0.475\linewidth]{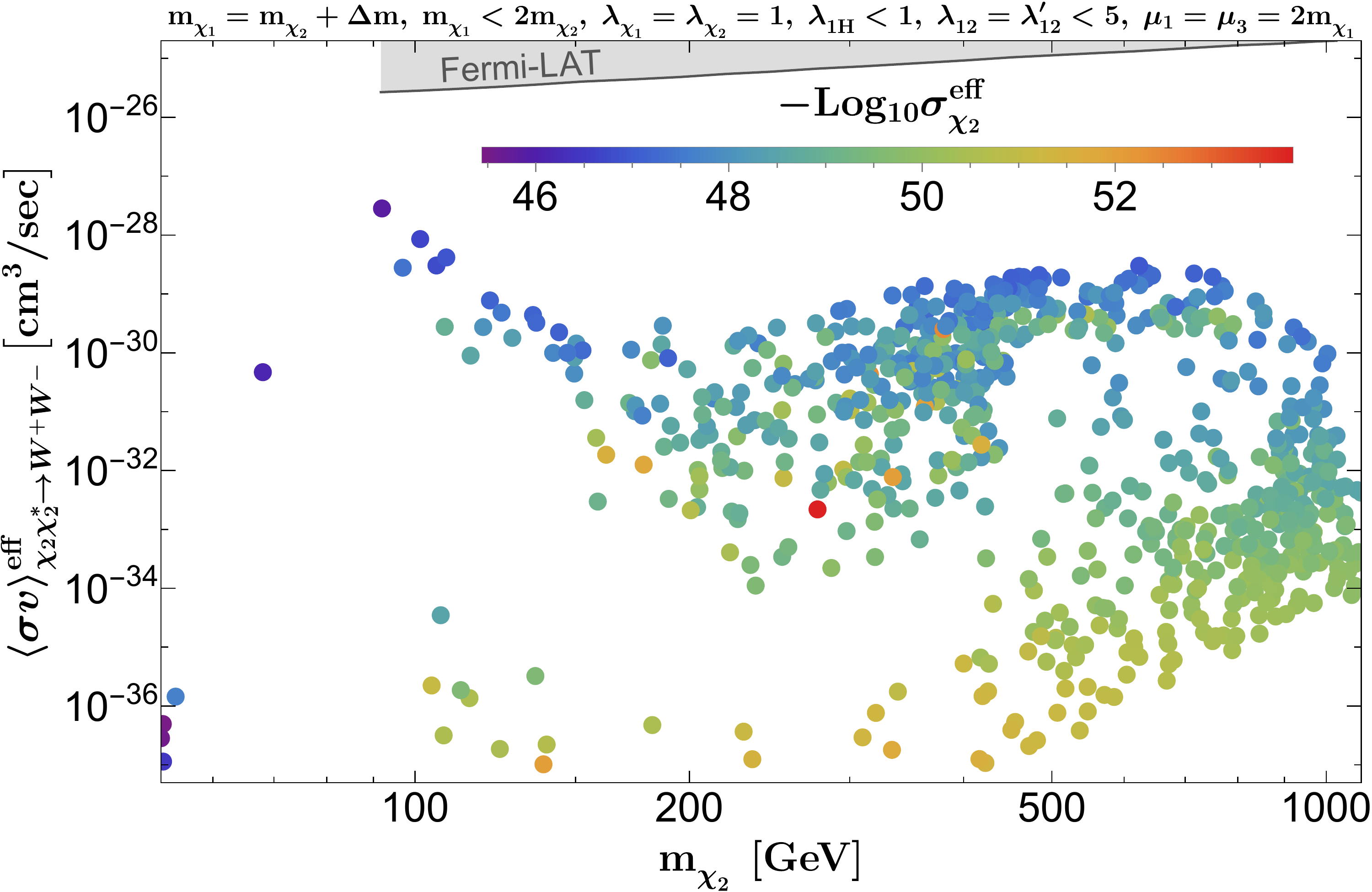}\label{fig:z3-twoscalar_chi2-idB_WW}}

\subfloat[]{\includegraphics[width=0.475\linewidth]{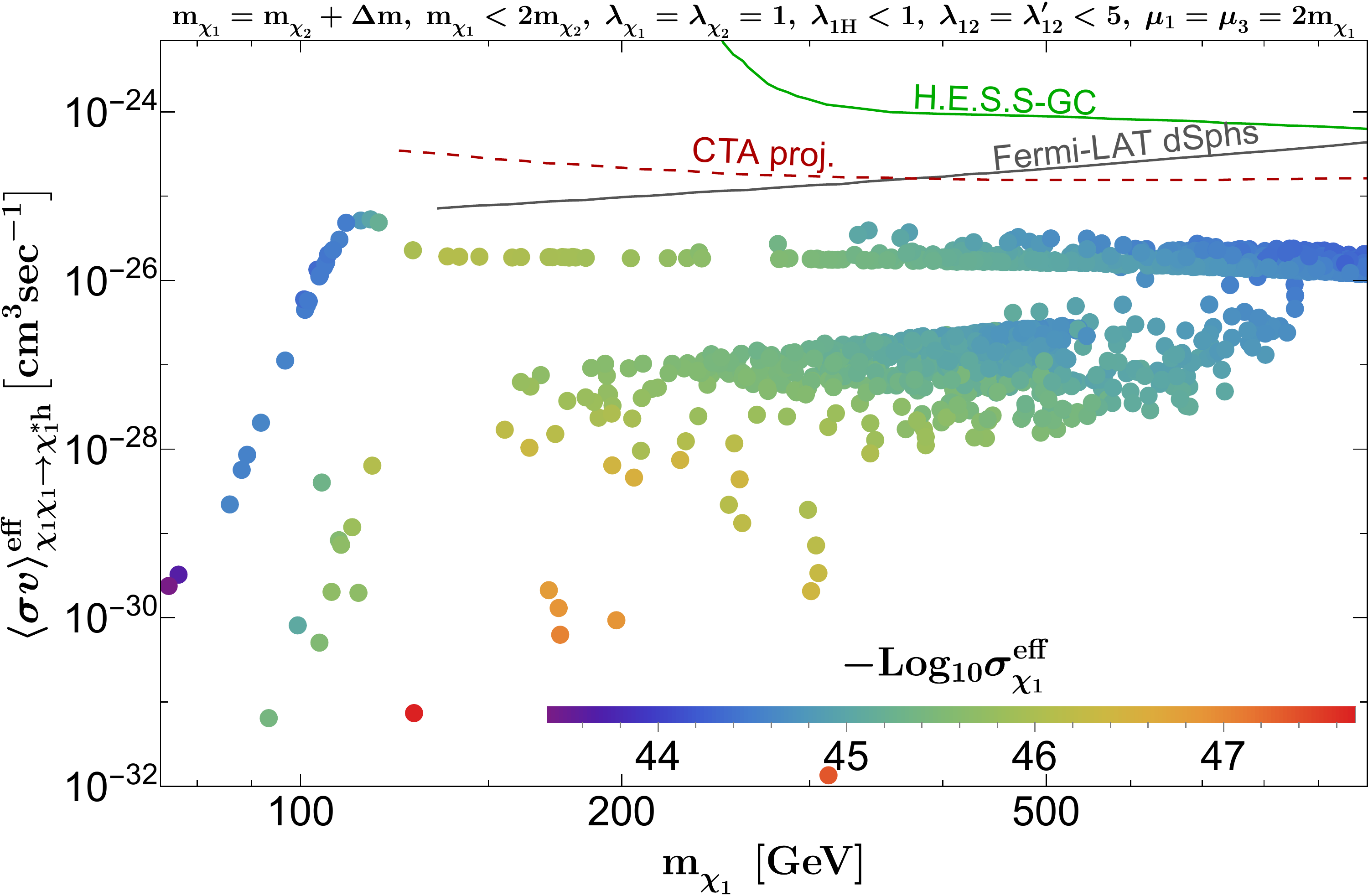}\label{fig:mchi1gmchi2_semi-chi1idB}}~~
\subfloat[]{\includegraphics[width=0.475\linewidth]{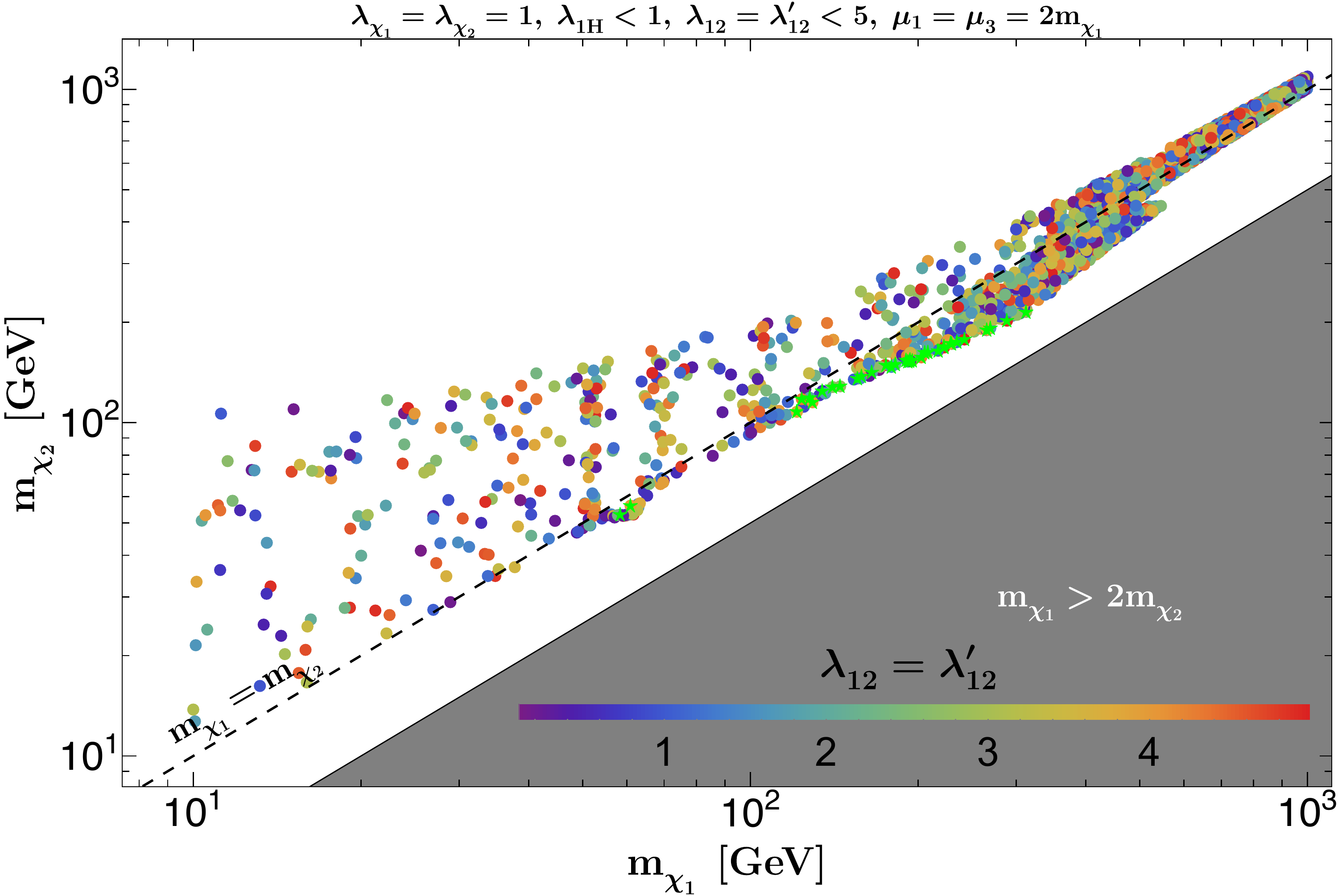}\label{fig:z3-wimp-pfimp_m1-m2-l12B}}
\caption{Figs\,.~\ref{fig:mchi1gmchi2_chi1idB}, \ref{fig:mchi1gmchi2_chi1idB_WW} and figs\,.~\ref{fig:z3-twoscalar_chi2-idB}, \ref{fig:z3-twoscalar_chi2-idB_WW} represent indirect detection limit on the self-annihilation of WIMP and pFIMP into $\to b\overline{b}~(W^+W^-)$ channel in the relic density allowed parameter space, respectively. \Cref{fig:mchi1gmchi2_semi-chi1idB} shows the indirect observational constraints on the WIMP semi-annihilation. In \cref{fig:z3-wimp-pfimp_m1-m2-l12B}, we show the relic density allowed parameter space in $m_{\chi_1}-m_{\chi_2}$ plane where WIMP and pFIMP DD and ID allowed points are indicated by green star. The thick and dashed lines correspond to the observed and projected limits of the experiments
as mentioned in the figure(s).}
\label{fig:z3-idB}
\end{figure}

The relic density in WIMP-pFIMP limit is calculated by solving cBEQ, using micrOMEGAs \cite{Belanger:2018ccd}; 
for relevant Feynman diagrams related to annihilation, semi-annihilation and conversion channels, see 
\cref{fig:z3-feynman3}. The WIMP and pFIMP relic density allowed parameter spaces are shown in \cref{fig:z3-ddB,fig:z3-idB}.
As before, the direct and indirect detection of WIMP is possible through the Higgs portal $\lambda_{1H}$, but the pFIMP detection is only possible via the 
WIMP loop-mediated diagrams, see \cref{fig:pfimp-detectionB}. The loop divergences have been taken care of using the on-shell renormalisation scheme, 
while the renormalization scale is chosen at $\sim 4m_{\chi_2}^2$.

\Cref{fig:z3-ddB} show the WIMP and pFIMP relic density allowed parameter space in $m_{\chi_1}-\sigma_{\chi_1}^{\rm eff}$, $m_{\chi_2}-\sigma_{\chi_2}^{\rm eff}$ 
planes, and the colour bar shows the variation of different relevant parameters as mentioned in figure insets. 
The different coloured lines correspond to the lower bounds from different direct detection 
experiments, while thick and dashed lines represent the observed and projected limits on the spin-independent DM-nucleon scattering cross-section. 
We see that along with the Higgs resonance, the semi-annihilation region is also available. One important point to note that we can also have 
$\sim$ 500 GeV WIMP mass, where the mass splitting $|\Delta m|$ can go up to $100$ GeV, see \cref{fig:z3-twoscalar_chi2-dd1B}. 
This is unlike $5$ GeV mass splitting for two real scalar WIMP-pFIMP DM scenario \cite{Bhattacharya:2022dco}. 
The reason for this is the presence of additional semi-conversion channel $\chi_2\chi_2\to\chi_1^* h$ (\cref{feyn:conv-6}) here, which is strongly active when $m_{\chi_1}\sim m_h$. 
\Cref{fig:z3-twoscalar_chi2-dd2B} show the relic density allowed parameter space in $m_{\chi_2}^{}-\sigma_{\chi_2}^{\rm eff}$, and the color bar show the variation of
WIMP-Higgs portal coupling $(\lambda_{1H})$. The pFIMP-nucleon cross-section is directly proportional to ($\mu_1$, $\lambda_{12}$, $\lambda_{1H}$), 
depends on the WIMP and pFIMP masses via loop factor, and on the effective relic density contribution.
Due to this reason, around the Higgs-resonance regime, the WIMP relic is very small compared to pFIMP, and the effective relic 
contribution decreases with larger pFIMP mass, as visible in \cref{fig:z3-twoscalar_chi2-dd1B}. 

\Cref{fig:z3-idB} shows indirect detection limit on the DM self-annihilation and semi-annihilation in the relic-allowed parameter space of the WIMP-pFIMP set up. 
The Fermi-LAT, H.E.S.S and CTA put limits on the annihilation of WIMP and pFIMP into $b\overline{b}$.
\Cref{fig:mchi1gmchi2_chi1idB,fig:z3-twoscalar_chi2-idB} show the relic allowed parameter space in 
$m_{\chi_1}-\langle\sigma v\rangle^{\rm eff}_{\chi_1\chi_1^*\to b\overline{b}}$ and $m_{\chi_2}-\langle\sigma v\rangle^{\rm eff}_{\chi_2\chi_2^*\to b\overline{b}}$ plane, 
while the color bar show the variation of $\rm -Log_{10}\sigma_{\chi_1}^{\rm eff}$ and $\rm -Log_{10}\sigma_{\chi_2}^{\rm eff}$ respectively.
For both the plots, near Higgs resonance, some points are disallowed by the Fermi-LAT data. Semi-annihilation bounds from indirect searches are not 
applicable to pFIMP due to the absence of the $\chi_2^3$ term. Fig\,.~\ref{fig:z3-wimp-pfimp_m1-m2-l12B} show relic density allowed points $m_{\chi_1}-m_{\chi_2}$ plane 
where the grey shaded region is excluded by the DM decay kinematics. The relic and DD allowed points, denoted by green stars lie in the vicinity of mass degenerate line. This is in contrast to the WIMP-WIMP situation described before (see fig\,.~\ref{fig:2wimp-m1-m2-l12}).
\section{Summary and Conclusion}
\label{sec:summary}
DM is well motivated from several observations, but yet undiscovered. Single component DM frameworks are often constrained heavily from relic density and DD/ID bounds, as the process which governs them are essentially the same, but one requires adequate depletion of weak interaction strength to satisfy correct relic density, but very tiny interaction for direct search from non observation of DM. Multipartite DM frameworks go beyond the simplistic assumption of being constituted of a single particle, and offer many dynamical possibilities and a larger parameter space to be explored, as the depletion process can be segregated from the direct/indirect/collider interactions. Achieving a stable DM candidate requires extra symmetries to be imposed under which DM transforms non trivially, but SM is a singlet. On the other hand, multi-component DM requires more than one stabilising symmetry most often.

The motivation of our study is to find the possible multicomponent DM frameworks when SM is extended with more than one scalar fields, but transforming under a single symmetry. Such efforts have already been done in many papers, however under what circumstances such scenarios evolve has not been elaborated systematically. The heavier component usually have decay terms to DM and to SM. In our paper, we show that the stabilisation of the heavier component depend strictly on the charges of the dark fields under symmetry transformation. For example, whenever, under $\mathcal{Z}_N$ symmetry, $q_1\neq q_2, q_1+q_2\neq N$, the heavier component decays only to dark sector particle, thus allowing it to be stabilized by imposing simple kinematical constraint. However, if the above condition on dark sector particle charges is not obeyed, then heavier component can decay to DM and SM both. Such decays at tree level, one loop and two loop level can only be stopped when we make some couplings vanishingly small. After that the scenario either resembles to $\mathcal{Z}_N \otimes \mathcal{Z}^{\prime}_N$ case, or $q_1\neq q_2, q_1+q_2\neq N^{\prime}$ case of a $\mathcal{Z}_{N^{\prime}}$ group, where $N^{\prime}>N$. We have explicitly demonstrated all the possibilities for having two DM components in $\mathcal{Z}_2$, $\mathcal{Z}_3$ and $\mathcal{Z}_4$ symmetric cases. We should note here that as there is no choice pertaining to $q_1\neq q_2, q_1+q_2\neq N$ for $\mathcal{Z}_2$ and $\mathcal{Z}_3$ symmetric cases; forcing to compromise some of the couplings of these models to make the heavier DM component stable.   

we specifically elaborate the phenomenology associated to $\mathcal{Z}_3$. Interestingly, the choices of terms that should be neglected (or assumed small) are not unique. This gives rise to different possible scenarios in $\mathcal{Z}_3$ symmetry after stabilising the heavier component. Like one can have  $\mathcal{Z}_3\otimes \mathcal{Z}^{\prime}_{3}$ or $\mathcal{Z}_6$ with $\{q_1,q_2\}=\{2,5\},\{4,1\}$ etc. They also provide different phenomenological implications. In this article, we have discussed two such examples. Conversion, semi conversion and semi-annihilations are some of the key processes that dictate the allowed parameter space of the two component DM model and their compatibility with DD/ID searches. Note that after the heavy particle becomes a stable DM candidate, co-annihilation process stops contributing. This reduces allowed parameter space of the model, particularly to comply with non-observation of DM in DD/ID experiments. 

Further, the two DM components can be WIMP or pFIMP depending on the strength of the corresponding Higgs portal couplings. Both WIMP-WIMP case, and WIMP-pFIMP cases are illustrated here and we show that WIMP-pFIMP case enjoys a larger parameter space as the pFIMP detection relies on loop level WIMP mediated interactions. There is another interesting feature that comes out of the specific WIMP-pFIMP analysis in $\mathcal{Z}_3$ symmetry that here pFIMP not only relies on conversion with WIMP to thermalise, but also semi-conversion plays a crucial role for the same, which makes a crucial distinction that one can choose a large mass splitting between WIMP and pFIMP unlike the $\mathcal{Z}_2$ case with two real scalars \cite{Bhattacharya:2022dco}. This in turn can help distinguishing WIMP and pFIMP in direct search and collider search experiments, via a kink or a double hump missing energy distribution, which is otherwise difficult with smaller mass splitting. We will discuss that possibility elsewhere.  
\acknowledgments
DP thanks Heptagon, IITG for useful discussions.
\appendix
\section{Kinetic mixing}
\label{app:kinetic-mixing}
The SM extended Lagrangian, containing two complex scalar fields, $\chi_1^{0}$ and $\chi_2^{0}$, which transform under the $\mathcal{Z}_3$ symmetry as $\chi_1^{0} \rightarrow \omega \chi_1^0$ and $\chi_2^0 \rightarrow \omega^2 \chi_2^0$, is given by
\begin{align}
\mathcal{L}=\mathcal{L}_{\rm SM}+\mathcal{L}^{\rm kin}_{\chi_1^0\chi_2^0}+\mathcal{L}^{\rm mass}(\chi_1^0,\chi_2^0)+\mathcal{L}^{\rm int}(\chi_1^0,\chi_2^0,H)\,.
\label{eq:lag}
\end{align}
The kinetic Lagrangian contains a non-canonical kinetic mixing term,
\begin{equation}
\mathcal{L}^{\rm kin}_{\chi_1^0\chi_2^0} =  (\partial_\mu\chi_1^0)^\dagger(\partial^\mu\chi_1^0) + (\partial_\mu\chi_2^0)^\dagger(\partial^\mu\chi_2^0) + \kappa \, (\partial_\mu\chi_1^0)(\partial^\mu\chi_2^0) + \kappa\, (\partial_\mu\chi_1^0)^\dagger(\partial^\mu\chi_2^0)^\dagger\,,
\label{eq:lkinetic}
\end{equation}
where $|\kappa|<1$ is required for the kinetic matrix to be positive definite. For simplicity, we also assume that $\kappa$ is real. The presence of the holomorphic kinetic mixing term prevents the kinetic matrix from being diagonal. Therefore, the first step is to identify a field basis in which the kinetic sector becomes diagonal\footnote{We refer to this as a holomorphic kinetic mixing term because it mixes two fields with identical gauge quantum numbers through a bilinear of the form $\chi_1^0 \chi_2^0$, rather than through a Hermitian combination involving a field and its conjugate. This terminology is analogous to the holomorphic bilinears that appear in supersymmetric theories \cite{Schmitz2009KineticMI, Bijnens:2018rqw, Boyanovsky:2017esz}.}.

The quadratic part of the scalar potential is given by
\begin{align}
\nonumber\mathcal{L}^{\rm mass}_{\chi_1^0\chi_2^0} = &
\underbrace{\left(\mu_{11}^2+\dfrac{1}{2}\lambda^0_{1H}\, v^2\right)}_{m_{11}^2} (\chi_1^0)^\dagger(\chi_1^0) + \underbrace{\left(\mu_{22}^2+\dfrac{1}{2}\lambda_{2H}^0\, v^2\right) (\chi_2^0)^\dagger(\chi_2^0)}_{m_{22}^2}  + \\\nonumber&\underbrace{\dfrac{1}{2}\left(\mu_{12}^2+\dfrac{1}{2}\lambda_{12H}\, v^2\right) (\chi_1^0\chi_2^0)}_{m_{12}^2} +\underbrace{\dfrac{1}{2}\left(\mu_{12}^2+\dfrac{1}{2}\lambda_{12H}\, v^2\right)(\chi_1^0)^\dagger (\chi_2^0)^\dagger}_{m_{12}^2}  \\
=& m_{11}^2(\chi_1^0)^\dagger(\chi_1^0) + m_{22}^2(\chi_2^0)^\dagger(\chi_2^0)+m_{12}^2(\chi_1^0\chi_2^0)+m_{12}^2(\chi_1^0)^\dagger (\chi_2^0)^\dagger\,,
\label{eq:lmass}
\end{align}
\begin{align}
\nonumber \mathcal{L}^{\rm int}(\chi_1^0,\chi_2^0,H)=\,&+\lambda^0_{\chi_1}|\chi^0_1|^4+\lambda^0_{\chi_2}|\chi^0_2|^4+\lambda^0_{1H}|\chi^0_1|^2({\rm H^{\dagger}H}-v^2/2)
+\lambda^0_{2H}|\chi^0_2|^2({\rm H^{\dagger}H}-v^2/2)\\\nonumber&
+\lambda^0_{12}|\chi^0_1|^2|\chi_2|^2+\dfrac{1}{2}\lambda_{12H}^0(\chi^0_1\chi^0_2+h.c.)({\rm H^{\dagger}H}-v^2/2)\\\nonumber&
+\dfrac{1}{2}\lambda_{12}^{0^\prime}\left[(\chi^0_1\chi^0_2)^2+h.c.\right]+\dfrac{1}{2}\left[\lambda^0_{3}~\chi^0_1\chi^0_2|\chi^0_1|^2+\lambda^0_{4}~\chi^0_1\chi^0_2|\chi^0_2|^2 + h.c. \right]\\&+\dfrac{1}{2}\left[\mu^0_{1}(\chi^0_1)^{*}(\chi^0_2)^2+\mu^0_{2}(\chi^0_1)^2(\chi^0_2)^{*}+\mu^0_{3}(\chi^0_1)^3+\mu^0_{4}(\chi^0_2)^3+h.c.\right]\,.
\label{eq:intold}
\end{align}
\section*{Kinetic diagonalization:}
\begin{align}
\begin{pmatrix}
\chi_1^0\\ (\chi_2^0)^\dagger
\end{pmatrix}
=
\dfrac{1}{\sqrt{2}}\begin{pmatrix}
1& -1\\
1& 1
\end{pmatrix}
\begin{pmatrix}
H_1\\ H_2
\end{pmatrix}
\end{align}

\noindent We introduce the field redefinitions \cite{Lebedev:2011aq,Ema:2017rqn,Boyanovsky:2017esz, vandeBruck:2009gp}:
\begin{equation}
\chi_1^0 =  \frac{H_1-H_2}{\sqrt{2}}, \qquad (\chi_2^0)^\dagger =\frac{H_1+H_2}{\sqrt{2}},
\label{eq:Hbasis}
\end{equation}
or equivalently
\begin{equation}
H_1 = \frac{\chi_1^0+(\chi_2^0)^\dagger}{\sqrt{2}}, \qquad  H_2 = \frac{(\chi_2^0)^\dagger-\chi_1^0}{\sqrt{2}}.
\end{equation}
Substituting Eq.~(\ref{eq:Hbasis}) into Eq.~(\ref{eq:lkinetic}) yields
\begin{equation}
\mathcal{L}^{\rm kin}_{H_1H_2} = (1+\kappa) \, \partial_\mu H_1^\dagger\partial^\mu H_1 + (1-\kappa)\, \partial_\mu  H_2^\dagger\partial^\mu H_2 .
\label{eq:Hkin}
\end{equation}
The kinetic mixing has therefore been eliminated. However, the kinetic terms are not yet canonically normalized.
\subsubsection*{Canonical normalization:}
To obtain canonical kinetic terms, we perform the rescaling 
\begin{equation}
H_1 =\sqrt{2}\kappa_{+}{h_1}, \qquad H_2 = \sqrt{2}\kappa_{-}{h_2}.
\label{eq:can}
\end{equation}
where $\kappa_{\pm}=\dfrac{1}{\sqrt{2(1\pm\kappa)}}$.
Substituting Eq.~(\ref{eq:can}) into Eq.~(\ref{eq:Hkin}) gives
\begin{equation}
\mathcal{L}^{\rm kin}_{h_1h_2} \equiv (\partial_\mu h_1)^\dagger\partial^\mu h_1 + (\partial_\mu h_2)^\dagger\partial^\mu h_2 ,
\end{equation}
which is the desired canonical form. At this stage, the kinetic sector has been completely diagonalised and canonically normalised.
\section*{Mass terms in the canonical basis:}
The next step is to rewrite the quadratic scalar potential in terms of the canonically normalized fields. Using Eqs.~(\ref{eq:Hbasis}) and (\ref{eq:can}), the potential in Eq.~(\ref{eq:lmass}) becomes
\begin{align}
\nonumber\mathcal{L}^{\rm mass}_{\chi_1^0\chi_2^0}\Rightarrow
\mathcal{L}^{\rm mass}_{h_1h_2} = &
\kappa_{+}^2(m_{11}^2+2m_{12}^2+m_{22}^2) \, h_1 h_1^\dagger \\\nonumber+&
\kappa_{-}^2(m_{11}^2-2m_{12}^2+m_{22}^2)\, h_2 h_2^\dagger \\+&
\kappa_{+}\kappa_{-}(m_{22}^2-m_{11}^2)\, (h_1h_2^\dagger+h_2h_1^\dagger) \,.
\label{eq:lmass1}
\end{align}

The mass-mixing matrix can be written as:
\begin{align}
\mathtt{M}_{h_1 h_2}=
\begin{pmatrix}
\kappa_{+}^2(m_{11}^2+2m_{12}^2+m_{22}^2)&\kappa_{+}\kappa_{-}(m_{22}^2-m_{11}^2)\\
\kappa_{+}\kappa_{-}(m_{22}^2-m_{11}^2)&\kappa_{-}^2(m_{11}^2-2m_{12}^2+m_{22}^2)
\end{pmatrix}
\label{eq:massmix1}
\end{align}
The mass terms can be written in matrix form as:
\begin{align}
\mathcal{L}^{\rm mass}_{h_1h_2} =
\begin{pmatrix}
h_1^\dagger &&h_2^\dagger
\end{pmatrix} 
\mathtt{M}_{h_1h_2} 
\begin{pmatrix}
h_1\\h_2
\end{pmatrix}
\end{align}
To diagonalise the mass matrix, we give the following rotation to the canonically normalised fields as:
\begin{align}
\begin{pmatrix}
h_1\\h_2
\end{pmatrix}=
\begin{pmatrix}
\cos\theta && -\sin\theta\\
\sin\theta &&\cos\theta
\end{pmatrix}\begin{pmatrix}
\chi_{1}\\\chi_{2}^\dagger
\end{pmatrix}\,,
\end{align}
where the mixing angle $\theta$ can be given by:
\begin{align}
\tan 2\theta=
\frac{2\kappa_{+}\kappa_{-}(m_{22}^2-m_{11}^2)}{(m_{11}^2+m_{22}^2)(\kappa_{+}^2-\kappa_{-}^2)+2m_{12}^2(\kappa_{+}^2+\kappa_{-}^2)}.
\end{align}
and the diagonalised fields have the physical mass:
\begin{align}
m_{\chi_{1,2}^{}}^2
=
\frac12
\left[
(\mathtt{M}_{h_1 h_2}^{11})^2+(\mathtt{M}_{h_1 h_2}^{22})^2
\mp
\sqrt{((\mathtt{M}_{h_1 h_2}^{11})^2-(\mathtt{M}_{h_1 h_2}^{22})^2)^2+4(\mathtt{M}_{h_1 h_2}^{12})^4}
\right].
\end{align}
Therefore, old basis in terms of new physical basis can be written as,
\begin{align}
\begin{pmatrix}
\chi_1^0\\
(\chi_2^0)^\dagger
\end{pmatrix}=\underbrace{\begin{pmatrix}
\kappa_+&&-\kappa_-\\
\kappa_+&&\kappa_-
\end{pmatrix}
\begin{pmatrix}
\cos\theta && -\sin\theta\\
\sin\theta &&\cos\theta
\end{pmatrix}}_{\mathbb{R}}\begin{pmatrix}
\chi_{1}\\\chi_{2}^\dagger
\end{pmatrix}\,.
\label{eq:rotation}
\end{align}
From eq.~\eqref{eq:rotation}, we get $\chi_1^0=\mathbb{R}_{11}\chi_1+\mathbb{R}_{12}\chi_2^\dagger$ and $(\chi_2^0)^\dagger=\mathbb{R}_{21}\chi_1+\mathbb{R}_{22}\chi_2^\dagger$, the Lagrangian can be expressed in terms of these new physical basis, $\chi_1$ and $\chi_2$, as:
\begin{align}
\mathcal{L}^{\rm phy}=\mathcal{L}_{\rm SM}+\mathcal{L}^{\rm kin}_{\chi_1\chi_2}+\mathcal{L}^{\rm mass}(\chi_1,\chi_2)+\mathcal{L}^{\rm int}(\chi_1,\chi_2,H)\,.
\label{eq:lagnew}
\end{align}
where 
\begin{gather}
\mathcal{L}^{\rm kin}_{\chi_1\chi_2} =  (\partial_\mu\chi_1)^\dagger(\partial^\mu\chi_1) + (\partial_\mu\chi_2)^\dagger(\partial^\mu\chi_2)\,,\\
\mathcal{L}^{\rm mass}_{\chi_1\chi_2} = m_{\chi_1}^2\chi_1^\dagger\chi_1 + m_{\chi_2}^2\chi_2^\dagger\chi_2\,,
\label{eq:lkineticnew}
\end{gather}
\begin{align}
\nonumber \mathcal{L}^{\rm int}(\chi_1,\chi_2,H)=\,&\lambda_{\chi_1}|\chi_1|^4+\lambda_{\chi_2}|\chi_2|^4+\lambda_{1H}|\chi_1|^2({\rm H^{\dagger}H}-v^2/2)
+\lambda_{2H}|\chi_2|^2({\rm H^{\dagger}H}-v^2/2)\\\nonumber+&
\lambda_{12}|\chi_1|^2|\chi_2|^2+\dfrac{1}{2}\lambda_{12H}(\chi_1\chi_2+h.c.)({\rm H^{\dagger}H}-v^2/2)\\\nonumber+&
\dfrac{1}{2}\lambda_{12}^{\prime}\left[(\chi_1\chi_2)^2+h.c.\right]+\dfrac{1}{2}\left[\lambda_{3}~\chi_1\chi_2|\chi_1|^2+\lambda_{4}~\chi_1\chi_2|\chi_2|^2 + h.c. \right]\\+&\dfrac{1}{2}\left[\mu_{1}\chi_2^2\chi_1^{*}+\mu_{2}\chi_1^2\chi_2^{*}+\mu_{3}\chi_1^3+\mu_{4}\chi_2^3+h.c.\right]\,,
\label{eq:intnew}
\end{align}
with 
\begin{align}
\begin{split}
\lambda_{\chi_1}=&2 \kappa_{+}^3 \kappa_{-}\sin\theta~\cos^3\theta \left(-2 \lambda^0_{\chi _1}+2 \lambda^0_{\chi _2}-\lambda^0_3+\lambda^0_4\right)\\&
+2 \kappa_+ \kappa_-^3 \sin^3\theta \cos\theta \left(-2 \lambda^0_{\chi _1}+2 \lambda^0_{\chi _2}+\lambda^0_3-\lambda^0_4\right)\\&
+\kappa_-^4 \sin^4\theta \left(\lambda^0_{\chi _1}+\lambda^0_{\chi _2}-\lambda^0_3-\lambda^0_4+\lambda^0_{12}+\lambda^{0^\prime}_{12}\right)\\&
+\kappa_+^4 \cos^4\theta \left(\lambda^0_{\chi _1}+\lambda^0_{\chi _2}+\lambda^0_3+\lambda^0_4+\lambda^0_{12}+\lambda^{0^\prime}_{12}\right)\\&
-2 \kappa_+^2 \kappa_-^2 \sin^2\theta \cos^2\theta \left(-3 \left(\lambda^0_{\chi _1}+\lambda^0_{\chi _2}\right)+\lambda^0_{12}+\lambda^{0^\prime}_{12}\right)\,,
\end{split}
\end{align}
\begin{align}
\begin{split}
\lambda_{\chi_2}=&2 \kappa _+ \kappa _-^3 \sin \theta \cos ^3\theta \left(2 \lambda _{\chi_1}^0-2 \lambda _{\chi _2}^0-\lambda _3^0+\lambda _4^0\right)\\+&2 \kappa _+^3
\kappa _- \sin ^3\theta \cos \theta \left(2 \lambda _{\chi _1}^0-2
\lambda _{\chi _2}^0+\lambda _3^0-\lambda _4^0\right)\\+&\kappa _+^4 \sin
^4\theta \left(\lambda _{\chi _1}^0+\lambda _{\chi _2}^0+\lambda
_3^0+\lambda _4^0+\lambda _{12}^0+\lambda_{12}^{0^\prime}\right)\\+&\kappa _-^4 \cos
^4\theta \left(\lambda _{\chi _1}^0+\lambda _{\chi _2}^0-\lambda
_3^0-\lambda _4^0+\lambda _{12}^0+\lambda_{12}^{0^\prime}\right)\\-&2 \kappa _+^2
\kappa _-^2 \sin ^2\theta \cos ^2\theta \left(-3 \left(\lambda _{\chi
_1}^0+\lambda _{\chi _2}^0\right)+\lambda _{12}^0+\lambda _{12}^{0^\prime}\right)\,,
\end{split}
\end{align}
\begin{align}
\begin{split}
\dfrac{1}{2}\lambda_{3}=-&\frac{1}{2} \kappa _+^3 \kappa _- (\cos (2 \theta )+\cos (4 \theta )) \left(2\lambda _{\chi _1}^0-2 \lambda _{\chi _2}^0+\lambda _3^0-\lambda
_4^0\right)\\+&\kappa _+ \kappa _-^3 \sin ^2\theta (2 \cos (2 \theta )+1)
\left(-2 \lambda _{\chi _1}^0+2 \lambda _{\chi _2}^0+\lambda _3^0-\lambda
_4^0\right)\\-&\frac{1}{2} \kappa _+^2 \kappa _-^2 \sin (4 \theta ) \left(-3
\left(\lambda _{\chi _1}^0+\lambda _{\chi _2}^0\right)+\lambda
_{12}^0+\lambda_{12}^{0^\prime}\right)\\-&2 \kappa _+^4 \sin \theta \cos ^3(\theta
) \left(\lambda _{\chi _1}^0+\lambda _{\chi _2}^0+\lambda _3^0+\lambda
_4^0+\lambda _{12}^0+\lambda_{12}^{0^\prime}\right)\\-&2 \kappa _-^4 \sin ^3\theta
\cos \theta \left(-\lambda _{\chi _1}^0-\lambda _{\chi _2}^0+\lambda
_3^0+\lambda _4^0-\lambda _{12}^0-\lambda _{12}^{0^\prime}\right)\,,
\end{split}
\end{align}
\begin{align}
\begin{split}
\dfrac{1}{2}\lambda_{4}=&\frac{1}{2}\kappa _+ \kappa _-^3 (\cos (2 \theta )+\cos (4 \theta ))
\left(-2 \lambda _{\chi _1}^0+2 \lambda _{\chi _2}^0+\lambda _3^0-\lambda
_4^0\right)\\-&
\kappa _+^3 \kappa _- \sin ^2\theta (2 \cos (2 \theta )+1)
\left(2 \lambda _{\chi _1}^0-2 \lambda _{\chi _2}^0+\lambda _3^0-\lambda
_4^0\right)\\+&
\frac{1}{2}\kappa _+^2 \kappa _-^2 \sin (4 \theta ) \left(-3 \left(\lambda_{\chi _1}^0+\lambda _{\chi _2}^0\right)+\lambda _{12}^0+\lambda_{12}^{0^\prime}\right)\\-&
2 \kappa _-^4 \sin \theta \cos ^3\theta \left(-\lambda_{\chi _1}^0-\lambda _{\chi _2}^0+\lambda _3^0+\lambda _4^0-\lambda_{12}^0-\lambda_{12}^{0^\prime}\right)\\-&
2 \kappa _+^4 \sin ^3\theta \cos \theta \left(\lambda _{\chi _1}^0+\lambda _{\chi _2}^0+\lambda _3^0+\lambda_4^0+\lambda _{12}^0+\lambda_{12}^{0^\prime}\right)\,,
\end{split}
\end{align}
\begin{align}
\begin{split}
\lambda_{12}=&\kappa _+ \kappa _-^3 \sin (4 \theta ) \left(-2 \lambda _{\chi _1}^0+2 \lambda_{\chi _2}^0+\lambda _3^0-\lambda _4^0\right)\\&
+\kappa _+^3 \kappa _- \sin (4\theta ) \left(2 \lambda _{\chi _1}^0-2 \lambda _{\chi _2}^0+\lambda_3^0-\lambda _4^0\right)\\&
+\kappa _-^4 \left(-\sin ^2(2 \theta )\right)\left(-\lambda _{\chi _1}^0-\lambda _{\chi _2}^0+\lambda _3^0+\lambda_4^0-\lambda _{12}^0-\lambda_{12}^{0^\prime}\right)\\&
+\kappa _+^4 \sin ^2(2 \theta )\left(\lambda _{\chi _1}^0+\lambda _{\chi _2}^0+\lambda _3^0+\lambda_4^0+\lambda _{12}^0+\lambda_{12}^{0^\prime}\right)\\&
-\kappa _+^2 \kappa _-^2\left(-(3 \cos (4 \theta )+1) \left(\lambda _{\chi _1}^0+\lambda _{\chi_2}^0\right)+\lambda _{12}^0 (\cos (4 \theta )-1)+(\cos (4 \theta )+3)\lambda_{12}^{0^\prime}\right)\,,
\end{split}
\end{align}
\begin{align}
\begin{split}
\dfrac{1}{2}\lambda_{12}^{\prime}=&\frac{1}{8} \biggl(2 \kappa _+ \kappa _-^3 \sin (4 \theta) \left(-2 \lambda_{\chi _1}^0+2 \lambda _{\chi _2}^0+\lambda _3^0-\lambda _4^0\right)\\&
+2 \kappa_+^3 \kappa _- \sin (4 \theta ) \left(2 \lambda _{\chi _1}^0-2 \lambda _{\chi_2}^0+\lambda _3^0-\lambda _4^0\right)\\    &
+2 \kappa _+^4 \sin ^2(2 \theta ) \left(\lambda _{\chi _1}^0+\lambda _{\chi _2}^0+\lambda _3^0+\lambda_4^0+\lambda _{12}^0+\lambda_{12}^{0^\prime}\right)\\    &
+\kappa _-^4 (\cos (4 \theta)-1) \left(-\lambda _{\chi _1}^0-\lambda _{\chi _2}^0+\lambda _3^0+\lambda_4^0-\lambda _{12}^0-\lambda_{12}^{0^\prime}\right)\\   &
-2 \kappa _+^2 \kappa _-^2\left(-(3 \cos (4 \theta )+1) \left(\lambda _{\chi _1}^0+\lambda _{\chi_2}^0\right)+\lambda _{12}^0 (\cos (4 \theta )+3)+(\cos (4 \theta )-5)\lambda_{12}^{0^\prime}\right)\biggr)\,,
\end{split}
\end{align}
\begin{align}
\begin{split}
\lambda_{1H}=&\kappa _+^2 \cos ^2\theta \left(\lambda _{1H}^0+\lambda _{2 H}^0+2 \lambda _{12H}^0\right)\\+&\kappa _- \sin \theta \left(\kappa _- \sin \theta\left(\lambda_{1H}^0+\lambda _{2 H}^0-2 \lambda _{12 H}^0\right)+2 \kappa _+\cos \theta \left(\lambda _{2 H}^0-\lambda_{1H}^0\right)\right)\,,
\end{split}
\end{align}
\begin{align}
\begin{split}
\lambda_{2H}=&\kappa _-^2 \cos ^2\theta \left(\lambda_{1H}^0+\lambda _{2 H}^0-2 \lambda _{12H}^0\right)\\+&\kappa _+ \sin \theta \left(\kappa _+ \sin \theta\left(\lambda_{1H}^0+\lambda _{2 H}^0+2 \lambda _{12 H}^0\right)+2 \kappa _-\cos \theta \left(\lambda_{1H}^0-\lambda _{2 H}^0\right)\right)\,,
\end{split}
\end{align}
\begin{align}
\begin{split}
\dfrac{1}{2}\lambda_{12H}=-&\kappa _- \kappa _+ \sin ^2\theta \left(\lambda _{1H}^0-\lambda _{2H}^0\right)+\kappa _- \kappa _+ \cos ^2\theta \left(\lambda _{2H}^0-\lambda _{1H}^0\right)\\+&\left(\kappa _--\kappa _+\right)\left(\kappa _-+\kappa _+\right) \sin \theta \cos \theta \left(\lambda_{1H}^0+\lambda _{2 H}^0\right)+\left(\kappa _-^2+\kappa _+^2\right)(-\sin (2 \theta )) \lambda _{12 H}^0\,,
\end{split}
\end{align}
\begin{align}
\begin{split}
\dfrac{1}{2}\mu_{1}=\frac{1}{2} \biggl(&\frac{1}{4} \kappa _+^2 \kappa _- \left(\mu _1^0-\mu _2^0-3\mu _3^0+3 \mu _4^0\right) (\sin \theta-3 \sin (3 \theta ))\\-&\frac{1}{4}\kappa _+ \kappa _-^2 \left(\mu _1^0+\mu _2^0-3 \left(\mu _3^0+\mu_4^0\right)\right) (\cos \theta+3 \cos (3 \theta ))\\-&3 \kappa _-^3\left(\mu _1^0-\mu _2^0+\mu _3^0-\mu _4^0\right) \sin \theta \cos^2\theta\\+&3 \kappa _+^3 \left(\mu _1^0+\mu _2^0+\mu _3^0+\mu _4^0\right)\sin ^2\theta \cos \theta\biggr)\,,
\end{split}
\end{align}
\begin{align}
\begin{split}
\dfrac{1}{2}\mu_2=\frac{1}{2} \biggl(&\frac{1}{4} \kappa _+ \kappa _-^2 \left(\mu _1^0+\mu _2^0-3\left(\mu _3^0+\mu _4^0\right)\right) (\sin \theta-3 \sin (3 \theta))\\+&\frac{1}{4} \kappa _+^2 \kappa _- \left(\mu _1^0-\mu _2^0-3 \mu _3^0+3 \mu_4^0\right) (\cos \theta+3 \cos (3 \theta ))\\-&3 \kappa _+^3 \left(\mu_1^0+\mu _2^0+\mu _3^0+\mu _4^0\right) \sin \theta \cos ^2\theta\\-&3\kappa _-^3 \left(\mu _1^0-\mu _2^0+\mu _3^0-\mu _4^0\right) \sin ^2\theta\cos \theta\biggr)\,,
\end{split}
\end{align}
\begin{align}
\begin{split}
\dfrac{1}{2}\mu_3=\frac{1}{2} \biggl(-&\left(\kappa _-^2 \sin ^2\theta-\kappa _+^2 \cos ^2\theta\right) \left(\kappa _- \left(\mu _1^0-\mu _2^0\right) \sin \theta+\kappa_+ \left(\mu _1^0+\mu _2^0\right) \cos \theta\right)\\+&\mu _3^0\left(-\left(\kappa _- \sin \theta-\kappa _+ \cos \theta\right){}^3\right)+\mu _4^0 \left(\kappa _- \sin \theta+\kappa _+ \cos\theta\right){}^3\biggr)\,,
\end{split}
\end{align}
\begin{align}
\dfrac{1}{2}\mu_{4}=\frac{1}{2} \biggl(&\mu _3^0 \left(-\left(\kappa _+ \sin \theta+\kappa _- \cos\theta\right){}^3\right)+\mu _4^0 \left(\kappa _- \cos \theta-\kappa _+\sin \theta\right){}^3\\\nonumber+&\left(\kappa _- \cos \theta-\kappa _+ \sin\theta\right) \left(\kappa _+ \sin \theta+\kappa _- \cos \theta\right) \left(\kappa _+ \left(\mu _1^0+\mu _2^0\right) \sin \theta-\kappa_- \left(\mu _1^0-\mu _2^0\right) \cos \theta\right)\biggr)\,.
\end{align}

\section{Heavy DM stability criteria from two and three-body decays}
\label{sec:decay}
\begin{figure}[htb!]\centering
\subfloat[]{
\begin{tikzpicture}
\begin{feynman}
\vertex (a1){\(A_0\)};
\vertex [right=1 cm of a1] (a2);
\vertex [above right=0.75 cm and 0.75 cm of a2] (a4){$A_1$};
\vertex [below right=0.75 cm and 0.75 cm of a2] (a5){$A_2$};
\diagram*{
(a1) -- [ line width=0.25mm,plain, arrow size=0.7pt, style=blue] (a2), 
(a2)-- [ line width=0.25mm,plain, style=black, arrow size=0.7pt] (a4) ,  
(a2)-- [ line width=0.25mm,plain, arrow size=0.7pt ] (a5)};
\end{feynman}
\end{tikzpicture}\label{fig:decay_exmp0}}
\subfloat[]{
\begin{tikzpicture}
\begin{feynman}
\vertex (a1){\(A_0\)};
\vertex [right=1 cm of a1] (a2);
\vertex [above right=0.75 cm and 0.75 cm of a2] (a4){$A_1$};
\vertex [right=0.75 cm of a2](a3) {$A_2$};
\vertex [below right=0.75 cm and 0.75 cm of a2] (a5){$A_3$};
\diagram*{
(a1) -- [ line width=0.25mm,plain, arrow size=0.7pt, style=blue] (a2), 
(a2)-- [ line width=0.25mm, plain, style=red, arrow size=0.7pt ] (a3) ,  
(a2)-- [ line width=0.25mm,plain, style=black, arrow size=0.7pt] (a4) ,  
(a2)-- [ line width=0.25mm,plain, arrow size=0.7pt ] (a5)};
\end{feynman}
\end{tikzpicture}\label{fig:decay_exmp1}}
\subfloat[]{
\begin{tikzpicture}
\begin{feynman}
\vertex (a1){\(A_0\)};
\vertex [right=1 cm of a1] (a2);
\vertex [above right=0.5 cm and 0.75 cm of a2] (a3);
\vertex [below right=0.5 cm and 0.75 cm of a2] (a4){$A_1$};
\vertex [above right=0.5 cm and 0.75 cm of a3] (a5){$A_2$};
\vertex [below right=0.5 cm and 0.75 cm of a3] (a6){$A_3$};
\diagram*{
(a1) -- [ line width=0.25mm,plain, arrow size=0.7pt, style=blue, edge label={\(\rm\color{black}{}\)}] (a2), 
(a2)-- [ line width=0.25mm, plain, style=red, arrow size=0.7pt , edge label'={\(\color{black}{\rm B}\)}] (a3) ,  
(a4)-- [ line width=0.25mm,plain, style=black, arrow size=0.7pt] (a2) ,  
(a5)-- [ line width=0.25mm,plain, arrow size=0.7pt ] (a3)-- [ line width=0.25mm,plain, arrow size=0.7pt] (a6)};
\end{feynman}
\end{tikzpicture}\label{fig:decay_exmp2}}
\subfloat[]{
\begin{tikzpicture}
\begin{feynman}
\vertex (a1){\(A_0\)};
\vertex [right=1.0 cm of a1] (a2);
\vertex [above right=0.5 cm and 0.75 cm of a2] (a3);
\vertex [below right=0.5 cm and 0.75 cm of a2] (a4){$A_1$};
\vertex [above right=0.5 cm and 0.75 cm of a3] (a5){$A_2$};
\vertex [below right=0.5 cm and 0.75 cm of a3] (a6){$A_3$};
\diagram*{
(a1) -- [ line width=0.25mm,plain, arrow size=0.7pt, style=blue, edge label={\(\rm\color{black}{}\)}] (a2), 
(a2)-- [ line width=0.25mm, plain, style=red, arrow size=0.7pt , edge label'={\(\color{black}{\rm B^{*}}\)}] (a3) ,  
(a4)-- [ line width=0.25mm,plain, style=black, arrow size=0.7pt] (a2) ,  
(a5)-- [ line width=0.25mm,plain, arrow size=0.7pt ] (a3)-- [ line width=0.25mm,plain, arrow size=0.7pt] (a6)};
\end{feynman}
\end{tikzpicture}\label{fig:decay_exmp3}}
\caption{Fig.~\ref{fig:decay_exmp0} show the two-body decay when $m_0>\sum\limits_{i=1}^2m_i$. The rest of the figures show the possible three-body decay: four-point~(\ref{fig:decay_exmp1}), off-shell $\rm B$ (\ref{fig:decay_exmp2}), on-shell $\rm B$ (\ref{fig:decay_exmp3}).}
\label{fig:decay_exmp}
\end{figure}
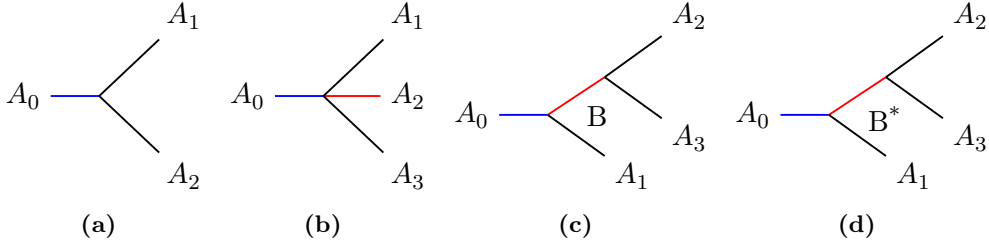

The two-body decay width in the rest frame of the decaying particle ($A_0$), $A_0(m_0)\to A_1(m_1)~A_2(m_2)$ (fig\,.~\ref{fig:decay_exmp0} ), is given by \cite{Pal:2014xrq},
\begin{gather}
\Gamma_{A_0\to A_1~A_2}=\dfrac{1}{64\pi^2m_0}\sqrt{\left[1-\left(\dfrac{m_1+m_2}{m_0}\right)^2\right]\left[1-\left(\dfrac{m_1-m_2}{m_0}\right)^2\right]}\int \overline{|\mathcal{M}|^2}d\Omega\,.
\end{gather}

Let us consider a process like $\rm A_0(m_0)\to A_1(m_1)~A_2(m_2)~A_3(m_3)$ where three particles can be produced by three different kinds of processes as shown in fig.~\ref{fig:decay_exmp}.
The three-body decay width in the rest frame of the decaying particle ($A_0$) is given by \cite{ParticleDataGroup:2022pth},
\bea
\Gamma_{A_0\to A_1A_2..A_n}=S_k\dfrac{1}{2m_0}\left(\prod_{i=1}^{n}\int \dfrac{d^3p_i^{\prime}}{(2\pi)^32E_i^{\prime}}\right)(2\pi)^4\delta^4(p-\sum_{i=1}^{n}p_i^{\prime})\overline{|\mathcal{M}|^2}_{A_0\to A_1A_2..A_n}\,,
\eea
$S_k=1/k!$ is the corresponding symmetry factor, with $k$ the number of identical final state particles. In the CM reference frame, the expression for the three body decay width \cite{Ilisie:2015wkg}, 

\bea
\Gamma_{A_0\to A_1~A_2~A_3}=S_k\dfrac{1}{(2\pi)^3}\dfrac{1}{32m_0^3}\int dm_{12}^2\int dm_{23}^2~\overline{|\mathcal{M}|^2}_{A_0\to A_1~A_2~A_3}\,,
\eea
where,
\begin{gather}
m_{23}^2|_{\rm max}=(E_2^*+E_3^*)^2-\left(\sqrt{{E_2^*}^2-m_2^2}-\sqrt{{E_3^*}^2-m_3^2}\right)^2\,,\\
m_{23}^2|_{\rm min}=(E_2^*+E_3^*)^2-\left(\sqrt{{E_2^*}^2-m_2^2}+\sqrt{{E_3^*}^2-m_3^2}\right)^2\,,
\end{gather}
\begin{gather}
m_{12}^2|_{\rm max}=(m_0-m_3)^2 {\rm~and ~}  m_{12}^2|_{\rm min}=(m_1+m_2)^2\,,\\
E_2^*=(m_{12}^2-m_1^2+m_2^2)/2m_{12}{\rm~and ~} E_3^*=(m^2_0-m_{12}^2-m_3^2)/2m_{12}\,.
\end{gather}

\paragraph{First alternative way}

\bea
\Gamma_{A_0\to A_1~A_2~A_3}=\dfrac{1}{(2\pi)^3}\frac{1}{8m_0}\int_{E^{\rm min}_{1}}^{E^{\rm max}_{1}}dE_1\int_{E^{\rm min}_{2}}^{E^{\rm max}_{2}}dE_2\overline{|\mathcal{M}|^2}_{A_0\to A_1~A_2~A_3}\,,
\eea
where the minimum and maximum energy of the particles in the CM frame are \cite{Ilisie:2015wkg, Choi:2021yps},
\begin{gather}
E_2^{\rm min}=\dfrac{1}{2m_{23}^2}\left[\left(m_0-E_1\right)m_{23}^2-\sqrt{\left(E_1^2-m_1^2\right)\lambda(m_{23}^2,m_2^2,m_3^2)}\right]\,,\\
E_2^{\rm max}=\dfrac{1}{2m_{23}^2}\left[\left(m_0-E_1\right)m_{23}^2+\sqrt{\left(E_1^2-m_1^2\right)\lambda(m_{23}^2,m_2^2,m_3^2)}\right]\,;
\end{gather}

with
\begin{gather}
m_{23}^2=m_0^2+m_1^2-2m_0E_1\,,\\
\lambda(a,b,c)\equiv a^2+b^2+c^2-2ab-2bc-2ca\,,\\
E_1^{\rm min}=m_{1}\,,\qquad E_1^{\rm max}=\dfrac{1}{2m_0}\left(m_0^2+m_1^2-4m_2^2\right)\,.
\end{gather}
and $\lambda$ is the K$\ddot{\rm a}$ll\'{e}n function.

\paragraph{Second alternative way \cite{Han:2005mu,Ilisie:2015wkg}}
\begin{align}
d\Gamma_{A_0\to A_1~A_2~A_3}=\frac{1}{64m_0^3} \frac{1}{(2\pi)^3}\frac{\sqrt{ \lambda(m_0^2,m_1^2,q^2)\lambda(q^2, m_2^2, m_3^2)} }{ q^2 } dq^2 d\cos \theta^*  |\mathcal{M}|^2_{A_0\to A_1~A_2~A_3}\,.
\end{align}
where, $q=p_0-p_1$, $q^2_{\rm min}=(m_2+m_3)^2$, $q^2_{\rm max}=(m_0-m_1)^2$.
\subsection{Tree and loop level decay of the heavy DM with $\mathcal{Z}_2$ symmetry}
\label{apndx:z2-decay}
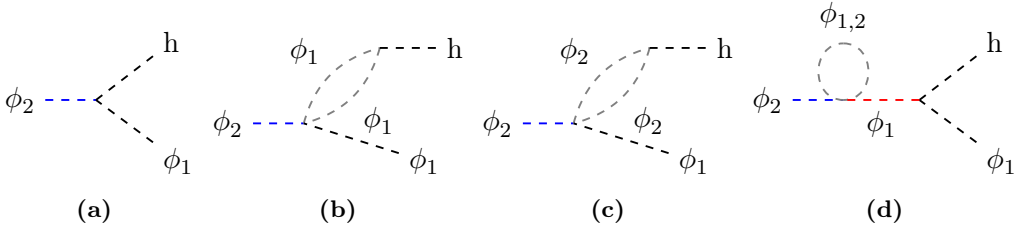
\begin{figure}[htb!]\centering
\subfloat[]{
\begin{tikzpicture}
\begin{feynman}
\vertex (a1){\(\phi_2\)};
\vertex [right=1 cm of a1] (a2);
\vertex [above right=0.5cm and 0.75 cm of a2] (a3){h};
\vertex [below right=0.5cm and 0.75 cm of a2] (a4){$\phi_1$};
\diagram*{
(a1) -- [ line width=0.25mm, scalar, arrow size=0.7pt, style=blue, edge label={\(\rm\color{black}{}\)}] (a2), 
(a2)-- [ line width=0.25mm, scalar, style=black, arrow size=0.7pt , edge label'={\(\color{black}{}\)}] (a3) ,  
(a4)-- [ line width=0.25mm, scalar, style=black, arrow size=0.7pt] (a2)};
\end{feynman}\end{tikzpicture}\label{fig:z2-decay1}}
\subfloat[]{\begin{tikzpicture}
\begin{feynman}
\vertex (a1){\(\phi_2\)};
\vertex [right=1cm of a1] (a2);
\vertex [above right=1cm and 1cm of a2] (a3);
\vertex [below right=0.5cm and 0.5cm of a2] (a4);
\vertex [right=0.75 cm of a3] (a30){h};
\vertex [right=0.75 cm of a4] (a40){$\phi_1$};
\diagram*{
(a1) -- [ line width=0.25mm,scalar, arrow size=0.7pt, style=blue, edge label={\(\rm\color{black}{}\)}] (a2), 
(a2) -- [ line width=0.25mm,scalar,  bend left, style=gray, edge label={\(\color{black}{\phi_{1}}\)},arrow size=0.7pt](a3),
(a3) -- [ line width=0.25mm,scalar, bend left, style=gray, arrow size=0.7pt, edge label={\(\color{black}{\phi_{1}}\)}] (a2), 
(a3) -- [ line width=0.25mm,scalar, style=black, arrow size=0.7pt , edge label'={\(\color{black}{}\)}] (a30), 
(a40) -- [ line width=0.25mm,scalar,style=black,  arrow size=0.7pt, edge label={\(\rm\color{black}{ }\)}] (a2)};
\end{feynman}\end{tikzpicture}\label{fig:z2-decay20}}
\subfloat[]{\begin{tikzpicture}
\begin{feynman}
\vertex (a1){\(\phi_2\)};
\vertex [right=1cm of a1] (a2);
\vertex [above right=1cm and 1cm of a2] (a3);
\vertex [below right=0.5cm and 0.5cm of a2] (a4);
\vertex [right=0.75 cm of a3] (a30){h};
\vertex [right=0.75 cm of a4] (a40){$\phi_1$};
\diagram*{
(a1) -- [ line width=0.25mm,scalar, arrow size=0.7pt, style=blue, edge label={\(\rm\color{black}{}\)}] (a2), 
(a2) -- [ line width=0.25mm,scalar,  bend left, style=gray, edge label={\(\color{black}{\phi_{2}}\)},arrow size=0.7pt](a3),
(a3) -- [ line width=0.25mm,scalar, bend left, style=gray, arrow size=0.7pt, edge label={\(\color{black}{\phi_{2}}\)}] (a2), 
(a3) -- [ line width=0.25mm,scalar, style=black, arrow size=0.7pt , edge label'={\(\color{black}{}\)}] (a30), 
(a40) -- [ line width=0.25mm,scalar,style=black,  arrow size=0.7pt, edge label={\(\rm\color{black}{ }\)}] (a2)};
\end{feynman}\end{tikzpicture}\label{fig:z2-decay21}}
\subfloat[]{\begin{tikzpicture}
\begin{feynman}
\vertex (a1){\(\phi_2\)};
\vertex [right=2 cm of a1] (a2);
\vertex [right=1 cm of a1] (a20);
\vertex [above=0.75 cm of a20] (a21);
\vertex[above = 0.01 cm of a21] {\( \phi_{1,2} \)};
\vertex [above=0.9 cm of a20] (a22);
\vertex [above right=0.5 cm and 0.75 cm of a2] (a3){h};
\vertex [below right=0.5 cm and 0.75 cm of a2] (a4){$\phi_1$};
\diagram*{
(a1) -- [ line width=0.25mm, scalar, arrow size=0.7pt, style=blue, edge label={\(\rm\color{black}{}\)}] (a20), 
(a2)-- [ line width=0.25mm, scalar, style=black, arrow size=0.7pt , edge label'={\(\color{black}{}\)}] (a3) ,  
(a4)-- [ line width=0.25mm, scalar, style=black, arrow size=0.7pt] (a2) ,
(a20) -- [ line width=0.25mm, scalar,  half left, style=gray] (a21),
(a21) -- [ line width=0.25mm, scalar,  half left, style=gray] (a20),
(a2) -- [ line width=0.25mm, scalar, style=red, edge label={\(\color{black}{\phi_1}\)}] (a20)};
\end{feynman}\end{tikzpicture}\label{fig:z2-decay3}}
\caption{\Cref{fig:z2-decay1,fig:z2-decay20,fig:z2-decay21} and \cref{fig:z2-decay3} are corresponding to the tree and 1-loop level decays of $\phi_2$ to $\phi_1$ plus on-shell Higgs, under the assumption $m_{\phi_2}>m_{\phi_1}$, respectively. However, the Higgs off-shell decay to di-photon or di-gluon would always be there for two non-degenerate DMs.}
\label{fig:z2-decay}
\end{figure}
The presence of the Higgs portal interaction for $\phi_1$ and $\phi_2$, as described in \cref{eq:z2-model}, allows the heavier particle decay to the lighter particle along with a Higgs boson.
The Higgs, whether on-shell or off-shell, can substantially decay into a pair of photons or gluons.
The decay of the heavier particle in a non-degenerate scenario imposes stringent constraints on the associated couplings, implying that the mass hierarchy is not limited to just the Higgs mass.
\Cref{fig:z2-decay} illustrates the decay of the heavier particle into the lighter one plus an on-shell Higgs.
The decay width involving an off-shell Higgs, which decays into light fermions or massless bosons, is significantly suppressed, so we consider only the on-shell Higgs in the total decay width calculation (tree-level + loop correction) to derive stringent limits on the couplings associated with the $\phi_2\to\phi_1 h$ decay.
We don't delve into the details of the complicated loop calculation; instead, we provide an approximate estimate of the coupling required to stabilise the heavier DM particle.
The vertex factor corresponding to the tree-level decay process, shown in \cref{fig:z2-decay1}, is $\lambda_{\phi_1\phi_2 H}v$.
We aim to determine the stringent limit on this coupling such that the decay time of the heavier particle, $\tau_{\phi_2}$, exceeds the age of the universe: $\tau_{\rm univ}=6.4\times 10^{41}\rm GeV^{-1}$. the decay time for the tree-level $\phi_2\to\phi_1 \rm h$ process is
\bea
\tau_{\phi_2}^{-1}=\dfrac{\lambda_{\phi_1\phi_2H}^2v^2}{16\pi m_{\phi_2}}\sqrt{\left[1-\left(\dfrac{m_{\phi_1}+m_{h}}{m_{\phi_2}}\right)^2\right]\left[1-\left(\dfrac{m_{\phi_1}-m_{h}}{m_{\phi_2}}\right)^2\right]}\,.
\label{eq:tree-phi2-decay}
\eea
The tree-level decay shown in \cref{fig:z2-decay1} indicates that a coupling of \(\lambda_{\phi_1\phi_2H} \lesssim 10^{-22}\) is sufficient to ensure that \(\tau_{\phi_2}\) exceeds the age of the universe, \(\tau_{\rm univ}\).
At the next order, a 1-loop decay is also possible. However, we disregard diagrams involving the \(h\phi_1\phi_2\) vertex in 1-loop decay, as these would be suppressed compared to the tree-level decay. The relevant decay diagrams are shown in \cref{fig:z2-decay20,fig:z2-decay21,fig:z2-decay3}.
Stringent limits on the \(\lambda_{112}\) and \(\lambda_{122}\) couplings are necessary to stabilize the heavier particle.
These couplings are expected to be \(\gtrsim \lambda_{\phi_1\phi_2H}\) due to the loop suppression factor of \(1/(16\pi^2)\).
However, a proper loop calculation is required to determine the upper limit accurately.
Ideally, the total decay width of \(\phi_2\) should include contributions from both the tree-level and all relevant 1-loop decays.
However, the presence of UV divergences in the loop diagrams adds complexity to this process, which we have not explored in detail here.
Instead, we focus on finding the appropriate combination of \(m_{\phi_1}\), \(m_{\phi_2}\), \(\lambda_{\phi_1\phi_2H}\), \(\lambda_{112}\), and \(\lambda_{122}\) that stabilises \(\phi_2\), ensuring that \(\tau_{\phi_2} > \tau_{\rm univ}\), while keeping the other parameters within a weak ordering regime.
\begin{table}
\centering
\renewcommand\arraystretch{1.5}
\begin{tabular}{>\ccw{c}>\ccg{c}>\ccw{c}>\ccg{c}>\ccw{c}>\ccg{c}>\ccw{c}>\ccg{c}>\ccw{c}>\ccg{c}>\ccw{c}} \hline \hline
 Decay channel & $\rm  m_{\phi_2}[GeV] $ & $\rm m_{\phi_1} [GeV] $ & $\lambda_{12H}$ & $\lambda_{112}$ & Life Time $\rm [GeV^{-1}] $ \\
\hline
$ \phi_2 \to \phi_1 h $ & 200 & 70 & $ 10^{-21} $ & - & $ 7.77 \times 10^{41} $ \\
$\rm \phi_2 \to \phi_1 ~SM ~SM $ & 200 & 70 & $ 10^{-21} $  & $ 10^{-19} $ & $ 1.30 \times 10^{42} $ \\
$\rm \phi_2 \to \phi_1 ~SM ~SM $ & 200 & 110 & $ 10^{-21} $ &  $ 10^{-19} $ & $ 2.91 \times 10^{47} $ \\
$\rm \phi_2 \to \phi_1 ~All $ & 200 & 70 & $ 10^{-21 } $  &$ 10^{-16} $ &   $ 7.77\times 10^{41} $ \\
$\rm \phi_2 \to \phi_1 ~All $ & 200 & 110 & $ 10^{-21} $  & $ 10^{-16} $ & $ 5.43 \times 10^{42} $ \\
\hline
\end{tabular}
\caption{A few example bound on the parameters to stabilise $ \phi_2 $ DM is also odd under $ \mathcal{Z}_2 $. Other parameters appearing in the decay width are taken as $ \lambda_{1H} =1 $. More importantly, the lifetime of the decaying DM has to be greater than the age of our universe $ \rm 6.4 \times 10^{41}~GeV^{-1}. $}
\label{tab:z2-benchmark}
\end{table}
\subsection{Tree and loop level decay of the heavy DM with $\mathcal{Z}_3$ symmetry}
\label{sec:decay-z3}
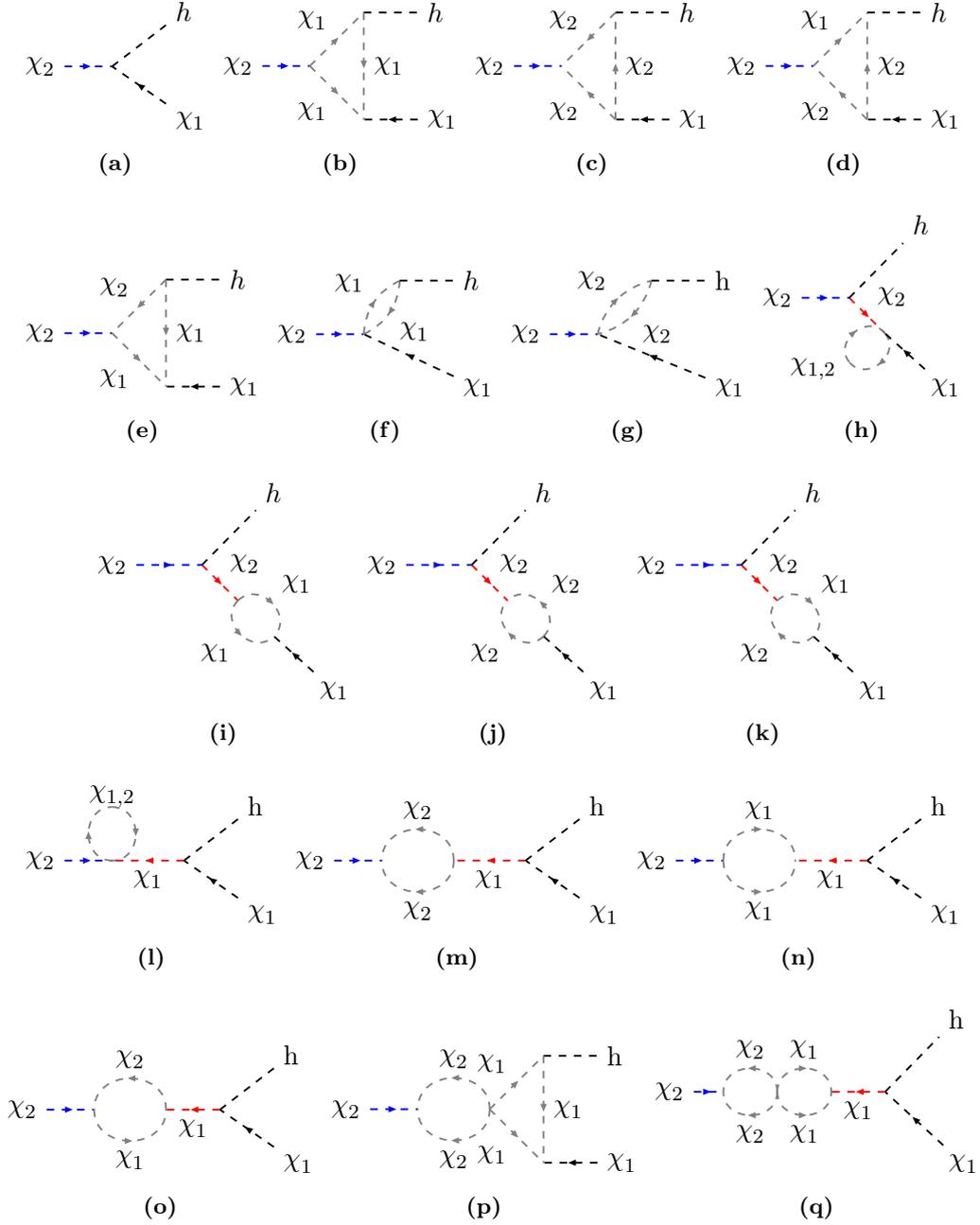
\begin{figure}[htb!]\centering
\subfloat[]{\begin{tikzpicture}
\begin{feynman}
\vertex (a1){\(\chi_2\)};
\vertex [right=1cm of a1] (a2);
\vertex [above right=0.5cm and 0.75cm of a2] (a3){$h$};
\vertex [below right=0.5cm and 0.75cm of a2] (a4){$\chi_1$};
\diagram*{
(a1) -- [ line width=0.25mm,charged scalar, arrow size=0.7pt, style=blue, edge label={\(\rm\color{black}{}\)}] (a2), 
(a2)-- [ line width=0.25mm, scalar, style=black, arrow size=0.7pt , edge label'={\(\color{black}{}\)}] (a3) ,  
(a4)-- [ line width=0.25mm,charged scalar, style=black, arrow size=0.7pt] (a2)};
\end{feynman}
\end{tikzpicture}\label{fig:z3-tree_decay1}}
\subfloat[]{\begin{tikzpicture}
\begin{feynman}
\vertex (a1){\(\chi_2\)};
\vertex [right=1cm of a1] (a2);
\vertex [above right=0.75cm and 0.75cm of a2] (a3);
\vertex [below right=0.75cm and 0.75cm of a2] (a4);
\vertex [ right=0.75cm of a3] (a30){$h$};
\vertex [ right=0.75cm of a4] (a40){$\chi_1$};
\diagram*{
(a1) -- [ line width=0.25mm,charged scalar, arrow size=0.7pt, style=blue, edge label={\(\rm\color{black}{}\)}] (a2), 
(a2) -- [ line width=0.25mm,charged scalar,  style=gray, edge label={\(\color{black}{\chi_1}\)},arrow size=0.7pt](a3),
(a2)  -- [ line width=0.25mm,charged scalar, style=gray, arrow size=0.7pt, edge label'={\(\color{black}{\chi_{1}}\)}] (a4), 
(a3)-- [ line width=0.25mm, scalar, style=black, arrow size=0.7pt , edge label'={\(\color{black}{}\)}] (a30) ,  
(a40)-- [ line width=0.25mm,charged scalar,style=black,  arrow size=0.7pt, edge label={\(\rm\color{black}{ }\)}] (a4), 
(a3)-- [ line width=0.25mm,charged scalar, style=gray, arrow size=0.7pt, edge label={\(\rm\color{black}{\chi_1}\)}] (a4)};
\end{feynman}
\end{tikzpicture}\label{fig:z3-loop_decay1}}
\subfloat[]{\begin{tikzpicture}
\begin{feynman}
\vertex (a1){\(\chi_2\)};
\vertex [right=1cm of a1] (a2);
\vertex [above right=0.75 cm and 0.75 cm of a2] (a3);
\vertex [below right=0.75 cm and 0.75 cm of a2] (a4);
\vertex [ right=0.75 cm of a3] (a30){$h$};
\vertex [ right=0.75 cm of a4] (a40){$\chi_1$};
\diagram*{
(a1) -- [ line width=0.25mm,charged scalar, arrow size=0.7pt, style=blue, edge label={\(\rm\color{black}{}\)}] (a2), 
(a3) -- [ line width=0.25mm,charged scalar,  style=gray, edge label'={\(\color{black}{\chi_2}\)},arrow size=0.7pt](a2),
(a4)  -- [ line width=0.25mm,charged scalar, style=gray, arrow size=0.7pt, edge label={\(\color{black}{\chi_{2}}\)}] (a2), 
(a3) -- [ line width=0.25mm, scalar, style=black, arrow size=0.7pt , edge label'={\(\color{black}{}\)}] (a30),  
(a40) -- [ line width=0.25mm,charged scalar,style=black,  arrow size=0.7pt, edge label={\(\rm\color{black}{ }\)}] (a4), 
(a4) -- [ line width=0.25mm,charged scalar, style=gray, arrow size=0.7pt, edge label'={\(\rm\color{black}{\chi_2}\)}] (a3)};
\end{feynman}
\end{tikzpicture}\label{fig:z3-loop_decay2}}
\subfloat[]{\begin{tikzpicture}
\begin{feynman}
\vertex (a1){\(\chi_2\)};
\vertex [right=1cm of a1] (a2);
\vertex [above right=0.75 cm and 0.75 cm of a2] (a3);
\vertex [below right=0.75 cm and 0.75 cm of a2] (a4);
\vertex [ right=0.75 cm of a3] (a30){$h$};
\vertex [ right=0.75 cm of a4] (a40){$\chi_1$};
\diagram*{
(a1) -- [ line width=0.25mm,charged scalar, arrow size=0.7pt, style=blue, edge label={\(\rm\color{black}{}\)}] (a2), 
(a2) -- [ line width=0.25mm,charged scalar,  style=gray, edge label={\(\color{black}{\chi_1}\)},arrow size=0.7pt](a3),
(a4)  -- [ line width=0.25mm,charged scalar, style=gray, arrow size=0.7pt, edge label={\(\color{black}{\chi_{2}}\)}] (a2), 
(a3) -- [ line width=0.25mm, scalar, style=black, arrow size=0.7pt , edge label'={\(\color{black}{}\)}] (a30),  
(a40) -- [ line width=0.25mm,charged scalar,style=black,  arrow size=0.7pt, edge label={\(\rm\color{black}{ }\)}] (a4), 
(a4) -- [ line width=0.25mm,charged scalar, style=gray, arrow size=0.7pt, edge label'={\(\rm\color{black}{\chi_2}\)}] (a3)};
\end{feynman}
\end{tikzpicture}\label{fig:z3-loop_decay3}}

\subfloat[]{\begin{tikzpicture}
\begin{feynman}
\vertex (a1){\(\chi_2\)};
\vertex [right=1cm of a1] (a2);
\vertex [above right=0.75 cm and 0.75 cm of a2] (a3);
\vertex [below right=0.75 cm and 0.75 cm of a2] (a4);
\vertex [ right=0.75 cm of a3] (a30){$h$};
\vertex [ right=0.75 cm of a4] (a40){$\chi_1$};
\diagram*{
(a1) -- [ line width=0.25mm,charged scalar, arrow size=0.7pt, style=blue, edge label={\(\rm\color{black}{}\)}] (a2), 
(a3) -- [ line width=0.25mm,charged scalar,  style=gray, edge label'={\(\color{black}{\chi_2}\)},arrow size=0.7pt](a2),
(a2)  -- [ line width=0.25mm,charged scalar, style=gray, arrow size=0.7pt, edge label'={\(\color{black}{\chi_{1}}\)}] (a4), 
(a3)-- [ line width=0.25mm, scalar, style=black, arrow size=0.7pt , edge label'={\(\color{black}{}\)}] (a30) ,  
(a40)-- [ line width=0.25mm,charged scalar,style=black,  arrow size=0.7pt, edge label={\(\rm\color{black}{ }\)}] (a4), 
(a3)-- [ line width=0.25mm,charged scalar, style=gray, arrow size=0.7pt, edge label={\(\rm\color{black}{\chi_1}\)}] (a4)};
\end{feynman}
\end{tikzpicture}\label{fig:z3-loop_decay4}}
\subfloat[]{\begin{tikzpicture}
\begin{feynman}
\vertex (a1){\(\chi_2\)};
\vertex [right=1 cm of a1] (a2);
\vertex [above right=0.75 cm and 0.5 cm of a2] (a3);
\vertex [below right=0.75 cm and 0.5 cm of a2] (a4);
\vertex [ right=0.75 cm of a3] (a30){$h$};
\vertex [ right=0.75 cm of a4] (a40){$\chi_1$};
\diagram*{
(a1) -- [ line width=0.25mm,charged scalar, arrow size=0.7pt, style=blue, edge label={\(\rm\color{black}{}\)}] (a2), 
(a2) -- [ line width=0.25mm,charged scalar, bend left,  style=gray, edge label={\(\color{black}{\chi_1}\)},arrow size=0.7pt](a3),
(a3) -- [ line width=0.25mm,charged scalar, bend left, style=gray, arrow size=0.7pt, edge label={\(\color{black}{\chi_{1}}\)}] (a2), 
(a3) -- [ line width=0.25mm, scalar, style=black, arrow size=0.7pt , edge label'={\(\color{black}{}\)}] (a30),  
(a40) -- [ line width=0.25mm,charged scalar,style=black,  arrow size=0.7pt, edge label={\(\rm\color{black}{ }\)}] (a2)};
\end{feynman}
\end{tikzpicture}\label{fig:z3-loop_decay5}}
\subfloat[]{\begin{tikzpicture}
\begin{feynman}
\vertex (a1){\(\chi_2\)};
\vertex [right=1 cm of a1] (a2);
\vertex [above right=0.75 cm and 0.75 cm of a2] (a3);
\vertex [below right=0.75 cm and 0.75 cm of a2] (a4);
\vertex [ right=0.75 cm of a3] (a30){h};
\vertex [ right=0.75 cm of a4] (a40){$\chi_1$};
\diagram*{
(a1) -- [ line width=0.25mm,charged scalar, arrow size=0.7pt, style=blue, edge label={\(\rm\color{black}{}\)}] (a2), 
(a2) -- [ line width=0.25mm,charged scalar, bend left, style=gray, edge label={\(\color{black}{\chi_2}\)},arrow size=0.7pt](a3),
(a3)  -- [ line width=0.25mm,charged scalar,bend left, style=gray, arrow size=0.7pt, edge label={\(\color{black}{\chi_{2}}\)}] (a2), 
(a3) -- [ line width=0.25mm, scalar, style=black, arrow size=0.7pt , edge label'={\(\color{black}{}\)}] (a30) ,  
(a40) -- [ line width=0.25mm,charged scalar,style=black,  arrow size=0.7pt, edge label={\(\rm\color{black}{ }\)}] (a2)};
\end{feynman}
\end{tikzpicture}\label{fig:z3-loop_decay6}}
\subfloat[]{\begin{tikzpicture}
\begin{feynman}
\vertex (a2){\(\chi_2\)};
\vertex [right=1 cm of a1] (a3);
\vertex [below right= 0.5 cm and 0.5 cm of a3] (a31);
\vertex [below= 0.9 cm of a3] (a32);
\vertex [below right= 1.0cm and -0.5 cm of a3] (a30);
\vertex [below right= 1 cm and 1 cm of a3] (a33){$\chi_1$};
\vertex [above right= 0.75 cm and 0.75 cm of a3] (a34){$h$};
\diagram*{
(a2)-- [ line width=0.25mm,charged scalar, style=blue, arrow size=0.7pt , edge label'={\(\color{black}{}\)}] (a3),  
(a3)-- [ line width=0.25mm, charged scalar, style=red, arrow size=0.7pt, edge label={\(\color{black}{\chi_2}\)}] (a31),
(a31) -- [ line width=0.25mm, charged scalar,arrow size=0.7pt ,  half left, style=gray] (a32),
(a32) -- [ line width=0.25mm, charged scalar, arrow size=0.7pt , half left, style=gray] (a31),
(a33) -- [ line width=0.25mm,charged  scalar, style=black, arrow size=0.7pt , edge label={\(\color{black}{}\)}] (a31),
(a3) -- [ line width=0.25mm,  scalar, style=black, arrow size=0.7pt , edge label'={\(\color{black}{}\)}] (a34)};
\node at (a30)[]{\(\chi_{1,2}\)};
\end{feynman}
\end{tikzpicture}\label{fig:z3-loop_decay7}}

\subfloat[]{\begin{tikzpicture}
\begin{feynman}
\vertex (a2){\(\chi_2\)};
\vertex [right=1.25 cm of a1] (a3);
\vertex [below right = 0.5 cm and 0.5 cm of a3] (a31);
\vertex [below right= 1 cm and 1 cm of a3] (a32);
\vertex [below right = 1.5 cm and 1.5 cm of a3] (a33){$\chi_1$};
\vertex [below right = 1.0 cm and -0.5 cm of a3] (a30);
\vertex [above right = 0.75 cm and 0.75 cm of a3] (a34){$h$};
\diagram*{
(a2)-- [ line width=0.25mm,charged scalar, style=blue, arrow size=0.7pt , edge label'={\(\color{black}{}\)}] (a3),  
(a3)-- [ line width=0.25mm,charged scalar, style=red, arrow size=0.7pt, edge label={\(\color{black}{\chi_2}\)}] (a31),
(a31) -- [ line width=0.25mm,charged scalar,arrow size=0.7pt, half left, style=gray, edge label={\(\color{black}{\chi_{1}}\)}] (a32),
(a31) -- [ line width=0.25mm,charged scalar, arrow size=0.7pt, half right, style=gray, edge label'={\(\color{black}{\chi_{1}}\)}] (a32),
(a33) -- [ line width=0.25mm,charged  scalar, style=black, arrow size=0.7pt , edge label={\(\color{black}{}\)}] (a32),
(a3) -- [ line width=0.25mm,scalar, style=black, arrow size=0.7pt , edge label'={\(\color{black}{}\)}] (a34)};
\end{feynman}
\end{tikzpicture}\label{fig:z3-loop_decay8}}
\subfloat[]{\begin{tikzpicture}
\begin{feynman}
\vertex (a2){\(\chi_2\)};
\vertex [right=1.25 cm of a1] (a3);
\vertex [below right = 0.5 cm and 0.5 cm of a3] (a31);
\vertex [below right= 1 cm and 1 cm of a3] (a32);
\vertex [below right = 1.5 cm and 1.5 cm of a3] (a33){$\chi_1$};
\vertex [below right = 1.0 cm and -0.5 cm of a3] (a30);
\vertex [above right = 0.75 cm and 0.75 cm of a3] (a34){$h$};
\diagram*{
(a2)-- [ line width=0.25mm,charged scalar, style=blue, arrow size=0.7pt , edge label'={\(\color{black}{}\)}] (a3),  
(a3)-- [ line width=0.25mm,charged scalar, style=red, arrow size=0.7pt, edge label={\(\color{black}{\chi_2}\)}] (a31),
(a32) -- [ line width=0.25mm,charged scalar,arrow size=0.7pt, half left, style=gray, edge label={\(\color{black}{\chi_{2}}\)}] (a31),
(a32) -- [ line width=0.25mm,charged scalar, arrow size=0.7pt, half right, style=gray, edge label'={\(\color{black}{\chi_{2}}\)}] (a31),
(a33) -- [ line width=0.25mm,charged  scalar, style=black, arrow size=0.7pt , edge label={\(\color{black}{}\)}] (a32),
(a3) -- [ line width=0.25mm,scalar, style=black, arrow size=0.7pt , edge label'={\(\color{black}{}\)}] (a34)};
\end{feynman}
\end{tikzpicture}\label{fig:z3-loop_decay9}}
\subfloat[]{\begin{tikzpicture}
\begin{feynman}
\vertex (a2){\(\chi_2\)};
\vertex [right=1.25 cm of a1] (a3);
\vertex [below right = 0.5 cm and 0.5 cm of a3] (a31);
\vertex [below right= 1 cm and 1 cm of a3] (a32);
\vertex [below right = 1.5 cm and 1.5 cm of a3] (a33){$\chi_1$};
\vertex [below right = 1.0 cm and -0.5 cm of a3] (a30);
\vertex [above right = 0.75 cm and 0.75 cm of a3] (a34){$h$};
\diagram*{
(a2)-- [ line width=0.25mm,charged scalar, style=blue, arrow size=0.7pt , edge label'={\(\color{black}{}\)}] (a3),  
(a3)-- [ line width=0.25mm,charged scalar, style=red, arrow size=0.7pt, edge label={\(\color{black}{\chi_2}\)}] (a31),
(a31) -- [ line width=0.25mm,charged scalar,arrow size=0.7pt, half left, style=gray, edge label={\(\color{black}{\chi_{1}}\)}] (a32),
(a32) -- [ line width=0.25mm,charged scalar, arrow size=0.7pt, half left, style=gray, edge label={\(\color{black}{\chi_{2}}\)}] (a31),
(a33) -- [ line width=0.25mm,charged  scalar, style=black, arrow size=0.7pt , edge label={\(\color{black}{}\)}] (a32),
(a3) -- [ line width=0.25mm,scalar, style=black, arrow size=0.7pt , edge label'={\(\color{black}{}\)}] (a34)};
\end{feynman}
\end{tikzpicture}\label{fig:z3-loop_decay10}}

\subfloat[]{\begin{tikzpicture}
\begin{feynman}
\vertex (a1){\(\chi_2\)};
\vertex [right=2 cm of a1] (a2);
\vertex [right=1 cm of a1] (a20);
\vertex [above=0.75 cm of a20] (a21);
\vertex [above=0.9 cm of a20] (a22);
\vertex [above right=0.5 cm and 0.75 cm of a2] (a3){h};
\vertex [below right=0.5 cm and 0.75 cm of a2] (a4){$\chi_1$};
\diagram*{
(a1) -- [ line width=0.25mm, charged scalar, arrow size=0.7pt, style=blue, edge label={\(\rm\color{black}{}\)}] (a20), 
(a2)-- [ line width=0.25mm, scalar, style=black, arrow size=0.7pt , edge label'={\(\color{black}{}\)}] (a3) ,  
(a4)-- [ line width=0.25mm, charged scalar, style=black, arrow size=0.7pt] (a2) ,
(a20) -- [ line width=0.25mm, charged scalar,arrow size=0.7pt ,  half left, style=gray] (a21),
(a21) -- [ line width=0.25mm, charged scalar, arrow size=0.7pt , half left, style=gray] (a20),
(a2) -- [ line width=0.25mm, charged scalar, style=red, arrow size=0.7pt , edge label={\(\color{black}{\chi_1}\)}] (a20)};
\node at (a22)[]{\(\chi_{1,2}\)};
\end{feynman}
\end{tikzpicture}\label{fig:z3-loop_decay11}}
\subfloat[]{\begin{tikzpicture}
\begin{feynman}
\vertex (a1){\(\chi_2\)};
\vertex [right=1 cm of a1] (a20);
\vertex [right=2 cm of a1] (a21);
\vertex [right=3 cm of a1] (a2);
\vertex [above right=0.5 cm and 0.75 cm of a2] (a3){h};
\vertex [below right=0.5 cm and 0.75 cm of a2] (a4){$\chi_1$};
\diagram*{
(a1) -- [ line width=0.25mm, charged scalar, arrow size=0.7pt, style=blue, edge label={\(\rm\color{black}{}\)}] (a20), 
(a2)-- [ line width=0.25mm, scalar, style=black, arrow size=0.7pt , edge label'={\(\color{black}{}\)}] (a3) ,  
(a4)-- [ line width=0.25mm, charged scalar, style=black, arrow size=0.7pt] (a2) ,
(a21) -- [ line width=0.25mm, charged scalar,arrow size=0.7pt , edge label'={\(\rm\color{black}{\chi_2}\)},  half right, style=gray] (a20),
(a21) -- [ line width=0.25mm, charged scalar, arrow size=0.7pt , edge label={\(\rm\color{black}{\chi_2}\)}, half left, style=gray] (a20),
(a2) -- [ line width=0.25mm, charged scalar, style=red, arrow size=0.7pt , edge label={\(\color{black}{\chi_1}\)}] (a21)};
\end{feynman}
\end{tikzpicture}\label{fig:z3-loop_decay12}}
\subfloat[]{\begin{tikzpicture}
\begin{feynman}
\vertex (a1){\(\chi_2\)};
\vertex [right=1 cm of a1] (a20);
\vertex [right=2 cm of a1] (a21);
\vertex [right=3 cm of a1] (a2);
\vertex [above right=0.5 cm and 0.75 cm of a2] (a3){h};
\vertex [below right=0.5 cm and 0.75 cm of a2] (a4){$\chi_1$};
\diagram*{
(a1) -- [ line width=0.25mm, charged scalar, arrow size=0.7pt, style=blue, edge label={\(\rm\color{black}{}\)}] (a20), 
(a2)-- [ line width=0.25mm, scalar, style=black, arrow size=0.7pt , edge label'={\(\color{black}{}\)}] (a3) ,  
(a4)-- [ line width=0.25mm, charged scalar, style=black, arrow size=0.7pt] (a2) ,
(a20) -- [ line width=0.25mm, charged scalar,arrow size=0.7pt , edge label'={\(\rm\color{black}{\chi_1}\)},  half right, style=gray] (a21),
(a20) -- [ line width=0.25mm, charged scalar, arrow size=0.7pt , edge label={\(\rm\color{black}{\chi_1}\)}, half left, style=gray] (a21),
(a2) -- [ line width=0.25mm, charged scalar, style=red, arrow size=0.7pt , edge label={\(\color{black}{\chi_1}\)}] (a21)};
\end{feynman}
\end{tikzpicture}\label{fig:z3-loop_decay13}}

\subfloat[]{\begin{tikzpicture}
\begin{feynman}
\vertex (a1){\(\chi_2\)};
\vertex [right=1 cm of a1] (a20);
\vertex [right=2 cm of a1] (a21);
\vertex [right=2.75 cm of a1] (a2);
\vertex [above right=0.5 cm and 0.75 cm of a2] (a3){h};
\vertex [below right=0.5 cm and 0.75 cm of a2] (a4){$\chi_1$};
\diagram*{
(a1) -- [ line width=0.25mm, charged scalar, arrow size=0.7pt, style=blue, edge label={\(\rm\color{black}{}\)}] (a20), 
(a2)-- [ line width=0.25mm, scalar, style=black, arrow size=0.7pt , edge label'={\(\color{black}{}\)}] (a3) ,  
(a4)-- [ line width=0.25mm, charged scalar, style=black, arrow size=0.7pt] (a2) ,
(a20) -- [ line width=0.25mm, charged scalar,arrow size=0.7pt , edge label'={\(\rm\color{black}{\chi_1}\)},  half right, style=gray] (a21),
(a21) -- [ line width=0.25mm, charged scalar, arrow size=0.7pt , edge label'={\(\rm\color{black}{\chi_2}\)}, half right, style=gray] (a20),
(a2) -- [ line width=0.25mm, charged scalar, style=red, arrow size=0.7pt , edge label={\(\color{black}{\chi_1}\)}] (a21)};
\end{feynman}
\end{tikzpicture}\label{fig:z3-loop_decay14}}
\subfloat[]{\begin{tikzpicture}
\begin{feynman}
\vertex (a1){\(\chi_2\)};
\vertex [right=1 cm of a1] (b);
\vertex [right=2 cm of a1] (a2);
\vertex [above right=0.75 cm and 0.75 cm of a2] (a3);
\vertex [below right=0.75 cm and 0.75 cm of a2] (a4);
\vertex [ right=0.75 cm of a3] (a30){h};
\vertex [ right=0.75 cm of a4] (a40){$\chi_1$};
\diagram*{
(a1) -- [ line width=0.25mm,charged scalar, arrow size=0.7pt, style=blue, edge label={\(\rm\color{black}{}\)}] (b), 
(a2) -- [ line width=0.25mm,charged scalar,  style=gray, edge label={\(\color{black}{\chi_1}\)},arrow size=0.7pt](a3),
(a2)  -- [ line width=0.25mm,charged scalar, style=gray, arrow size=0.7pt, edge label'={\(\color{black}{\chi_{1}}\)}] (a4), 
(a3) -- [ line width=0.25mm,scalar, style=black, arrow size=0.7pt , edge label'={\(\color{black}{}\)}] (a30),  
(a40) -- [ line width=0.25mm,charged scalar,style=black,  arrow size=0.7pt, edge label={\(\rm\color{black}{ }\)}] (a4),
(a3) -- [ line width=0.25mm,charged scalar, style=gray, arrow size=0.7pt, edge label={\(\rm\color{black}{\chi_1}\)}] (a4),
(a2) -- [ line width=0.25mm,charged scalar,arrow size=0.7pt , edge label'={\(\rm\color{black}{\chi_2}\)},  half right, style=gray] (b),
(a2) -- [ line width=0.25mm,charged scalar, arrow size=0.7pt , edge label={\(\rm\color{black}{\chi_2}\)}, half left, style=gray] (b)};
\end{feynman}
\end{tikzpicture}\label{fig:z3-2loop_decay1}}
\subfloat[]{\begin{tikzpicture}
\begin{feynman}
\vertex (a){\(\chi_2\)};
\vertex [right=0.75 cm of a1] (b);
\vertex [right=1.5 cm of a1] (c);
\vertex [right=2.25 cm of a1] (d);
\vertex [right=0.75 cm of d] (d0);
\vertex [above right=0.75 cm and 0.75 cm of d0] (d1){h};
\vertex [below right=0.75 cm and 0.75 cm of d0] (d2){$\chi_1$};
\diagram*{
(a) -- [ line width=0.25mm,charged scalar, arrow size=0.7pt, style=blue, edge label={\(\rm\color{black}{}\)}] (b), 
(c) -- [ line width=0.25mm,charged scalar,arrow size=0.7pt , edge label'={\(\rm\color{black}{\chi_2}\)},  half right, style=gray] (b),
(c) -- [ line width=0.25mm,charged scalar, arrow size=0.7pt , edge label={\(\rm\color{black}{\chi_2}\)}, half left, style=gray] (b),
(c) -- [ line width=0.25mm,charged scalar,arrow size=0.7pt , edge label'={\(\rm\color{black}{\chi_1}\)},  half right, style=gray] (d),
(c) -- [ line width=0.25mm,charged scalar, arrow size=0.7pt , edge label={\(\rm\color{black}{\chi_1}\)}, half left, style=gray] (d),
(d0) -- [ line width=0.25mm, scalar,  style=black, arrow size=0.7pt](d1),
(d2)  -- [ line width=0.25mm,charged scalar, style=black, arrow size=0.7pt] (d0),
(d0) -- [ line width=0.25mm,charged scalar, style=red, edge label={\(\rm\color{black}{\chi_1}\)}, arrow size=0.7pt] (d)};
\end{feynman}
\end{tikzpicture}\label{fig:z3-2loop_decay2}}
\caption{\Cref{fig:z3-tree_decay1}, is the Feynman diagram corresponding to the tree level decay: $\chi_2\to\chi_1 ~h$. 
After ignoring the diagrams involves $h\chi_1\chi_2$ vertex, the remaining Feynman diagrams: \cref{fig:z3-loop_decay1,fig:z3-loop_decay2,fig:z3-loop_decay3,fig:z3-loop_decay4,fig:z3-loop_decay5,fig:z3-loop_decay6,fig:z3-loop_decay7,fig:z3-loop_decay8,fig:z3-loop_decay9,fig:z3-loop_decay10,fig:z3-loop_decay11,fig:z3-loop_decay12,fig:z3-loop_decay13,fig:z3-loop_decay14} represented the 1-loop mediated decay. After the appropriate choice, see \cref{tab:z3}, of sacrificing coupling associated with the 1-loop decay, and there are still possible 2-loop decay processes, see \cref{fig:z3-2loop_decay1,fig:z3-2loop_decay2}.}
\label{fig:z3_decay}
\end{figure}
Similar to the case with \(\mathcal{Z}_2\) symmetry, there are decay channels for the heavier DM particle under \(\mathcal{Z}_3\) symmetry as well.
The interactions between the DM particles and visible sectors, as shown in eq\,.~\eqref{eq:model-potentialz3}, give rise to tree-level, 1-loop, and 2-loop decay processes.
Let's analyse these decay processes step by step, assuming \(m_{\chi_2} > m_{\chi_1}\).
However, the decay constraints are equally applicable to the opposite mass hierarchy.

\begin{itemize}
\item The tree-level decay process \(\chi_2 \to \chi_1^* + h\) (see \cref{fig:z3-tree_decay1}) is mediated by the coupling \(\lambda_{12H}\). According to \cref{eq:tree-phi2-decay}, to ensure \(\tau_{\chi_2} > \tau_{\rm univ}\) in the presence of this tree-level decay, we require \(\lambda_{12H} \lesssim 4 \times 10^{-22}\).
\item If we ignore the diagrams involving the \(\lambda_{12H}\) coupling, we obtain 1-loop decay diagrams, as shown in \cref{fig:z3-loop_decay1,fig:z3-loop_decay2,fig:z3-loop_decay3,fig:z3-loop_decay4,fig:z3-loop_decay5,fig:z3-loop_decay6,fig:z3-loop_decay7,fig:z3-loop_decay8,fig:z3-loop_decay9,fig:z3-loop_decay10,fig:z3-loop_decay11,fig:z3-loop_decay12,fig:z3-loop_decay13,fig:z3-loop_decay14}.
The associated interaction terms and their corresponding couplings, whose small values contribute to the stability of \(\chi_2\), present various scenarios, as summarised in \cref{tab:z3}.
\item In specific cases, there may also be 2-loop decay processes, such as those depicted in \cref{fig:z3-2loop_decay1,fig:z3-2loop_decay2}. To stabilise \(\chi_2\), the couplings associated with these decays must be minimised, although the constraints will be less stringent than the 1-loop limits due to the suppression factor \((16\pi^2)^{-2}\).
\end{itemize}
Although we have imposed constraints on the couplings associated with each decay process, the total decay width is the sum of contributions from tree-level, 1-loop, and 2-loop processes. Calculating the total decay width is a highly complicated task, and while the precise limits on the couplings may shift slightly with a more accurate computation, the overall phenomenology remains unchanged.
\begin{table}[htb!]
\centering	
\renewcommand\arraystretch{1.5}
\begin{tabular}{>\ccw{c}>\ccg{c}>\ccw{c}>\ccg{c}>\ccw{c}>\ccg{c}>\ccw{c}>\ccg{c}>\ccw{c}>\ccg{c}>\ccw{c}} \hline\hline
Decay channel  &$ m_{\chi_2} [\rm GeV] $ & $ m_{\chi_1}[\rm GeV] $ & $ \mu_{2} [\rm GeV] $&$\lambda_{12H}$ &  $\lambda_{3}$ & Life Time $\rm[GeV^{-1}] $\\\hline
$ \chi_2 \to \chi_1 {\rm~ h}$ & 200 & 70 & $ 10^{-17} $& $ 10^{-21} $& $10^{-19}$  & $ 1.70 \times 10^{42} $ \\
$ \chi_2 \to \chi_1 {\rm~ h}$ & 200 & 70 & $ 10^{-19} $ & $ 10^{-21} $& $10^{-19}$  & $ 1.58 \times 10^{42} $ \\
$ \chi_2 \to \chi_1 {\rm~ h}$ & 200 & 50 & $ 10^{-17} $ & $ 10^{-21} $& $10^{-19}$  & $ 6.40 \times 10^{42} $ \\
$ {\rm \chi_2 \to\chi_1 ~SM~SM} $ & 200 & 70 & $ 10^{-15} $ & $ 10^{-21} $ & $ 10^{-15} $ & $ 6.50 \times 10^{42} $ \\
$ {\rm \chi_2 \to\chi_1 ~SM~SM} $ & 200 & 110 & $ 10^{-15} $ & $ 10^{-21} $ & $ 10^{-15} $ & $ 1.74 \times 10^{48} $ \\
$ {\rm \chi_2 \to\chi_1 ~All} $ & 200 & 110 & $ 10^{-15} $ & $ 10^{-21} $ & $ 10^{-15} $ & $ 1.74 \times 10^{48} $ \\
$ {\rm \chi_2 \to\chi_1~ ~All} $ & 200 & 70 & $ 10^{-15} $ & $ 10^{-21} $ & $ 10^{-15} $ & $ 6.50 \times 10^{42} $ \\\hline\hline
\end{tabular}
\caption{The bound on particular parameters to have the lifetime of the $ \mathcal{Z}_3 $ charged DM greater than the age of the universe $6.4\times 10^{41}\rm GeV^{-1}$. The other parameter values that will appear in the decay width of the DM  are taken as $\mu_1=\mu_3=\mu_4=2m_{\chi_1}$, $\lambda_{1H}=\lambda_{12}=\lambda_{12}^{\prime}=\lambda_4=1$.}
\label{tab:z3-benchmark}
\end{table}
\begin{table}[htb!]
\centering
\renewcommand\arraystretch{1.5}
\begin{tabular}{c c}
\begin{tabular}{>\ccw{c}>\ccg{c}}
\hline\hline
Diagram & contributing operators \\\hline
\multicolumn{2}{c}{\cellcolor{cyan!10} tree-level}\\\hline
Fig.~\ref{fig:z3-tree_decay1} & $\chi_1 \chi_2 H^{\dagger} H $ \\\hline
\multicolumn{2}{c}{\cellcolor{cyan!10} one-loop}\\\hline
Fig.~\ref{fig:z3-loop_decay1} & $\abs{\chi_1}^2 \abs{H}^2$,~ {\color{blue}$\chi_1^2 \chi_2^*$},~{\color{magenta}$\chi_1^3$} \\\hline
Fig.~\ref{fig:z3-loop_decay2} & $\abs{\chi_2}^2 \abs{H}^2$,~{\color{orange}$\chi_1^* \chi_2^2 $},~ {\color{violet}$\chi_2^3$} \\\hline
Fig.~\ref{fig:z3-loop_decay3} & $\chi_1 \chi_2 H^{\dagger} H $,~ {\color{orange}$\chi_1^* \chi_2^2 $},~{\color{orange}$\chi_1 \chi_2^{*2} $}\\\hline
Fig.~\ref{fig:z3-loop_decay4} & $\chi_1 \chi_2 H^{\dagger} H $,~{\color{orange}$\chi_1^* \chi_2^2$},~{\color{magenta}$\chi_1^3$}\\\hline
Fig.~\ref{fig:z3-loop_decay5} & $\abs{\chi_1}^2 \abs{H}^2 $,~ {\color{cyan}$\abs{\chi_1}^2 \chi_1 \chi_2 $} \\\hline
Fig.~\ref{fig:z3-loop_decay6} & $\abs{\chi_2}^2 \abs{H}^2 $,~ {\color{teal}$\chi_1 \chi _2 \abs{\chi_2}^2 $} \\\hline
Fig.~\ref{fig:z3-loop_decay7} & $\abs{\chi_2}^2 \abs{H}^2 $,~ {\color{cyan}$\abs{\chi_1}^2 \chi_1 \chi_2$}/ {\color{teal}$\chi_1 \chi_2 \abs{\chi_2}^2$} \\\hline
\end{tabular}
&
\begin{tabular}{>\ccw{c}>\ccg{c}}
\hline\hline
Diagram & contributing operators \\\hline
\multicolumn{2}{c}{\cellcolor{cyan!10} one-loop (contd.)}\\\hline
Fig.~\ref{fig:z3-loop_decay8} & $\abs{\chi_2}^2 \abs{H}^2 $,~ {\color{blue}$\chi_1^2 \chi_2^*$},~ {\color{magenta}$ \chi_1^3$}\\\hline
Fig.~\ref{fig:z3-loop_decay9} & $\abs{\chi_2}^2 \abs{H}^2 $,~ {\color{orange}$\chi_1^* \chi_2^2 $},~{\color{violet}$\chi_2^3$}\\\hline
Fig.~\ref{fig:z3-loop_decay10} & $\abs{\chi_2}^2 \abs{H}^2 $,~{\color{orange}$\chi_1^*\chi_2^2$},~ {\color{blue}$\chi_1^2 \chi_2^*$}\\\hline
Fig.~\ref{fig:z3-loop_decay11} & $\abs{\chi_{1}}^2 \abs{H}^2 $,~{\color{teal}$\chi_1 \chi_2 \abs{\chi_2}^2$}/{\color{cyan}$\abs{\chi_1}^2 \chi_1 \chi_2 $} \\\hline
Fig.~\ref{fig:z3-loop_decay12} & $\abs{\chi_{1}}^2 \abs{H}^2 $,~ {\color{orange}$\chi_1^* \chi_2^2 $},~{\color{violet}$\chi_2^3 $} \\\hline
Fig.~\ref{fig:z3-loop_decay13} & $\abs{\chi_{1}}^2 \abs{H}^2 $,~{\color{blue}$\chi_1^2 \chi_2^* $},~ {\color{magenta} $\chi_1^3$} \\\hline
Fig.~\ref{fig:z3-loop_decay14} & $\abs{\chi_{1}}^2 \abs{H}^2 $,~{\color{orange}$\chi_1^* \chi_2^2 $},~ {\color{blue}$\chi_1^2 \chi_2^*$} \\\hline
\multicolumn{2}{c}{\cellcolor{cyan!10} two-loop}\\\hline
Fig.~\ref{fig:z3-2loop_decay1} & $\abs{\chi_1}^2 \abs{H}^2 $,~ {\color{green}$\chi_1^2 \chi_2^2$},~{\color{magenta}$\chi_1^3 $},~{\color{violet}$\chi_2^3$} \\\hline
Fig.~\ref{fig:z3-2loop_decay2} & $\abs{\chi_1^2} \abs{H}^2 $,~{\color{green}$\chi_1^2 \chi_2^2$},~{\color{magenta}$\chi_1^3$},~{\color{violet}$\chi_2^3$} \\\hline
\end{tabular}
\end{tabular}
\caption{Diagram-wise operators contributing to the process of the decay of the LLP dark matter component for the mass hierarchy $m_{\chi_2} > m_{\chi_1}$. The corresponding colored interactions are sacrificed in different scenarios corresponding to the table \ref{tab:z3}.}
\label{tab:z3_scenario_wise_couplings}
\end{table}

In tables~\ref{tab:z3} and \ref{tab:z3_q1_neq_q2}, we present six different scenarios arising from a $\mathcal{Z}_3$ symmetry, categorized based on the choices of $q_1$ and $q_2$. Scenarios A, B, and C correspond to the case where $q_1 \neq q_2$ and $q_1 + q_2 = N (=3)$, while scenarios D, E, and F represent the case where $q_1 = q_2$ but $q_1 + q_2 \neq N$. In each scenario, certain operators or interaction strengths must be suppressed to obtain a stable dark matter candidate, one being absolutely stable, and the other is an LLP. The corresponding decay diagrams at tree-level, one-loop and two-loop level, for the mass hierarchy $m_{\chi_2} > m_{\chi_1}$, are shown in fig\,.~\ref{fig:z3_decay}. For the opposite hierarchy, $m_{\chi_1} > m_{\chi_2}$, similar/inverted diagrams contribute to the decay of $\chi_1$, and the required operator suppression would remain the same.

Table \ref{tab:z3_scenario_wise_couplings} lists all relevant couplings associated with the decay diagrams in fig\,.~\ref{fig:z3_decay}, organized by loop order. We have highlighted, using color, the specific operators sacrificed in each scenario. The tree-level diagrams vanish if the coupling $\chi_1 \chi_2 H^\dagger H$ is negligible. Suppressing this operator allows us to further reduce one-loop contributions by taking couplings like $\chi_1^* \chi_2^2$, $\chi_1^2 \chi_2^*$, $\chi_1^3$, and $\chi_2^3$ to be small. Additionally, making the $\chi_1^2 \chi_2^2$ coupling small suppresses the two-loop decay channels, which helps ensure that $\chi_2$ behaves as an LLP. Once the two-loop decay is sufficiently suppressed, higher-loop decays (three-loop and beyond) are automatically negligible, thereby protecting the LLP's longevity.

\section{Unitarity Bounds on the model parameters }
\label{appndx:unitarity_bounds}
In order to study the tree-level unitarity bounds on the quartic couplings of our model at high energies, we consider the generic structure of $2 \to 2$ scattering amplitudes. These amplitudes can be expanded in terms of Legendre polynomials as~\cite{Lee:1977eg}  
\begin{equation}
\mathcal{M}^{2 \to 2} = 16\pi \sum_{l=0}^\infty a_l (2l+1) P_l(\cos\theta),
\label{eq:A1}
\end{equation}
where $\theta$ is the scattering angle and $P_l(\cos\theta)$ denotes the Legendre polynomial of order $l$.  
In the high-energy limit, the dominant contribution comes from the s-wave ($l=0$), with the partial amplitude $a_0$ governing the leading behaviour of the process.  
The unitarity condition then requires
\begin{equation}
|\text{Re}(a_0)| \leq \tfrac{1}{2}.
\label{eq:A2}
\end{equation}
The above constraint can be equivalently expressed as a bound on the full scattering amplitude $\mathcal{M}$~\cite{Kanemura:2015ska,DeCurtis:2016scv,Goodsell:2018tti}:
\begin{equation}
|\mathcal{M}| \leq 8\pi.
\label{eq:A3}
\end{equation}
When multiple $2 \to 2$ processes are possible, one needs to construct an S-scattering matrix $\mathcal{M}^{2 \to 2}_{i,j} = \mathcal{M}_{i \to j}$ by taking into account all possible two-particle states. 
In the high-energy limit, the SM Higgs doublet is expressed as $ H = \left( G^+ \ \ \dfrac{h + iG^0}{\sqrt{2}} \right)^T, $ and the interaction Lagrangian in eqs\,.\eqref{eq:z3-model} and \eqref{eq:model-potentialz3} gives rise to $22$ two-particle neutral states:
\begin{align}
\begin{split}
\bigg\{\dfrac{h\, h}{\sqrt{2}}\,,\ \ 
\dfrac{G^0\,G^0}{\sqrt{2}}\,,\ \
G^+\,G^-\,,\ \ h\,G^0\,,\qquad\qquad\qquad\\
h\, \chi_i\,, \ \ h\, \chi_i^*\,, \ \ G^0\, \chi_i\,, \ \ G^0\, \chi_i^*\,,\qquad\qquad\quad\quad\\ 
\dfrac{\chi_i \, \chi_i}{\sqrt{2}}\,, \ \  \dfrac{\chi_i^* \, \chi_i^*}{\sqrt{2}}\,, 
\chi_i \, \chi_i^*\,, \ \ 
\chi_1 \, \chi_2\,, \ \ \chi_1^* \, \chi_2^*\,, \ \ 
\chi_1\, \chi_2^*\,,\ \ \chi_1^*\, \chi_2\,\bigg\};
\end{split}
\end{align}
where $i=1,~2$. The single charge six two-particle states:
\begin{equation}
\sum_{i=1,2} \bigg\{ h\, G^{\pm},\ \ 
G^0\,G^{\pm},\ \
\chi_i\,G^{\pm}\,, \ \ \chi_i^*\,G^{\pm}\,\bigg\}
\end{equation}
and only one double charge two-particle state is,
\begin{equation}
\dfrac{G^{\pm}\, G^{\pm}}{\sqrt{2}}\,.
\end{equation}
Therefore, the scattering amplitude matrix $(\mathcal{M})$ can be expressed in block-diagonal form by decomposing it into the neutral (NC), singly charged (SC), and double charged states as \cite{Bento:2022vsb},
\bea
\mathcal{M} = \begin{pmatrix}
\mathcal{M}^{\rm NC}_{22\times 22} & 0 & 0 \\
0 & \mathcal{M}^{\rm SC}_{6\times 6} & 0 \\
0 & 0 & \mathcal{M}^{\rm DC}_{1\times 1}
\end{pmatrix}\,,
\label{eq:unitary}
\eea
with the submatrices given by

\bea
\mathcal{M}^{\rm NC}_{22\times 22}=
\begin{pmatrix}
\mathcal{M}^{\rm HH}_{\rm HH^{}_{4\times 4}}& \mathcal{M}^{\rm HH}_{\rm H\chi^{}_{4\times 8}}&\mathcal{M}^{\rm HH}_{\rm \chi\chi^{}_{4\times 10}}\\
\mathcal{M}^{\rm H\chi}_{\rm HH^{}_{8\times 4}}& \mathcal{M}^{\rm H\chi}_{\rm H\chi^{}_{8\times 8}}&\mathcal{M}^{\rm H\chi}_{\rm \chi\chi^{}_{8\times 10}}\\
\mathcal{M}^{\chi\chi}_{\rm HH^{}_{10\times 4}}& \mathcal{M}^{\chi\chi}_{\rm H\chi^{}_{10\times 8}}&\mathcal{M}^{\chi\chi}_{\rm \chi\chi^{}_{10\times 10}}\\
\end{pmatrix}
=
\begin{pmatrix}
\mathcal{M}^{\rm HH}_{\rm HH^{}_{4\times 4}}& 0_{4\times8} &\mathcal{M}^{\rm HH}_{\rm \chi\chi^{}_{4\times 10}}\\
0_{8\times4}& \mathcal{M}^{\rm H\chi}_{\rm H\chi^{}_{8\times 8}}&0_{8\times 10}\\
\mathcal{M}^{\chi\chi}_{\rm HH^{}_{10\times 4}}& 0_{10\times 8}&\mathcal{M}^{\chi\chi}_{\rm \chi\chi^{}_{10\times 10}}\\
\end{pmatrix},
\label{mat:chiH88}
\eea
where
\begin{align}
\mathcal{M}^{\rm HH}_{\rm HH^{}_{4\times 4}} =
\begin{blockarray}{ccccc}
 & h\,h/\sqrt{2} & G^0\,G^0/\sqrt{2} & G^+\,G^-&h\,G^0\\
\begin{block}{c(cccc)}
h\,h/\sqrt{2}       & 3\lambda_H & \lambda_H & \sqrt{2}\lambda_H& 0\\
G^0\,G^0/\sqrt{2}   & \lambda_H & 3\lambda_H & \sqrt{2}\lambda_H& 0\\
G^+\,G^-            & \sqrt{2}\lambda_H & \sqrt{2}\lambda_H & 4\lambda_H& 0\\
h\,G^0              & 0 & 0 & 0 & 2\lambda_H\\
\end{block}
\end{blockarray}\,,
\end{align}

\begin{align}
\mathcal{M}^{\rm H\chi}_{\rm H\chi^{}_{8\times 8}} =
\begin{blockarray}{ccccccccc}
                & h\,\chi_1 &  h\,\chi_1^* &  h\,\chi_2 &  h\,\chi_2^* & G^0\,\chi_1 &  G^0\,\chi_1^* &  G^0\,\chi_2 &  G^0\,\chi_2^*\\
\begin{block}{c(cccccccc)}
h\,\chi_1       & 0 & \lambda_{1H} & \lambda_{12H}/2 & 0 & 0 & 0 & 0 & 0\\
h\,\chi_1^*     & \lambda_{1H} & 0 & 0 & \lambda_{12H}/2 & 0 & 0 & 0 & 0\\
h\,\chi_2       & \lambda_{12H}/2 & 0 & 0 & \lambda_{2H} & 0 & 0 & 0 & 0\\
h\,\chi_2^*     & 0  & \lambda_{12H}/2 & \lambda_{2H} & 0 & 0 & 0 & 0 & 0\\
G^0\,\chi_1     & 0 & 0 & 0 & 0 & 0 & \lambda_{1H} & \lambda_{12H}/2 & 0\\
G^0\,\chi_1^*   & 0 & 0 & 0 & 0 & \lambda_{1H} & 0 & 0 & \lambda_{12H}/2\\
G^0\,\chi_2     & 0 & 0 & 0 & 0 & \lambda_{12H}/2 & 0 & 0 & \lambda_{2H}\\
G^0\,\chi_2^*   & 0 & 0 & 0 & 0 & 0  & \lambda_{12H}/2 & \lambda_{2H} & 0\\
\end{block}
\end{blockarray}\,,
\end{align}

\begin{align}
\mathcal{M}^{\rm \chi\chi}_{\rm HH^{}_{10\times 4}} =
\begin{blockarray}{ccccc}
                              & h\,h/\sqrt{2} & G^0\,G^0/\sqrt{2} & G^+\,G^-&h\,G^0\\
\begin{block}{c(cccc)}
\chi_1\,\chi_1/\sqrt{2}       &  0  &  0  &  0   &  0 \\
\chi_1^*\,\chi_1^*/\sqrt{2}   &  0  &  0  &  0   &  0 \\
\chi_2\,\chi_2/\sqrt{2}       &  0  &  0  &  0   &  0 \\
\chi_2^*\,\chi_2^*/\sqrt{2}   &  0  &  0  &  0   &  0 \\
\chi_1\,\chi_1^*    & \lambda_{1H}/\sqrt{2} & \lambda_{1H}/\sqrt{2} & \lambda_{1H} & 0  \\
\chi_2\,\chi_2^*    & \lambda_{2H}/\sqrt{2} & \lambda_{2H}/\sqrt{2} & \lambda_{2H} & 0 \\
\chi_1\,\chi_2  & \lambda_{12H}/(2\sqrt{2}) & \lambda_{12H}/(2\sqrt{2}) & \lambda_{12H}/2 & 0 \\
\chi_1^*\,\chi_2^* & \lambda_{12H}/(2\sqrt{2}) & \lambda_{12H}/(2\sqrt{2}) & \lambda_{12H}/2 & 0 \\
\chi_1\,\chi_2^*              & 0 & 0 & 0 & 0 \\
\chi_1^*\,\chi_2              & 0 & 0 & 0 & 0 \\
\end{block}
\end{blockarray}\,,
\end{align}

{\scriptsize
\begin{align}\nonumber
\mathcal{M}^{\rm \chi\chi}_{\rm \chi\chi^{}_{10\times 10}} =
\begin{blockarray}{ccccccccccc}
 & \chi_1\,\chi_1/\sqrt{2}  & \chi_1^*\,\chi_1^*/\sqrt{2}  & \chi_2\,\chi_2/\sqrt{2}  & \chi_2^*\,\chi_2^*/\sqrt{2} & \chi_1\,\chi_1^* & \chi_2\,\chi_2^* & \chi_1\,\chi_2 & \chi_1^*\,\chi_2^* & \chi_1\,\chi_2^* & \chi_1^*\,\chi_2  \\
\begin{block}{c(cccccccccc)}
\chi_1\,\chi_1/\sqrt{2}       		& 0 & 2\lambda_{\chi_1} & \lambda_{12}^{\prime} & 0 & 0 & 0 & 0 & 0 & 0 & \lambda_{3}/\sqrt{2} \\
\chi_1^*\,\chi_1^*/\sqrt{2}   	& 2\lambda_{\chi_1} & 0 & 0 & \lambda_{12}^{\prime} & 0 & 0 & 0 & 0 & \lambda_{3}/\sqrt{2} & 0\\
\chi_2\,\chi_2/\sqrt{2}       		& \lambda_{12}^{\prime} & 0 & 0 & 2\lambda_{\chi_2} & 0 & 0 & 0 & 0 & \lambda_{4}/\sqrt{2} & 0\\
\chi_2^*\,\chi_2^*/\sqrt{2}   	& 0  & \lambda_{12}^{\prime} & 2\lambda_{\chi_2} & 0 & 0 & 0 & 0 & 0 & 0 & \lambda_{4}/\sqrt{2}\\
\chi_1\,\chi_1^*              		& 0 & 0 & 0 & 0 & 4\lambda_{\chi_1} & \lambda_{12} & \lambda_{3} & \lambda_{3} & 0 & 0\\
\chi_2\,\chi_2^*              		& 0 & 0 & 0 & 0 & \lambda_{12} & 4\lambda_{\chi_2} & \lambda_{4} & \lambda_{4} & 0 & 0\\
\chi_1\,\chi_2                		& 0 & 0 & 0 & 0 & \lambda_{3} & \lambda_{4} & 2\lambda_{12}^{\prime} & \lambda_{12} & 0 & 0\\
\chi_1^*\,\chi_2^*            		& 0 & 0 & 0 & 0 & \lambda_{3} & \lambda_{4} & \lambda_{12} & 2\lambda_{12}^{\prime} & 0 & 0\\
\chi_1\,\chi_2^*              		& 0 & \dfrac{\lambda_3}{\sqrt{2}} & \dfrac{\lambda_4}{\sqrt{2}} & 0 & 0 & 0 & 0 & 0 & 0 & \lambda_{12}\\
\chi_1^*\,\chi_2              		& \dfrac{\lambda_3}{\sqrt{2}}  & 0 & 0 &  \dfrac{\lambda_4}{\sqrt{2}} & 0 & 0 & 0 & 0 & \lambda_{12} & 0\\
\end{block}
\end{blockarray}\,,
\end{align}}

\begin{gather}
\label{mat:sc}
\mathcal{M}^{\rm SC}_{6\times 6} =
\begin{blockarray}{ccccccc}
 & \big|h\,G^+\rangle & \big|G^0\,G^+\rangle & \big|\chi_1\,G^+\rangle&\big|\chi_1^*\,G^+\rangle& \big|\chi_2\,G^+\rangle&\big|\chi_2^*\,G^+\rangle\\
\begin{block}{c(cccccc)}
\big|h\,G^+\rangle              & 2\lambda_H & 0 & 0 & 0 & 0 & 0 \\
\big|G^0\,G^+\rangle            & 0 & 2\lambda_H & 0 & 0 & 0 & 0 \\
\big|\chi_1\,G^+ \rangle        & 0 & 0 & 0 & \lambda_{1H} & \lambda_{12H}/2 & 0 \\
\big|\chi_1^*\,G^+\rangle       & 0 & 0 & \lambda_{1H} & 0 & 0 & \lambda_{12H}/2 \\
\big|\chi_2\,G^+ \rangle        & 0 & 0 & \lambda_{12H}/2 & 0 & 0 & \lambda_{2H} \\
\big|\chi_2^*\,G^+\rangle       & 0 & 0 & 0 & \lambda_{12H}/2 & \lambda_{2H} & 0 \\
\end{block}
\end{blockarray}\,,
\end{gather}

\begin{align}
\mathcal{M}^{\rm DC}_{1\times 1} =
\begin{blockarray}{cc}
 &\dfrac{1}{\sqrt{2}}\big|G^+\,G^+\rangle \\
\begin{block}{c(c)}
\dfrac{1}{\sqrt{2}}\big|G^+\,G^+\rangle  &  2\lambda_{H}     \\
\end{block}
\end{blockarray}\,,
\end{align}

After determining the eigenvalues of eq\,.~\eqref{eq:unitary}, we conclude that the tree-level unitarity constraints in this setup are as follows:
\begin{gather}
|\lambda_H|\leq 4\pi\,,\\
|\lambda_{12}-2\lambda_{12}^{\prime}|\leq 8\pi\,,\\
|\lambda_{1H}+\lambda_{2H}\pm \sqrt{(\lambda_{1H}-\lambda_{2H})^2+\lambda_{12H}^2}|\leq 16\pi
\end{gather}
The other ten distinct eigenvalues, along with their corresponding unitarity constraints, can be obtained by solving the remaining eigenvalue equations of the matrix in eq\,.~\eqref{eq:unitary}. These lengthy results are omitted here, but all unitarity constraints have been carefully evaluated and numerically verified in our analysis.

\section{Relevant Feynmann diagrams for two-component DM in $\mathcal{Z}_3$ scenario}
\label{fig:z3-feynman}
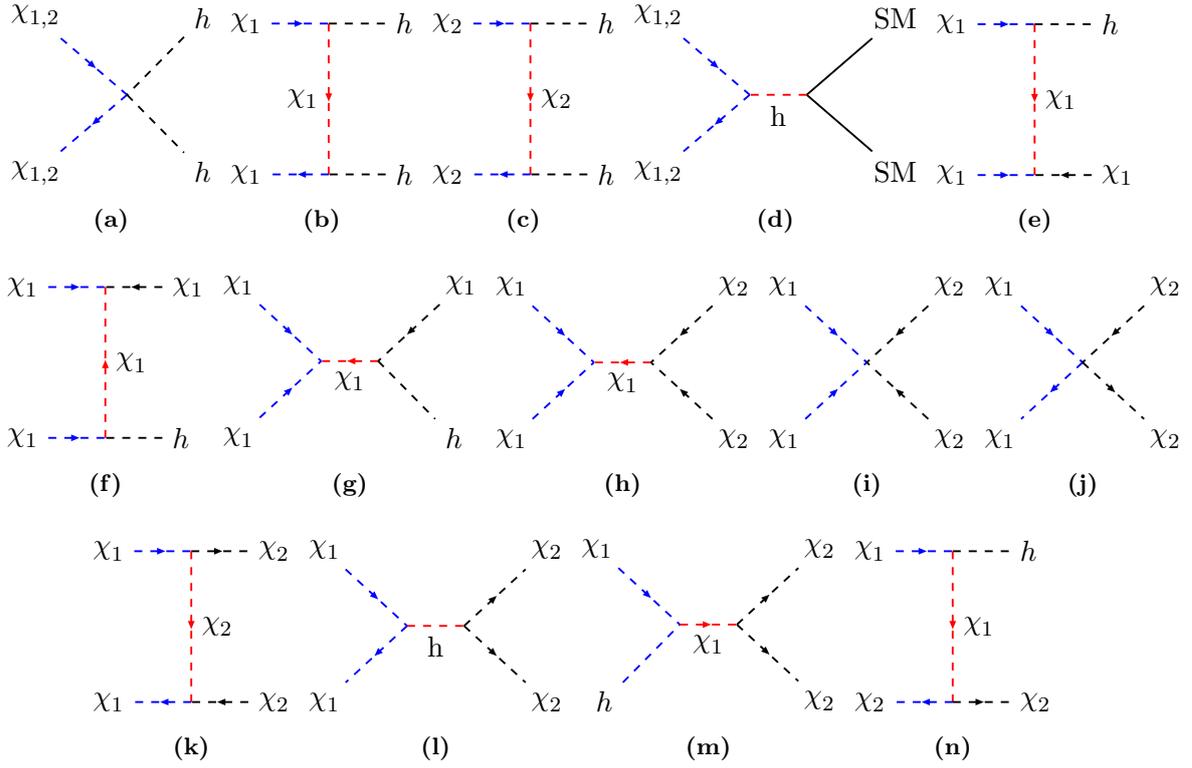
\begin{figure}[htb!]
\centering
\subfloat[]{\begin{tikzpicture}
\begin{feynman}
\vertex (a);
\vertex[above left=0.75cm and 0.75cm of a] (a1){\(\chi_{1,2}\)};
\vertex[below left=0.75cm and 0.75cm of a] (a2){\(\chi_{1,2}\)}; 
\vertex[above right=0.75cm and 0.75cm of a] (b1){\(h\)};
\vertex[below right=0.75cm and 0.75cm of a] (b2){\(h\)}; 
\diagram* {
(a1) -- [ line width=0.25mm,charged scalar, arrow size=0.7pt, style=blue] (a),
(a) -- [ line width=0.25mm,charged scalar, arrow size=0.7pt, style=blue] (a2), 
(b1) -- [ line width=0.25mm, scalar, arrow size=0.7pt, style=black] (a),
(a) -- [ line width=0.25mm, scalar, arrow size=0.7pt, style=black] (b2)};
\end{feynman}
\end{tikzpicture}\label{feyn:ann-1}}
\subfloat[]{\begin{tikzpicture}
\begin{feynman}
\vertex (a);
\vertex[left=0.75cm and 0.75cm of a] (a1){\(\chi_{1}\)};
\vertex[right=0.75cm and 0.75cm of a] (a2){\(h\)}; 
\vertex[below=2cm of a] (b); 
\vertex[left=0.75cm and 0.75cm of b] (b1){\(\chi_{1}\)};
\vertex[right=0.75cm and 0.75cm of b] (b2){\(h\)}; 
\diagram* {
(a1) -- [ line width=0.25mm,charged scalar, arrow size=0.7pt, style=blue] (a),
(a) -- [ line width=0.25mm, scalar, arrow size=0.7pt, style=black] (a2),
(a) -- [ line width=0.25mm,charged scalar, arrow size=0.7pt, edge label'={\(\rm \color{black}{\chi_{1} }\)},style=red] (b),
(b2) -- [ line width=0.25mm, scalar, arrow size=0.7pt, style=black] (b), 
(b)-- [ line width=0.25mm,charged scalar, arrow size=0.7pt, style=blue] (b1)};
\end{feynman}
\end{tikzpicture}\label{feyn:ann-2}}
\subfloat[]{\begin{tikzpicture}
\begin{feynman}
\vertex (a);
\vertex[left=0.75cm and 0.75cm of a] (a1){\(\chi_{2}\)};
\vertex[right=0.75cm and 0.75cm of a] (a2){\(h\)}; 
\vertex[below=2cm of a] (b); 
\vertex[left=0.75cm and 0.75cm of b] (b1){\(\chi_{2}\)};
\vertex[right=0.75cm and 0.75cm of b] (b2){\(h\)}; 
\diagram*{
(a1) -- [ line width=0.25mm,charged scalar, arrow size=0.7pt, style=blue] (a),
(a) -- [ line width=0.25mm, scalar, arrow size=0.7pt, style=black] (a2), 
(a) -- [ line width=0.25mm,charged scalar, arrow size=0.7pt, edge label={\(\rm \color{black}{\chi_{2}}\)},style=red] (b),
(b2) -- [ line width=0.25mm, scalar, arrow size=0.7pt, style=black] (b), 
(b)-- [ line width=0.25mm,charged scalar, arrow size=0.7pt, style=blue] (b1)};
\end{feynman}
\end{tikzpicture}\label{feyn:ann-3}}
\subfloat[]{\begin{tikzpicture}
\begin{feynman}
\vertex (a);
\vertex[above left=0.75cm and 0.75cm of a] (a1){\(\chi_{1,2}\)};
\vertex[below left=0.75cm and 0.75cm of a] (a2){\(\chi_{1,2}\)}; 
\vertex[right=0.75cm of a] (b); 
\vertex[above right=0.75cm and 0.75cm of b] (b1){\(\rm SM\)};
\vertex[below right=0.75cm and 0.75cm of b] (b2){\(\rm SM\)}; 
\diagram* {(a1) -- [ line width=0.25mm,charged scalar, arrow size=0.7pt, style=blue] (a) -- [ line width=0.25mm,charged scalar, arrow size=0.7pt, style=blue] (a2), (b1) -- [ line width=0.25mm, plain, arrow size=0.7pt, style=black] (b)-- [ line width=0.25mm, plain, arrow size=0.7pt, style=black] (b2), (b) -- [ line width=0.25mm, scalar, arrow size=0.7pt, edge label={\(\rm \color{black}{h}\)},style=red] (a) };
\end{feynman}
\end{tikzpicture}\label{feyn:ann-4}}
\subfloat[]{\begin{tikzpicture}
\begin{feynman}
\vertex (a);
\vertex[left=0.75cm and 0.75cm of a] (a1){\(\chi_1\)};
\vertex[right=0.75cm and 0.75cm of a] (a2){\(h\)}; 
\vertex[below=2cm of a] (b); 
\vertex[left=0.75cm and 0.75cm of b] (b1){\(\chi_1\)};
\vertex[right=0.75cm and 0.75cm of b] (b2){\(\chi_1\)}; 
\diagram* {(a1) -- [ line width=0.25mm,charged scalar, arrow size=0.7pt, style=blue] (a)-- [ line width=0.25mm, scalar, arrow size=0.7pt, style=black] (a2), (a) -- [ line width=0.25mm,charged scalar, arrow size=0.7pt, edge label={\(\rm \color{black}{\chi_1 }\)},style=red] (b),(b2) -- [ line width=0.25mm,charged scalar, arrow size=0.7pt, style=black] (b), (b1)-- [ line width=0.25mm,charged scalar, arrow size=0.7pt, style=blue] (b)};
\end{feynman}
\end{tikzpicture}\label{feyn:semi-1}}

\subfloat[]{\begin{tikzpicture}
\begin{feynman}
\vertex (a);
\vertex[left=0.75cm and 0.75cm of a] (a1){\(\chi_1\)};
\vertex[right=0.75cm and 0.75cm of a] (a2){\(\chi_1\)}; 
\vertex[below=2cm of a] (b); 
\vertex[left=0.75cm and 0.75cm of b] (b1){\(\chi_1\)};
\vertex[right=0.75cm and 0.75cm of b] (b2){\(h\)}; 
\diagram* {(a1) -- [ line width=0.25mm,charged scalar, arrow size=0.7pt, style=blue] (a), (a2) -- [ line width=0.25mm,charged scalar, arrow size=0.7pt, style=black] (a) , (b) -- [ line width=0.25mm,charged scalar, arrow size=0.7pt, edge label'={\(\rm \color{black}{\chi_1 }\)},style=red] (a), (b2) -- [ line width=0.25mm,scalar, arrow size=0.7pt, style=black] (b), (b1)-- [ line width=0.25mm,charged scalar, arrow size=0.7pt, style=blue] (b)};
\end{feynman}
\end{tikzpicture}\label{feyn:semi-2}}
\subfloat[]{\begin{tikzpicture}
\begin{feynman}
\vertex (a);
\vertex[above left=0.75cm and 0.75cm of a] (a1){\(\chi_1\)};
\vertex[below left=0.75cm and 0.75cm of a] (a2){\(\chi_1\)}; 
\vertex[right=0.75cm of a] (b); 
\vertex[above right=0.75cm and 0.75cm of b] (b1){\(\chi_1\)};
\vertex[below right=0.75cm and 0.75cm of b] (b2){\(h\)}; 
\diagram* {(a1) -- [ line width=0.25mm,charged scalar, arrow size=0.7pt, style=blue] (a), (a2) -- [ line width=0.25mm,charged scalar, arrow size=0.7pt, style=blue] (a), (b) -- [ line width=0.25mm,charged scalar, arrow size=0.7pt, edge label={\(\rm \color{black}{\chi_1}\)},style=red] (a) , (b1) -- [ line width=0.25mm,charged scalar, arrow size=0.7pt, style=black] (b)-- [ line width=0.25mm, scalar, arrow size=0.7pt, style=black] (b2)};
\end{feynman}
\end{tikzpicture}\label{feyn:semi-3}}
\subfloat[]{\begin{tikzpicture}
\begin{feynman}
\vertex (a);
\vertex[above left=0.75cm and 0.75cm of a] (a1){\(\chi_1\)};
\vertex[below left=0.75cm and 0.75cm of a] (a2){\(\chi_1\)}; 
\vertex[right=0.75cm of a] (b); 
\vertex[above right=0.75cm and 0.75cm of b] (b1){\(\chi_2\)};
\vertex[below right=0.75cm and 0.75cm of b] (b2){\(\chi_2\)}; 
\diagram* {(a1) -- [ line width=0.25mm,charged scalar, arrow size=0.7pt, style=blue] (a), (a2) -- [ line width=0.25mm,charged scalar, arrow size=0.7pt, style=blue] (a), (b) -- [ line width=0.25mm,charged scalar, arrow size=0.7pt, edge label={\(\rm \color{black}{\chi_1}\)},style=red] (a) , (b1) -- [ line width=0.25mm,charged scalar, arrow size=0.7pt, style=black] (b), (b2)-- [ line width=0.25mm, charged scalar, arrow size=0.7pt, style=black] (b)};
\end{feynman}
\end{tikzpicture}\label{feyn:conv-1}}
\subfloat[]{\begin{tikzpicture}
\begin{feynman}
\vertex (a);
\vertex[above left=0.75cm and 0.75cm of a] (a1){\(\chi_1\)};
\vertex[below left=0.75cm and 0.75cm of a] (a2){\(\chi_1\)}; 
\vertex[above right=0.75cm and 0.75cm of a] (b1){\(\chi_2\)};
\vertex[below right=0.75cm and 0.75cm of a] (b2){\(\chi_2\)}; 
\diagram* {(a1) -- [ line width=0.25mm,charged scalar, arrow size=0.7pt, style=blue] (a), (a2) -- [ line width=0.25mm,charged scalar, arrow size=0.7pt, style=blue] (a), (b1) -- [ line width=0.25mm, charged scalar, arrow size=0.7pt, style=black] (a),(b2) -- [ line width=0.25mm, charged scalar, arrow size=0.7pt, style=black] (a)};
\end{feynman}
\end{tikzpicture}\label{feyn:conv-2}}
\subfloat[]{\begin{tikzpicture}
\begin{feynman}
\vertex (a);
\vertex[above left=0.75cm and 0.75cm of a] (a1){\(\chi_1\)};
\vertex[below left=0.75cm and 0.75cm of a] (a2){\(\chi_1\)}; 
\vertex[above right=0.75cm and 0.75cm of a] (b1){\(\chi_2\)};
\vertex[below right=0.75cm and 0.75cm of a] (b2){\(\chi_2\)}; 
\diagram* {(a1) -- [ line width=0.25mm,charged scalar, arrow size=0.7pt, style=blue] (a) -- [ line width=0.25mm,charged scalar, arrow size=0.7pt, style=blue] (a2), (b1) -- [ line width=0.25mm, charged scalar, arrow size=0.7pt, style=black] (a) -- [ line width=0.25mm, charged scalar, arrow size=0.7pt, style=black] (b2)};
\end{feynman}
\end{tikzpicture}\label{feyn:conv-3}}

\subfloat[]{\begin{tikzpicture}
\begin{feynman}
\vertex (a);
\vertex[ left=0.75cm and 0.75cm of a] (a1){\(\chi_1\)};
\vertex[ right=0.75cm and 0.75cm of a] (a2){\(\chi_2\)}; 
\vertex[below=2cm of a] (b); 
\vertex[ left=0.75cm and 0.75cm of b] (b1){\(\chi_1\)};
\vertex[ right=0.75cm and 0.75cm of b] (b2){\(\chi_2\)}; 
\diagram* {(a1) -- [ line width=0.25mm,charged scalar, arrow size=0.7pt, style=blue] (a), (a) -- [ line width=0.25mm,charged scalar, arrow size=0.7pt, style=black] (a2) , (a) -- [ line width=0.25mm,charged scalar, arrow size=0.7pt, edge label={\(\rm \color{black}{\chi_2 }\)},style=red] (b), (b2) -- [ line width=0.25mm,charged scalar, arrow size=0.7pt, style=black] (b), (b)-- [ line width=0.25mm,charged scalar, arrow size=0.7pt, style=blue] (b1)};
\end{feynman}
\end{tikzpicture}\label{feyn:conv-4}}
\subfloat[]{\begin{tikzpicture}
\begin{feynman}
\vertex (a);
\vertex[above left=0.75cm and 0.75cm of a] (a1){\(\chi_1\)};
\vertex[below left=0.75cm and 0.75cm of a] (a2){\(\chi_1\)}; 
\vertex[right=0.75cm of a] (b); 
\vertex[above right=0.75cm and 0.75cm of b] (b1){\(\chi_2\)};
\vertex[below right=0.75cm and 0.75cm of b] (b2){\(\chi_2\)}; 
\diagram* {
(a1) -- [ line width=0.25mm,charged scalar, arrow size=0.7pt, style=blue] (a), 
(a) -- [ line width=0.25mm,charged scalar, arrow size=0.7pt, style=blue] (a2),
(a) -- [ line width=0.25mm, scalar, arrow size=0.7pt, edge label'={\(\rm \color{black}{h}\)},style=red] (b) , 
(b) -- [ line width=0.25mm,charged scalar, arrow size=0.7pt, style=black] (b1), 
(b)-- [ line width=0.25mm,charged scalar, arrow size=0.7pt, style=black] (b2)};
\end{feynman}
\end{tikzpicture}\label{feyn:conv-5}}
\subfloat[]{\begin{tikzpicture}
\begin{feynman}
\vertex (a);
\vertex[above left=0.75cm and 0.75cm of a] (a1){\(\chi_1\)};
\vertex[below left=0.75cm and 0.75cm of a] (a2){\(h\)}; 
\vertex[right=0.75cm of a] (b); 
\vertex[above right=0.75cm and 0.75cm of b] (b1){\(\chi_2\)};
\vertex[below right=0.75cm and 0.75cm of b] (b2){\(\chi_2\)}; 
\diagram* {(a1) -- [ line width=0.25mm,charged scalar, arrow size=0.7pt, style=blue] (a), (a2) -- [ line width=0.25mm,scalar, arrow size=0.7pt, style=blue] (a), (a) -- [ line width=0.25mm,charged scalar, arrow size=0.7pt, edge label'={\(\rm \color{black}{\chi_1}\)},style=red] (b) , (b) -- [ line width=0.25mm,charged scalar, arrow size=0.7pt, style=black] (b1), (b)-- [ line width=0.25mm,charged scalar, arrow size=0.7pt, style=black] (b2)};
\end{feynman}
\end{tikzpicture}\label{feyn:conv-6}}
\subfloat[]{\begin{tikzpicture}
\begin{feynman}
\vertex (a);
\vertex[left=0.75cm and 0.75cm of a] (a1){\(\chi_1\)};
\vertex[right=0.75cm and 0.75cm of a] (a2){\(h\)}; 
\vertex[below=2cm of a] (b); 
\vertex[left=0.75cm and 0.75cm of b] (b1){\(\chi_2\)};
\vertex[right=0.75cm and 0.75cm of b] (b2){\(\chi_2\)}; 
\diagram* {(a1) -- [ line width=0.25mm,charged scalar, arrow size=0.7pt, style=blue] (a)-- [ line width=0.25mm, scalar, arrow size=0.7pt, style=black] (a2), (a) -- [ line width=0.25mm,charged scalar, arrow size=0.7pt, edge label={\(\rm \color{black}{\chi_1 }\)},style=red] (b),(b) -- [ line width=0.25mm,charged scalar, arrow size=0.7pt, style=black] (b2), (b)-- [ line width=0.25mm,charged scalar, arrow size=0.7pt, style=blue] (b1)};
\end{feynman}
\end{tikzpicture}\label{feyn:conv-7}}
\caption{The Feynman diagrams, \cref{feyn:ann-1,feyn:ann-2,feyn:ann-3,feyn:ann-4}, represents the self-annihilation of $\chi_1$ and $\chi_2$ DM where $\rm SM=\{Higgs,~quarks,~leptons,~W^{\pm}~and~Z~boson\}$. The Feynman diagrams, \cref{feyn:semi-1,feyn:semi-2,feyn:semi-3}, \cref{feyn:conv-1,feyn:conv-2,feyn:conv-3,feyn:conv-4,feyn:conv-5}, and \cref{feyn:conv-6,feyn:conv-7} represent the semi-annihilation, conversion and semi-conversion channels of $\chi_1$ relevant for Scenario-A and B.}
\label{fig:z3-feynman3}
\end{figure}

\newpage
\bibliographystyle{JHEP}
\bibliography{Z3-Z2}
\end{document}